\documentclass[aps,pre,11pt,a4paper,nofootinbib]{revtex4-1}

\usepackage{graphicx}
\usepackage{times}
\usepackage{amssymb}
\usepackage{amsmath}
\usepackage{subfig}
\usepackage{stmaryrd}
\usepackage{hyperref}
\usepackage{cleveref}

\newcommand{\nn}{\nonumber}
\newcommand{\vphi}{\varphi}
\newcommand{\veps}{\varepsilon}
\newcommand{\ea}{eqnarray}

\begin{document}

\title{Recent advances in liquid mixtures in electric fields}

\author{Yael Katsir}

\author{Yoav Tsori}
\affiliation{Department of Chemical Engineering and the Ilse Katz Institute for 
Nanoscale
Science and Technology, Ben-Gurion University of the Negev, 84105 Beer-Sheva, 
Israel.}

\date{\today}

\begin{abstract}
When immiscible liquids are subject to electric fields interfacial forces 
arise due to a difference in the permittivity or the conductance 
of the liquids, and these forces lead to shape change in droplets or to 
interfacial instabilities.
In this Topical Review we discuss recent advances in the theory and experiments of 
liquids in electric fields with an emphasis on liquids which are initially miscible and 
demix under the influence of an external field. In purely dielectric liquids
demixing occurs if the electrode geometry leads to sufficiently large field gradients. In 
polar liquids field gradients are prevalent due to screening by dissociated ions 
irrespective of the electrode geometry. We examine the conditions for these ``electro 
prewetting'' transitions and highlight few possible systems where they might be 
important, such as in stabilization of colloids and in gating of pores in membranes. 

\end{abstract}

\maketitle
\tableofcontents

\section{Introduction} \label{chap1}

Electrostatic forces are ubiquitous and their effect is important in many soft matter 
systems involving liquids bounded by hard or soft walls. They arise on purpose and are 
easily controlled when water or other solvents flow in microfluidics channels in contact 
with a metallic electrode whose potential is externally controlled. They are less easily 
controlled when the solvent is nearby a charged nonmetallic surface which can induce or 
impede the flow. In biological settings electrostatic forces determine whether proteins 
or other molecules bind to other molecules or to cellular structure which are often 
charged. When colloids are suspended in solvents the competition between entropic and 
electrostatic forces may lead to inter-colloidal attraction and eventually to coagulation 
and sedimentation of the colloids, or to repulsion between the colloids and to 
stabilization of the suspension. The interplay between shear forces, surface tension and 
electrostatic also plays a vital role in many industrial process where liquid droplets 
are 
transported and ejected via small orifices, as occurs for example in pesticide 
spraying in agriculture or in ink-jet printing. 

This paper gives a concise overview of interfacial instabilities that occur when electric 
fields are applied in a direction perpendicular to an initially flat interface between 
two liquids. Sec. \ref{chap2_dielectric} discusses this normal-field instability 
in purely dielectric liquids where the electrostatic forces destabilizing the interface 
are proportional to the difference between the liquids' permittivities squared. The 
situation is more complex when residual conductivity exists in the liquid phases and in 
this case mobile dissociated ions exert shear forces on the interface and modify its 
shape, Sec. \ref{chap2_leaky_dielectric}.

Section \ref{chap3} then poses a more fundamental question: what if electric fields could 
affect the relative miscibility of the two liquids? Namely, not only alter the interface 
but destroy it? Sec. \ref{chap3_landau} shows the Landau theory that addressed 
this question and proved that indeed such possibility exists. The experiments supporting 
and contradicting the Landau theory are summarized in Sec. 
\ref{chap3_experiments}.

Section \ref{chap4} goes one step further and examines situations where electric field 
gradients act on dielectric liquids. In these systems a dielectrophoretic force 
acts on the liquids and, if strong enough, it may lead to demixing of the liquids from 
each other. In those cases the shape, size and location of the electrodes producing the 
fields are crucial for the understanding of the statics and dynamics of the phase 
transitions. Peculiarly, an interfacial instability exists where 
the electric field stabilizes the interface while surface tension destabilizes it, in 
contrast to the normal-field instability of Secs. \ref{chap2_dielectric} and 
\ref{chap2_leaky_dielectric}. New experimental results of phase separation dynamics and 
equilibrium are shown and analyzed.

Demixing occurs also in mixtures of polar solvents, but this time due to screening of the 
field which always exist irrespective of the electrodes. The ``electro-prewetting'' 
transitions described in Sec. \ref{chap5} have a specific dependence on the salt content, 
temperature and relative composition of the mixture. The relative miscibility of the ions 
in the solvents plays a crucial role.

After surveying the basic physical concepts two ``applications'' are considered: Sec. 
\ref{chap6_colloids} gives an account of the electrostatic and van der Waals forces 
between two 
colloids immersed in a polar solution. The section details the complex interplay 
between these forces that depends on the relative adsorption of the liquids at the 
surface of the colloids, in addition to the temperature and mixture composition. Contrary 
to the regular Derjaguin, Landau, Verwey, and Overbeek (DLVO) behavior in simple 
liquids here the addition of ions leads to a  repulsion between the colloids in a 
certain window of parameters. Sec. 
\ref{chap7_pore} considers another situation where polar liquids are found in contact 
with hard surfaces: porous membranes. Pore gating between two states can be achieved by 
controlling the surface potential of the membrane. This gating of membranes to small 
molecules by external potentials could be advantageous over other methods. Finally Sec. 
\ref{chap8} is a summary and outlook.

\section{Force and stress in liquids in electric fields}

When a liquid is placed under the influence of an electric field ${\bf E}$ stress 
develops. This stress originates from the electrostatic free energy density $-(1/2)
{\bf E}\cdot{\bf D}$, where ${\bf D}=\veps{\bf E}$ is the displacement field and $\veps$ 
is the local dielectric constant. Due to the vectorial nature of the field, 
the stress $\overleftrightarrow{T}$ is 
tensorial. For a unit surface whose normal is $\hat{n}$, 
the force acting on that surface is given by $-\overleftrightarrow{T}\cdot\hat{n}$ (the 
i'th component is
$-\overleftrightarrow{T}_{\rm ij}n_j$ where we have used the summation convention on the 
index $j$). 
The electric field has diagonal and non-diagonal contributions to the stress tensor 
\cite{LL_book_elec,melcher_book,saville_arfm_1997}
\begin{equation}\label{stress_tensor}
\overleftrightarrow{T}_{\rm ij}=-p_0(c,T)\delta_{\rm ij}+\frac12 
\veps E^2\left(-1+\frac{c}{\veps}\left(\frac{\partial\veps}{\partial 
c}\right)_T\right)\delta_{\rm ij}+\veps E_iE_j
\end{equation}
Here $p_0(c,T)$ is the equation of state of the liquid in the absence of field, where 
$c$ is the density and $T$ is the temperature. In fluids the ``regular'' pressure has a 
diagonal contribution to $\overleftrightarrow{T}_{\rm ij}$. 

The body force ${\bf f}$ is given as a divergence of the this stress:
$f_i=\partial \overleftrightarrow{T}_{\rm ij}/\partial x_j$, and is given by 
\begin{\ea}\label{body_force}
{\bf f}=-\nabla p_0+\frac12\nabla\left(E^2 c\frac{\partial \veps}{\partial 
c}\right)_T-\frac12E^2\nabla\veps+\rho{\bf E}
\end{\ea}
where $\rho$ is the charge density. The second and third terms describe electrostriction 
and dielectrophoretic forces whereas the last term reflects the force that is transferred 
to the liquid by free moving charges. 

The discontinuity of the normal field across the interface is obtained as 
\begin{\ea}\label{bc_D}
\llbracket {\bf D}\rrbracket\cdot\hat{n}=\sigma
\end{\ea}
where $\llbracket {\bf D}\rrbracket\equiv {\bf D}^{(2)}-{\bf D}^{(1)}$ is the 
discontinuity of the displacement field across the interface, $\sigma$ is the surface 
charge density, and the surface unit vector $\hat{n}$ points from region 1 to region 2.
The continuity of the tangential field across the interface is given by 
\begin{\ea}\label{bc_E}
\llbracket {\bf E}\rrbracket\cdot\hat{t_i}=0
\end{\ea}
where $\hat{t}_i$ ($i=1$, $2$) are the two orthogonal unit vector lying in the plane of 
the interface. 
At the interface between two regions of different permittivity the force is 
discontinuous. The i'th component of the net force per unit area of the interface, ${\bf 
f}_s$, is given by
\begin{\ea}
{\bf f}_{{\rm s,}i}=\llbracket \overleftrightarrow{T}_{\rm ij}\rrbracket n_j
\end{\ea}
where $\llbracket \overleftrightarrow{T}_{\rm ij}\rrbracket = \overleftrightarrow{T}_{\rm 
ij}^{(2)}-\overleftrightarrow{T}_{\rm ij}^{(1)}$. When the isotropic parts of the force 
can be neglected 
[first and second terms in Eq. (\ref{body_force})], the electric field bisects the angel 
between $\hat{n}$ and the direction
of the resultant force acting on the surface. This can be seen by choosing the 
x-axis to be parallel to ${\bf E}$ and by noting that $\hat{\bf f}_{\rm s}\cdot\hat{\bf 
E}$ equals $\hat{\bf f}_{\rm s}\cdot\hat{n}$ \cite{panofsky_phillips_book}.

The net force per unit area has three components: one in the direction 
perpendicular to the surface (parallel to $\hat{n}$) and two in directions parallel to 
$\hat{t}_i$. They are \cite{saville_arfm_1997}
\begin{\ea}\label{stress_components}
\llbracket \overleftrightarrow{T}\cdot\hat{n}\rrbracket \cdot\hat{n}&=&\frac12\llbracket 
\left({\bf 
E}\cdot\hat{n}\right)^2-\left({\bf E}\cdot\hat{t}_1\right)^2-\left({\bf 
E}\cdot\hat{t}_2\right)^2-p_0+c\frac{\partial\veps}{\partial c}E^2\rrbracket\nn\\
\llbracket \overleftrightarrow{T}\cdot\hat{n}\rrbracket \cdot\hat{t}_i&=&\sigma{\bf 
E}\cdot\hat{t}_i~,~~~i=1,2
\end{\ea}
In the second equation we used Eqs. (\ref{bc_D}) and  (\ref{bc_E}).

A body force induces flow in the liquid. The 
Navier-Stokes equation for the the flow velocity ${\bf u}$ in incompressible 
liquids is
\begin{\ea}\label{navier_stokes}
c\left[\frac{\partial{\bf u}}{\partial t}+\left({\bf u}\cdot\nabla\right){\bf 
u}\right]=-\nabla p_0+\frac12\nabla\left(E^2 c\frac{\partial \veps}{\partial 
c}\right)_T-\frac12E^2\nabla\veps+\rho{\bf E}+\eta\nabla^2{\bf u}.
\end{\ea}
Here $\eta$ is the fluid's viscosity and the $i$th component of $\nabla^2{\bf u}$ is 
$\nabla^2u_i$. The term $\left({\bf u}\cdot\nabla\right){\bf u}$ is a vector
whose $i$th component is ${\bf u}\cdot\nabla u_i$.

\subsection{Normal field instability in two immiscible dielectric 
liquids}\label{chap2_dielectric}

\begin{figure}[!tb]
\includegraphics[width=12cm,viewport=100 590 520 725,clip]{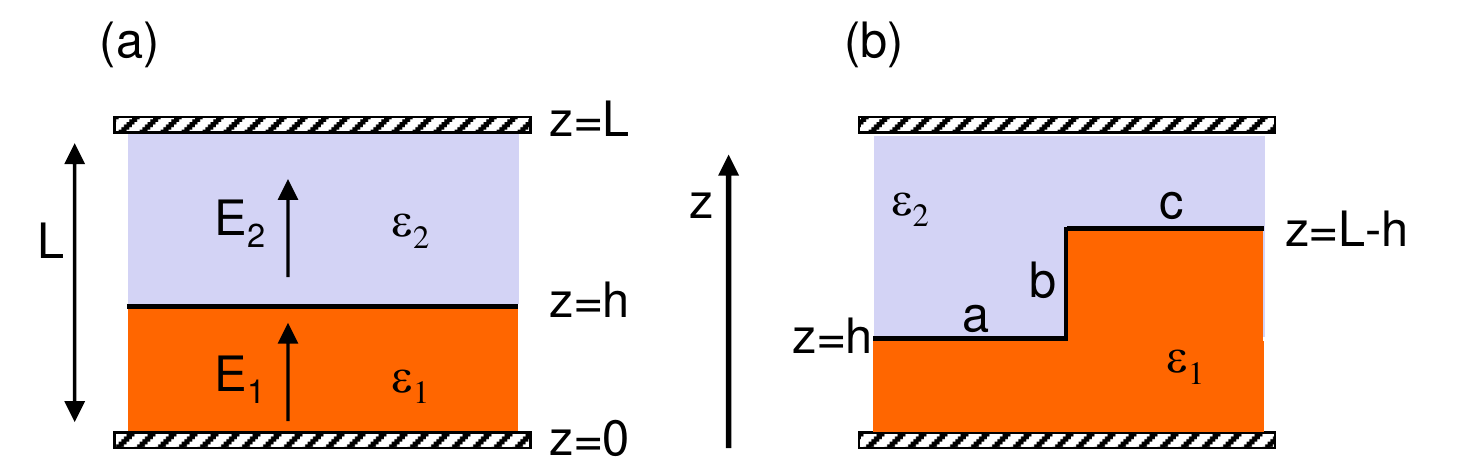}
\caption{\small Two liquids in electric field. (a) Schematic illustration of a bilayer of 
two 
liquids 1 and 2 with permittivities $\veps_1$ and $\veps_2$, respectively, confined by a 
parallel-plates capacitor whose plates are at $z=0$ and $z=L$. The thickness of the
liquid layers are $h$ and $L-h$. The fields ${\bf E}_1$ and ${\bf E}_2$ are oriented in 
the $z$-direction. (b) Idealized configuration where the bilayer breaks into two parts, 
with small (left side) and large (right side) values of $h$. `a', `b', and `c' mark the 
three interfaces.
}
\label{fig_liquid_interfaces}
\end{figure}

Let us illustrate the force and stress in a simple example -- a bilayer of two purely 
dielectric liquids, 1 and 2, with dielectric constants $\veps_1$ and $\veps_2$, 
respectively, sandwiched inside a parallel-plate capacitor, see 
Fig.~\ref{fig_liquid_interfaces}a. The 
distance between the plates is $L$ and the thickness of the first liquid is $h$. In this 
geometry the electric fields ${\bf E}_1$ and ${\bf E}_2$ are oriented in the 
$z$-direction 
and are constant within the two regions. They are found from the boundary conditions on 
the interface $\veps_1E_1=\veps_2E_2$ (Eq. (\ref{bc_D}) with $\sigma=0$) and from 
$E_1h+E_2(L-h)=E_0L$, where $E_0$ is the 
average electric field imposed by the capacitor. One thus finds that
\begin{\ea}\label{E1E2}
{\bf E}_1&=&\frac{\veps_2 E_0}{\veps_1(1-h/L)+\veps_2 h/L}\hat{z}~,~~~~~~{\bf 
E}_2=\frac{\veps_1 E_0}{\veps_1(1-h/L)+\veps_2 h/L}\hat{z}~.
\end{\ea} 
From Eq. (\ref{stress_tensor}), when $\partial\veps/\partial c=0$ the stresses just 
``below'' and just ``above'' the 
interface, $\overleftrightarrow{T}_{\rm zz}^{(1)}$ and $\overleftrightarrow{T}_{\rm 
zz}^{(2)}$, respectively, are then 
given by
\begin{\ea}
\overleftrightarrow{T}_{\rm zz}^{(1)}=\frac12\frac{\veps_1\veps_2^2 
E_0^2}{\left(\veps_1(1-h/L)+\veps_2 h/L\right)^2}~,~~~~~~~
\overleftrightarrow{T}_{\rm zz}^{(2)}=\frac12\frac{\veps_2\veps_1^2 
E_0^2}{\left(\veps_1(1-h/L)+\veps_2 h/L\right)^2}
\end{\ea} 
Since these stresses are constant throughout the bulk of the liquids there is no body 
force of electrostatic origin. The difference 
\begin{\ea}\label{Dsigma}
\overleftrightarrow{T}_{\rm zz}^{(2)}-\overleftrightarrow{T}_{\rm 
zz}^{(1)}=-\frac12\frac{\veps_1\veps_2 \Delta\veps E_0^2}{\left(\veps_1+\Delta\veps 
h/L\right)^2}  
\end{\ea}
gives the net stress on the interface. Here we used $\Delta\veps\equiv 
\veps_2-\veps_1$. If $\Delta\veps$ is positive the interface is pushed downwards so as to 
decrease $h$, if $\Delta\veps$ is negative then the interface is pushed upwards.

Under sufficiently large electric field an interfacial instability may occur and this can 
be seen as follows. Assume the bilayer divides into two parts, one with small 
value of $h$ and one with a large value, as is depicted in 
Fig.~\ref{fig_liquid_interfaces}b. In this 
idealized picture all three interfaces, marked by `a', `b', `c', are either parallel or 
perpendicular to the electrodes. Far from interface `b' the fringe field can be 
ignored and the field is still in the $z$-direction. In each domain 
the expressions for the fields stay the same as in Eq. (\ref{E1E2}). 

As Eq. (\ref{Dsigma}) shows the stress is largest when $h$ is smallest; for 
incompressible liquids this means that if the interface `a'
pushes downwards interface `c' will ``cede'' and will 
move upwards to conserve the volume of liquid 1. The conclusion is that the interface 
illustrated in Fig.~\ref{fig_liquid_interfaces}b is not stable; in the long time the 
film will be 
divided to two liquid domains 1 and 2 with an interface perpendicular to the electrodes 
(and parallel to the field). In this equilibrium state the fields in both liquids are
equal: ${\bf E}_1={\bf E}_2=E_0\hat{z}$. The stress tensor $\overleftrightarrow{T}_{\rm 
ij}$ from Eq. 
(\ref{stress_tensor}) is then diagonal and continuous across the interface, hence no 
net surface force acts to displace the interface. 

At early times the destabilization of an initially flat interface is characterized by a 
fastest growing $q$-mode modulation of the surface. Let $h(x,t)$ be the thickness 
of the layer of the first liquid and for simplicity assume the second liquid is gas. For 
thin films a Poiseuille flow is assumed where the $x$-component of the flow velocity 
vanishes at $z=0$ and is maximal at $z=h$. The integration of ${\bf u}$ in Eq. 
(\ref{navier_stokes}) along the $z$ coordinate gives a flux
\begin{\ea}\label{flux}
-\frac{h^3}{3\eta}\frac{\partial p}{\partial x}
\end{\ea} 
The pressure has three contributions \cite{hermin_prl_1999}: one is the disjoining 
pressure given by $A/6h^3$ 
due to van der Waals forces, where $A$ is the effective Hamaker constant of the system, 
the second occurs in curved interfaces where surface tension plays a role: $-\gamma 
h''(x)$, where $\gamma$ is the surface tension between the two layers. The third 
contribution to the pressure is electrostatic. 

At the initial destabilization state the 
interface is only weakly perturbed and $h(x)$ can be written as $h(x)=h_0+\delta h(x,t)$, 
where $h_0$ is the average film thickness and $\delta h\ll h_0$ is the small 
spatially-dependent perturbation growing in time. In the long wavelength approximation 
$\delta h'\ll 1$ and to lowest (linear) order in $\delta h$ one can write the pressure as
\begin{\ea}\label{pressure} 
p(x)=-\frac{A}{2h_0^4}\delta h-\gamma \delta 
h''-\frac{\veps_1\veps_2\left(\Delta\veps\right)^2E_0^2}{L\left(\veps_1+\Delta\veps 
h_0/L\right)^3}\delta h+{\rm const.}
\end{\ea} 
The set of equations for $\delta h$ is complete when one uses Eq. (\ref{pressure}) and 
Eq. (\ref{flux}) together with the ``continuity'' equation for $h$: $\partial h/\partial 
t+\partial \left(-h^3/(3\eta)\partial p/\partial x\right)/\partial x=0$. In 
this linear approximation one may substitute a 
sinusoidal ansatz with q-number $q$ and growth rate $\omega$: $\delta h=e^{iqx+\omega t}$ 
to obtain the dispersion relation between $\omega$ and $q$:
\begin{\ea}\label{dispersion}
\omega(q)=\frac{\gamma h_0^3}{3\eta}\left(\xi_e^{-2}q^2-q^4\right)
\end{\ea} 
where 
\begin{\ea}\label{healing_length}
\xi_e^{-2}=\frac{A}{2\gamma 
h_0^4}+\frac{\veps_1\veps_2\left(\Delta\veps\right)^2E_0^2}{
\gamma L\left(\veps_1+\Delta\veps 
h_0/L\right)^3}
\end{\ea} 
is the healing length having two contributions, from van der Waals and from 
electrostatics, both weighed against surface tension 
\cite{hermin_prl_1999,russel_jnnfm_2002}. 

In Eq. (\ref{dispersion}) 
the dependence of $\omega$ on $q$ has a positive contribution scaling as $q^2$ and a 
negative contribution proportional to $-q^4$ and thus for small $q$ values
$\omega(q)$ is
positive and increases with increasing $q$. The growth rate $\omega(q)$ for all $q$'s 
smaller than $\xi_e^{-1}$ is positive and they are unstable; modulations with large 
enough $q$'s, $q>\xi_e^{-1}$, are stable and diminish exponentially with time.
The fastest growing $q$-mode obeys 
$\partial\omega (q)/\partial q=0$ and hence
\begin{\ea}\label{fastest_growing}
q_{\rm fastest}=\frac{1}{\sqrt{2}} \xi_e^{-1}~,~~~~~~\omega_{\rm fastest}=\frac{\gamma 
h_0^3}{12\eta}\xi_e^{-4}~.
\end{\ea} 

Which of the two forces is more dominant, the dispersion or 
electrostatic force? The van der Waals pressure scales as $A/h_0^3$ whereas the 
electrostatic pressure is $\sim \veps E_0^2$. If we take the field to be $E_0\simeq 
1$V/$\mu$m, $\veps\simeq 
\veps_0$ ($\veps_0$ is the vacuum permittivity), and $A\simeq 10^{-20}$J we find 
that for film thicknesses $h_0$ larger than $\sim 10$nm the electrostatic force is the 
dominant force. For such relatively thick films the pattern period $2\pi/q_{\rm fastest}$ 
observed in experiments scales as $\gamma^{1/2}/(\Delta\veps E_0)$ and can be thus 
reduced if the surface tension is decreased or if the ``dielectric contrast'' 
$\Delta\veps$ or electric fields are increased 
\cite{russell_nature_2000,russel_mm_2011,russell_jcp_2001,russell_mm_2002a,
russel_afm_2006,steiner_small_2010}.

\subsection{The role of a small residual conductivity}\label{chap2_leaky_dielectric}

In the classical experiments with liquid droplets embedded in an immiscible liquid under 
the 
influence of an external field the droplets elongated in the direction of the field, as 
expected \cite{okonski_jpc_1953,mason_prsl_1962}. In some cases, however, droplets became 
oblate rather than prolate. Taylor and Melcher realized the importance of shear 
stress due to a small number 
of dissolved ions \cite{taylor_prsl_1966,taylor_arfd_1969}. In perfectly conducting 
liquids the electric fields are always perpendicular to the interfaces and hence no shear 
force exists. In the other extreme, that of perfect dielectrics ($\sigma=0$), the force 
density is perpendicular to the surface and the shear component vanishes as well, as can 
be seen from the component of the force parallel to the interface, Eq. 
(\ref{stress_components}). 

In Taylor's ``leaky dielectric'' model the Maxwell shear stress 
of the residual charge must be balanced by a stress due to liquid flow inside and outside 
of the droplet. For a field alternating with angular frequency $\omega$ much larger than 
the typical inverse ion relaxation time $\Sigma/\veps$, where $\Sigma$ is the electrical 
conductivity, the behavior of the liquid is similar to that of a pure dielectric since 
the ions move very little about their place. When the frequency is reduced below this 
threshold, $\omega<\Sigma/\veps$, ions oscillation are large and they move about more 
significantly as the frequency is further reduced. In the limit of a DC field ($\omega\to 
0$) clearly even a vanishingly small amount of ions can lead to a very strong response, 
recalling that $\Sigma$ is proportional to the ion number density.

Taylor analyzed the electrohydrodynamics problem and his derivation lead to a function 
$\Phi$ given by
\begin{\ea}
\Phi=R\left(D^2+1\right)-2+3\left(RD-1\right
)\frac{2M+3}{5M+5}
\end{\ea} 
The parameters appearing $M$, $R$, and $D$ are the ratios of the values of viscosity, 
resistivity, and dielectric constant of the outer medium to that of the drop, 
respectively. Prolate 
drops are predicted when $\Phi>0$ while oblate drops correspond to $\Phi<0$. 
Spherical drops, $\Phi=0$, thus occur as a special case.

Based on this understanding one may ask how does residual conductivity affect the normal 
field instability described above? Namely how do the dispersion relation Eq. 
(\ref{dispersion}) and the fastest-growing $q$-mode Eq. (\ref{fastest_growing}) 
change when conductivity is taken into account? 
It turns out that the existence of ions leaves the general shape of the curve $\omega(q)$ 
intact but the maximum shifts to larger values -- both the fastest-growing $q$-mode and 
its growth rate are increased \cite{russel_jnnfm_2002}.

\begin{figure}[!tb]
\includegraphics[width=8cm,viewport=40 305 520 740,clip]{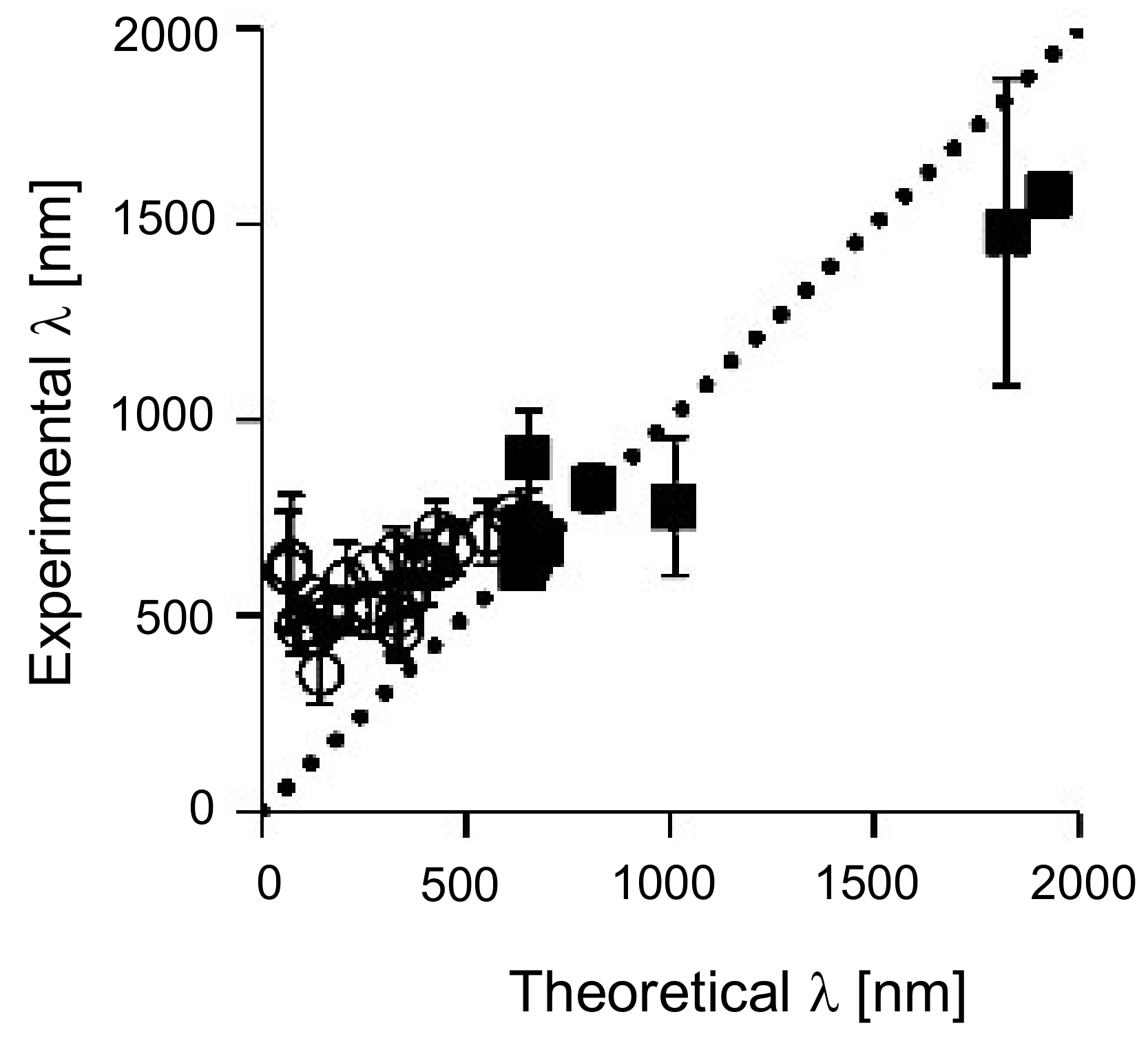}
\caption{\small Experimental vs theoretical fastest growing wavelength in various 
experiments 
with varying film thicknesses and voltages. The liquids used were polystyrene and an 
ionic liquid. The average field is smaller than $139$V/$\mu$m (filled squares) and larger 
than $158$V/$\mu$m (open circles). The dotted line is the expected $\lambda_{\rm 
th}=\lambda_{\rm exp}$ relation. The deviation from this line occurs for high field 
strengths. Clearly the size reduction stops at $\simeq 400$nm. Adapted from Ref. 
\cite{russel_mm_2011}.
}
\label{fig_fastest_growing}
\end{figure}

Patterning of films using the normal-field instability is an appealing concept for 
nanotechnological applications because of its simplicity and small number of 
processing steps \cite{russell_epl_2001,russell_nature_2000,russell_nature_mat_2003}.
A typical setup 
involves a polymer film of thickness $\approx 50$-$700$nm placed on a substrate, and a 
gap 
with varying thickness between the polymer and the mask. The idea is to quench the 
polymer structure at a specific time, at the onset of the instability, where the 
most unstable mode is dominant, or at a later time, where nonlinear structures with 
additional periodicities develop. Ideally one could increase the voltage and field across 
the substrate and mask to decrease the period of unstable mode indefinitely. However, 
when 
the electric field increases above $\sim 100$V/$\mu$m (depending on the polymer used and 
the overall sample geometry) dielectric breakdown marked by a spark occurs, and current 
flows between the two electrodes. A possible route to decrease the length-scale 
associated 
with the unstable mode, $\lambda\sim q_{\rm fastest}^{-1}$, is to decrease the surface 
tension $\gamma$, since $\lambda\sim\gamma^{1/2}$ when van der Waals forces can be 
neglected. A smart strategy is to fill the air gap between the polymer film and the mask 
with a second polymer, thereby creating a bilayer polymer system. The surface tension 
between the two polymers was indeed reduced this way but the dielectric contrast 
$\Delta\veps$ was reduced too, and this had a detrimental effect. In addition, the leaky 
dielectric model has prompted researchers to use polymers with a small conductivity 
or even to replace one of the polymers by an ionic liquid \cite{russel_mm_2011}. A 
comparison between the theoretical and experimental values of the fastest growing 
wavelength is shown in Fig.~\ref{fig_fastest_growing}. 
This has led to a decrease in the feature size, though only down to a 
limit set by the dielectric breakdown of the thin polymer layer.

%

\section{Changes in the relative miscibility of dielectric liquids}\label{chap3}

The preceding section described the interfacial instability that occurs when the stress 
by the electric field opposes the stress by the surface tension between two 
existing phases. But a more fundamental question arises: can the external field create or 
destroy an interface between two phases? That is, what is 
the effect of an electric field on the liquid-vapor coexistence of a pure component or 
the liquid-liquid coexistence for binary mixtures, and how is the critical point changed? 
A treatise on this problem was given by Landau.

\subsection{Landau theory of critical effects of external fields on partially 
miscible dielectric liquids}\label{chap3_landau}

In the book of Landau and Lifshitz \cite{LL_book_elec} the effect of a uniform electric 
field on the critical point was given as a short solved 
problem. Unfortunately it appeared only in the first edition of the book and was 
removed from the second edition by the Editors presumably because it was considered 
as ``unimportant''. This unimportant problem has caught considerable attention in recent 
years. 

We illustrate Landau's reasoning for a binary mixture of two liquids, A and B. The phase 
diagram is given by $T$ and $\phi$, the volume fraction of A component 
($0\leq \phi\leq 1$). 
The mixture's free energy density $f_{\rm m}(\phi,T)$ includes the enthalpic 
contributions, 
favoring separation, and the entropic force, favoring mixing. The electrostatic energy 
density is $f_{\rm es}=-(1/2)\veps(\phi)E^2$, where $\veps(\phi)$ is a 
constitutive equation relating the local permittivity with the local composition.
Close enough to the critical point $(\phi_c,T_c)$ one can expand the free energies in 
a Taylor series in the small deviation $\vphi\equiv\phi-\phi_c$
\begin{\ea}\label{taylor_m}
f_{\rm m}&\simeq& f_{\rm m}(\phi_c,T)+\frac{\partial 
f_{\rm m}(\phi_c,T)}{\partial\phi}\vphi+\frac12\frac{\partial^2 
f_{\rm m}(\phi_c,T)}{\partial\phi^2}\vphi^2+~.~.~.\\
f_{\rm 
es}&\simeq&-\frac12\veps(\phi_c,T)E^2-\frac12\frac{\partial\veps(\phi_c,T)}{
\partial\phi }
\vphi E^2-\frac14\frac{\partial^2\veps(\phi_c,T)}{\partial\phi^2}\vphi^2 E^2
\label{taylor_es}
\end{\ea} 
The terms linear in $\vphi$ are unimportant to the thermodynamic state since they can be 
expressed as a chemical potential. 
In Landau's phenomenological theory of phase transitions $\partial^2 
f_{\rm m}(\phi_c,T)/\partial\phi^2\simeq (k_B/v_0)(T-T_c)$, where $k_B$ is the 
Boltzmann's 
constant and $v_0$ is a molecular 
volume. Hence we see that the quadratic term proportional to $\vphi^2$ in the second 
line can be lumped into the first line as an effective critical temperature. The 
resulting field-induced shift to the critical temperature is
\begin{\ea}\label{landau_DT}
\Delta T_c=\frac{v_0}{2k_B}\frac{\partial^2\veps}{\partial\phi^2} E^2
\end{\ea} 
Here $\veps$ is calculated at the critical point. In the original formulation, given for 
the liquid-vapor critical point of a pure substance of density $c$ the shift in $T_c$ 
is analogously given by
\begin{\ea}
\Delta 
T_c=\frac{c}{2}\frac{\left(\partial^2\veps/\partial c^2\right)_T}{
\partial^2p/\partial c\partial T} E^2
\end{\ea} 
In both cases the shift is proportional to the second derivative of $\veps$ with respect 
to composition and to the electric field squared. As long as the deviation 
of $\phi$ (or $c$) from the critical value is small the whole binodal curve $T_b(\phi)$ 
is simply shifted upwards or downwards, depending on the sign of 
$\partial^2\veps/\partial\phi ^2$.

\subsection{Experiments that followed in simple liquids and in block 
copolymers}\label{chap3_experiments}

Debye and Kleboth were the first to investigate experimentally the effect of electric 
field on the 
critical point. They worked on binary mixtures since the experiments on liquid-vapor 
coexistence require high pressures and are considerably more difficult. The liquids they 
chose were isooctane and nitrobenzene. They measured a reduction of $T_c$ by $15$mK under 
a field of $4$-$5$V$/\mu$m \cite{debye_jcp_1965}. Orzechowski repeated and verified their 
measurement 
later \cite{orzech_chemphys_1999}. Wirtz and Fuller measured a reduction of $T_c$ by 
$20$mK for a mixture of nitroethane and n-hexane \cite{wirtz_fuller_prl_1993}. Other 
researchers found results similar in sign and magnitude 
\cite{beaglehole_jcp_1981,early_jcp_1992}, with the exception of Reich and Gordon who 
worked 
with a mixture of polystyrene and poly(vinyl methyl ether), (PS/PVME, the mixture
has a lower critical solution temperature), who observed large immiscibility. The large 
magnitude of the shift was due to the increased molecular volume and the subsequent 
reduction of entropy \cite{reich_jpspp_1979}. 
A recent work by 
Kriisa and Roth presented a presumably improved measurement utilizing a fluorescence 
technique and obtained a positive value of $\Delta T_c$ (improved miscibility) in the 
same system \cite{roth_jcp_2014}.

In the above studies, although the order of magnitude of $\Delta T_c$ was consistent with 
Eq. (\ref{landau_DT}) the sign was not -- the researchers observed enhanced {\it 
miscibility} in the electric field ($\veps''$ was positive).
In Landau's theory $E$ stands for the magnitude of the {\it average} electric field. 
However, composition variations mean variations in the dielectric constant, and these 
lead, via Laplace's equation, to variations in the electric field. For a plane wave 
composition variation $\vphi({\bf r})=\vphi_{\bf q}e^{i{\bf q}\cdot{\bf r}}$, where ${\bf 
q}$ is the wavevector and $\vphi_q$ is the amplitude of the wave, the additional 
contribution to the free energy due to dielectric anisotropy is proportional to 
\cite{ah_mm_1994,tsori_rmp_2009,onuki_mm_1995}
\begin{\ea}\label{ah}
v_0(\partial\veps/\partial\phi)^2\cos^2(\theta)E^2|\vphi_{\bf q}|^2
\end{\ea} 
where $\theta$ is the angle between the electric field and ${\bf q}$. This energy is
quadratic in $\vphi$ and in $E$ as in 
Eq. (\ref{taylor_es}) but it contains a dependence on $(\partial\veps/\partial\phi)^2$ 
and 
on the relative angle between the wavenumber of the composition variation and the 
average electric field. Evidently, the energy due to dielectric anisotropy is always 
positive, favoring mixing of the liquids and reduction of $T_c$ \cite{tsori_jpcb_2014}. 

\begin{figure}
\centering
\begin{minipage}{12cm}
~~~\includegraphics[keepaspectratio=true,width=0.4\textwidth]{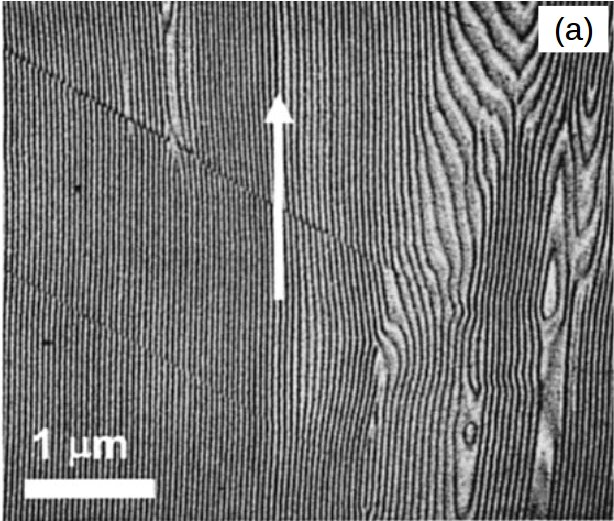}
\end{minipage}
\\ \vspace{0.3cm}
\begin{minipage}{12cm}
\includegraphics[keepaspectratio=true,
viewport=55 540 510 790,clip,width=0.6\textwidth]{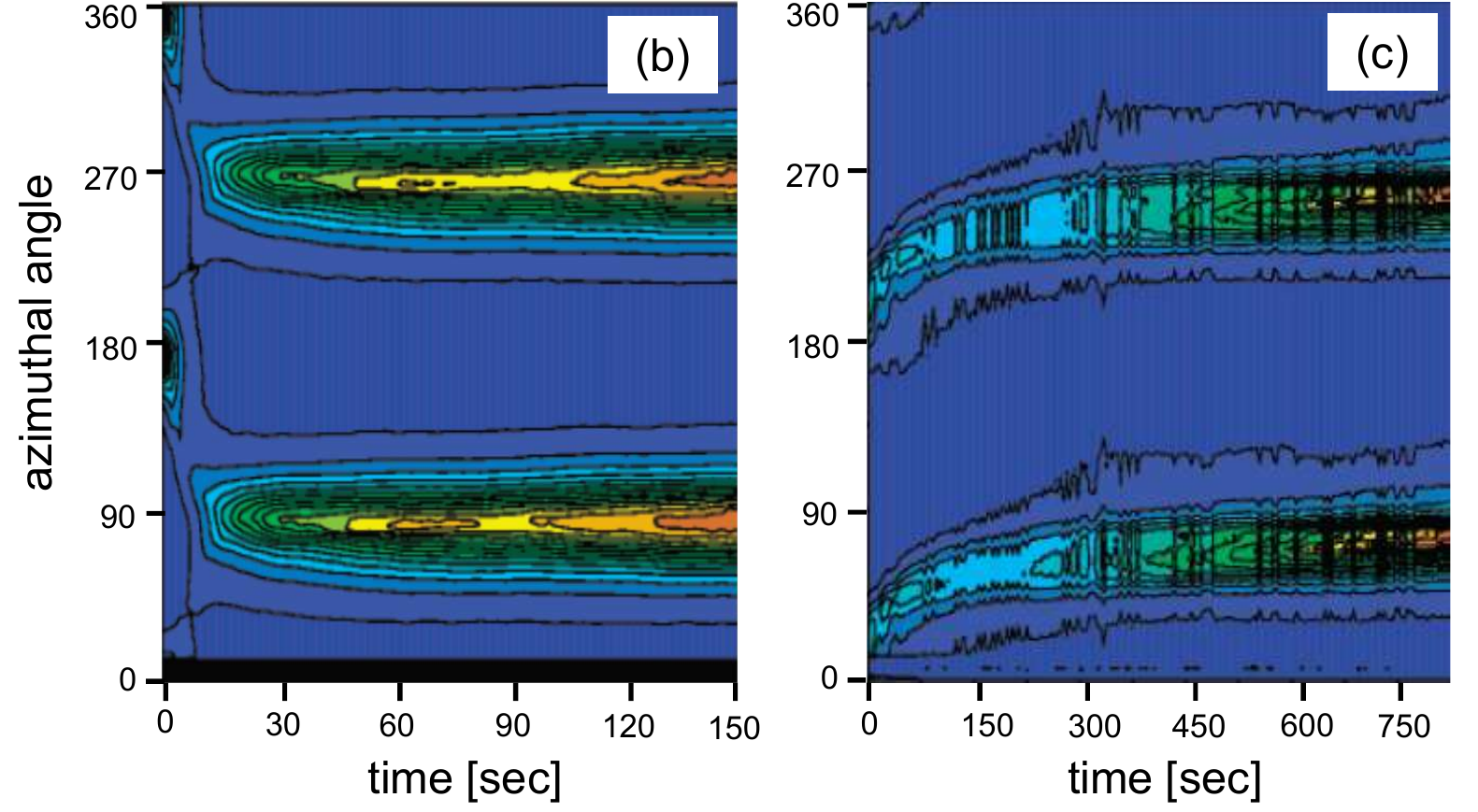}
\end{minipage}
\caption{\small (a) TEM image of PS-b-PHEMA-b-PMMA triblock copolymer aligned from a THF 
solution in electric field of $E=1.8$ V$/\mu$m. Arrow indicates the field's direction. 
Adapted from Ref. \cite{boker_mm_2002}. 
(b) in-situ SAXS time-sequence pattern from a copolymer solution as a function of
azimuthal angle in the plane perpendicular to the beam's direction. The electric field is
perpendicular to the beam. Angles of $0$ and $180$ degrees indicate lamellae
perpendicular to the field; lamellae are parallel to ${\bf E}$ at $90$ and $270$
degrees. (b) and (c) are for system close and far from ODT point, respectively.
Adapted from Ref. \cite{boker_mm_2003}.}
\label{fig_bcp_orientation}
\end{figure}

The energy penalty in Eq. (\ref{ah}) has a special importance in orientation 
of {\it ordered} phases by external fields. It underlies the orientation of lamellar 
grains of block copolymers, as first demonstrated by 
Amundson and co-workers \cite{ah_mm_1991,ah_mm_1993,ah_mm_1994}.
In those experiments the lamellae feel a torque that orients them parallel to the 
field (the lamellae normal orients in the direction perpendicular to the field, ${\bf 
q}\cdot{\bf E}=0$) leading to the lowest energy state with $\theta=\pi/2$ in Eq. 
(\ref{ah}). Figure \ref{fig_bcp_orientation} (a) is a Transmission Electron Microscopy 
(TEM) 
image showing 
that indeed alignment of lamellar block copolymers by an electric field is possible. 
Note that in polymers the volume 
$v_0$ is much larger than in simple liquids and the electrostatic energy is 
proportionally amplified compared to the thermal energy (this is true in both Eqs. 
(\ref{ah}) and (\ref{landau_DT})). In subsequent years this orientation 
mechanism has been the subject of extensive theoretical 
\cite{per_williams_mm_1999,TA_mm_2002,fraaije_mm_2002,tsori_mm_2003, tsori_mm_2006}
and experimental investigation both in the melt 
\cite{russell_sci_1996,russell_mm_2002b,russell_mm_2003,russell_mm_2004a} and 
in solutions \cite{boker_prl_2002,boker_mm_2003,boker_polymer_2006,boker_langmuir_2005,
boker_softmatter_2007}.

The orientation dynamics has been studied as well. For a structure confined in a thin 
film, experiments \cite{russell_mm_2005,russell_mm_2006,boker_softmatter_2012} and 
numerical calculations \cite{sevink_mm_2006,fraaije_jcp_2006} showed that 
lamellae or cylinders first form parallel to the substrate due to interfacial 
interactions. For a strong enough electric field they orient parallel to the field; they 
do so via two main pathways shown in the scattering patterns in 
Fig.~\ref{fig_bcp_orientation} (b)-(c): the 
first one is partial melting, where small regions melt and then recrystallize and grow 
with the preferred orientation (Fig.~\ref{fig_bcp_orientation} (b)). This mechanism is 
dominant for 
temperatures or compositions close enough to the order-disorder boundary in the phase 
diagram. The second mechanism is grain orientation, where large grains orient as a whole 
in the direction parallel to the field without melting first. This occurs far from the 
coexistence line and in systems that are not too viscous (Fig.~\ref{fig_bcp_orientation} 
(c)).

Due to dielectric anisotropy ordered phases in electric field tend to orient such that 
dielectric interfaces perpendicular to the field are minimized, reducing the energy 
penalty in Eq. (\ref{ah}). A perfect lamellar stack or an hexagonal array of cylinders 
can reduce this energy to zero. But in other phases, such as the BCC lattice 
of spheres or the gyroid phase, this cannot happen. Once the soft crystal attains its 
optimal orientation, an 
increase of the field leads to stretching of domains in the direction parallel to the 
field. The actual deformation is dictated as a balance between electrostatic and elastic 
forces. At a critical field the energy of the deformed phase is too high and an 
order-order phase transition occurs accompanied by a change in the crystal's symmetry. In 
block copolymers a sphere-to-cylinders was predicted theoretically by Tsori {\it et al.} 
\cite{tsori_prl_2003} and was verified experimentally by Xu {\it et al.}
\cite{russell_mm_2004a} and by Giacomelli {\it el.} \cite{stepanek_mm_2010}. Ensuing work 
considered the bulk phase diagram \cite{tsori_mm_2006} and the sphere-to-cylinder 
transition in thin films \cite{matsen_jcp_2006}. Other order-order transitions are 
possible and may be technologically interesting, for example the gyroid-to-cylinder 
transition. This transition was studied theoretically 
\cite{zvelin_mm_2007,zvelin_softmatter_2008}, and experimentally with emphasis on surface 
effects \cite{steiner_softmatter_2010}.

\section{Dielectric liquids in electric field gradients}\label{chap4}


As we mentioned above, variations in the composition of a mixture lead to variations of 
the electric field. We therefore define nonuniform fields as the fields that the 
electrodes confining the system would produce if the dielectric constant were spatially 
uniform. Such fields occur in fact only in highly idealized systems where the region of 
interest is infinitely smaller than the electrodes; field gradients are inevitable 
when the electrode size is comparable to the system size under investigation. When the 
field has gradient, Eq. (\ref{body_force}) shows that there is a dielectrophoretic force 
which tends to pull the liquid with high value of $\veps$ to regions with high value of 
$E$ \cite{pohl_book,chaikin_jpcm_2003,chaikin_prl_2006,chaikin_jcp_2008}. It is not 
surprising therefore that field gradients lead to composition gradients in an 
initially-homogeneous system. The composition gradients occur on the same scale as the 
electric field, which is usually macroscopic. It may be said that the ``composition 
follows the field'' since $\nabla\phi$ is roughly proportional to $\nabla E^2$. 

The interesting phenomenon is that there is a {\it critical} field above which a sharp 
composition gradients appear \cite{tsori_nature_2004}. In this state the composition 
variations can be much steeper than the lengthscale characterizing the field. To see this 
imagine the liquid mixture whose free energy 
density is $f=f_{\rm m}+f_{\rm es}$ confined by the wedge capacitor. The wedge is a 
hypothetical system comprised of two flat conductors tilted with an opening angle $\beta$ 
between them and potential difference $V$. $r$ is the distance from the imaginary meeting 
point of the two electrodes; its minimal and maximal values are $R_1$ and $R_2$, 
respectively. The mixture is confined to a region far from 
the edges of the electrodes but field gradients exist due to opening of the electrodes. 
In this simple system the field is in the azimuthal 
direction and its magnitude is given by $E=V/(\beta r)$. Close to the critical point one 
may use the expansions in small $\vphi=\phi-\phi_c$ as in Eqs. (\ref{taylor_m}) and 
(\ref{taylor_es}). In equilibrium the mixture's profile is given by the Euler-Lagrange 
equation 
\begin{\ea}\label{wedge_gov_eqn}
f_{\rm m}'(\vphi)=\frac12 \veps'\left(\frac{V}{\beta r}\right)^2+\mu.
\end{\ea} 
%
Here we used a prime to denote denote differentiation with respect to $\vphi$ and assumed 
that $\veps''=0$ to demonstrate a demixing phase transition that would not occur if the 
field were uniform. 

\begin{figure}[h!]
\includegraphics[scale=0.55,viewport=140 265 455 775,clip]{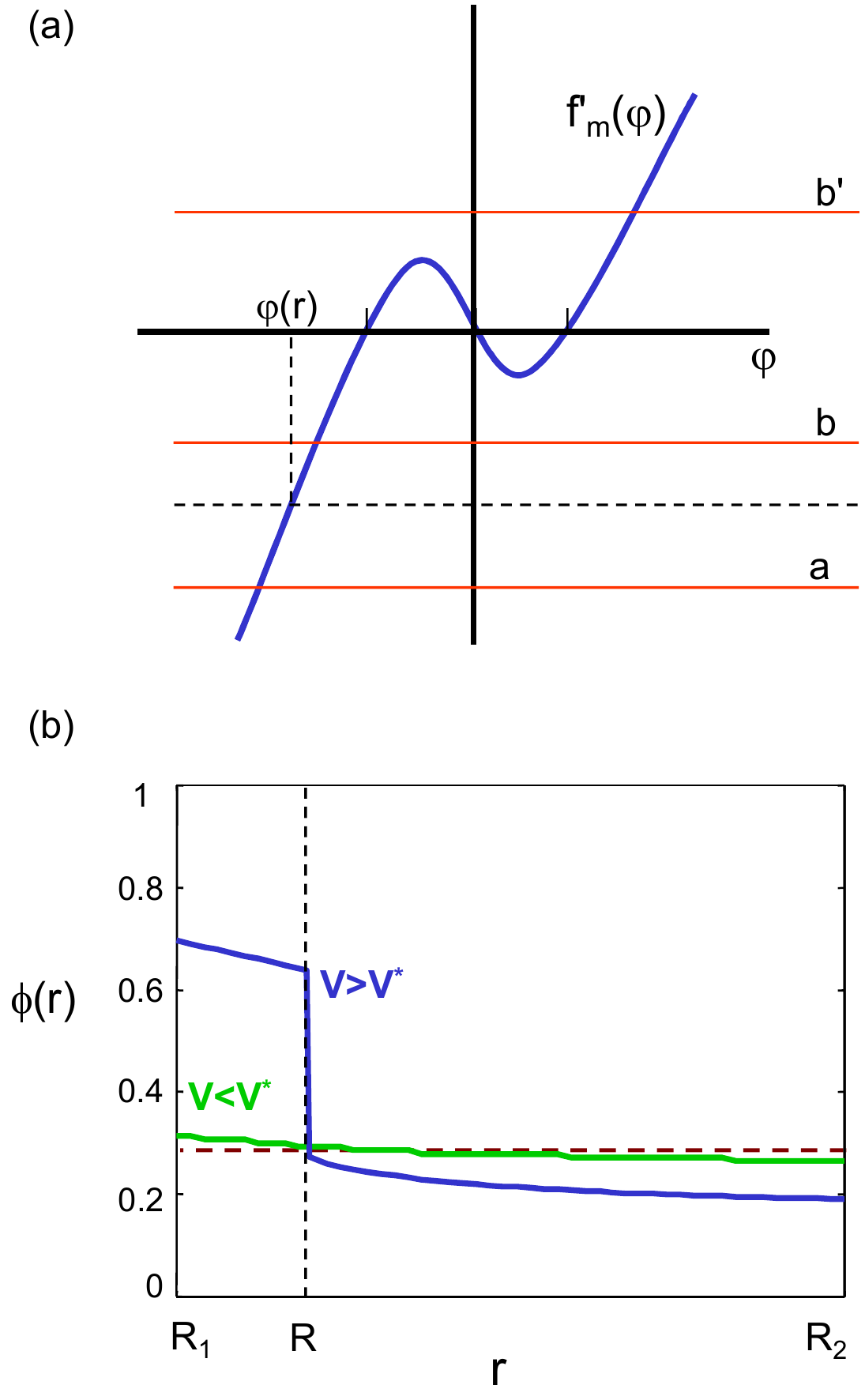}
\caption{\small (a) Graphical solution to Eq. (\ref{wedge_gov_eqn}). Solid curve is 
$f_{\rm m}'(\vphi)$.
Its roots are the transition (binodal) compositions. 
The intersection between $f_{\rm m}'(\vphi)$ and the horizontal dashed line gives the 
solution
$\vphi(r)$ to Eq. (\ref{wedge_gov_eqn}). For voltages $V$ below the critical value $V^*$, 
the dashed line is bounded by lines a and b, corresponding to the the maximal
and minimal values 
of the right-hand side of Eq. (\ref{wedge_gov_eqn}), giving rise to a continuous profile
$\vphi(r)$. At $V>V^*$, line b is displaced to b', and the intersection is at $\vphi<0$
for large $r$'s and at $\vphi>0$ at small $r$'s. (b) Qualitative composition profiles 
$\phi(r)$.
Horizontal dashed line is the average composition $\phi_0$ in the absence of field. 
$\phi(r)$ varies smoothly when $V<V^*$, and has a sharp jump at $r=R$ when $V>V^*$.
}
\label{fig_qualit_expl}
\end{figure}

The solution to this equation can be obtained graphically with the simplifying assumption 
that $f_{\rm m}$ from the expansion in Eq. (\ref{taylor_m}) has only quadratic and 
quartic 
terms in $\vphi$. When the temperature is above $T_c$, $f'_{\rm m}(\vphi)$ 
behaves similar $\vphi+\vphi^3$, namely it is monotonically increasing. 
Therefore, as the 
coordinate $r$ moves from large to small value the right hand side of the equation, being 
independent on $\vphi$, increases monotonically and $\vphi(r)$ accordingly changes 
continuously with $r$. 
\begin{figure}[h!]
\centering
\begin{minipage}{0.47\textwidth}
\includegraphics[scale=0.65,viewport=170 180 420 595,clip]{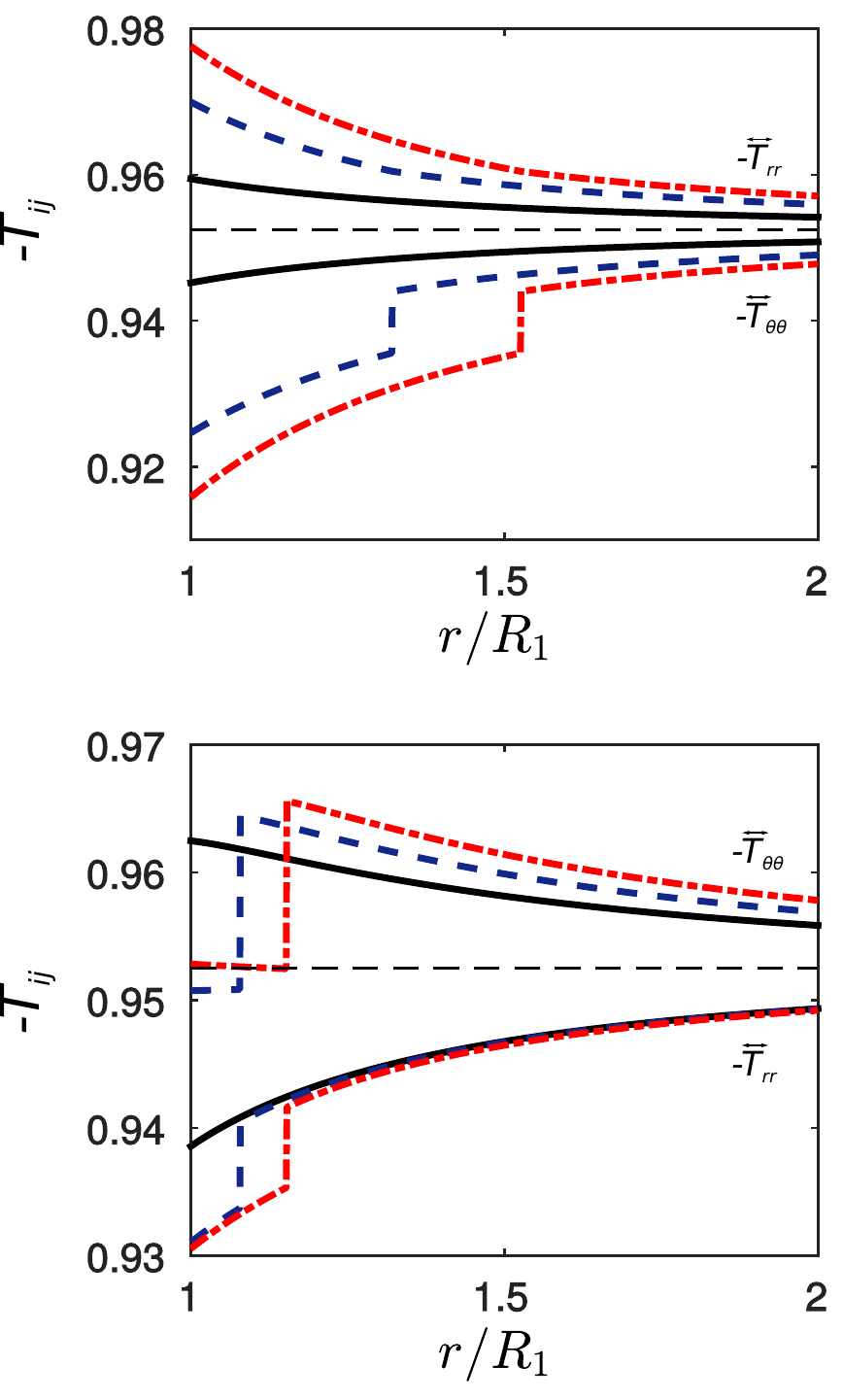}
\end{minipage}
\begin{minipage}{0.5\textwidth}
\caption{\small Diagonal elements of the stress tensor 
$T_{\rm rr}$ and $T_{\rm \theta\theta}$ for 
vapor-liquid coexistence in electric field gradients in (a) wedge 
capacitor and (b) charged wire. In both (a) and (b) the smallest radius is $R_1$. 
Horizontal dashed line corresponds to the pressure $p_0$ in the absence 
of external potential. At small potentials the density of the fluid is larger close to 
the inner radius and the stress is smoothly varying (black lines). At potentials above 
the critical value a demixing transition occurs and the pressures become discontinuous 
across the vapor-liquid interface (blue dashed lines). At even a larger potential the 
interface moves to larger values of $r$ (red dash-dot curves). Adapted from Ref. 
\cite{tsori_jpcb_2011}.
}
\label{fig_pressure_profiles}
\end{minipage}
\end{figure}

The situation is different at temperatures $T$ smaller than $T_c$. Here $f_{\rm m}$ 
behaves qualitatively as $-\vphi+\vphi^3$ as is shown in Fig.~{\ref{fig_qualit_expl}a. 
When the field, or external potential, is too small, the right-hand side of Eq.
(\ref{wedge_gov_eqn}) is a horizontal line bounded between lines a and b, 
corresponding to the smallest and largest electric fields in the wedge. The resultant 
profile $\vphi(r)$ is then similar to the $T>T_c$ case. But his behavior is true only 
for small potentials; there exist a critical potential $V^*$ above which b is displaced 
to b'. In this case the solution to Eq. (\ref{wedge_gov_eqn}) jumps from the left branch 
of $f'_m(\vphi)$  (negative values of $\vphi$) to the right branch (positive values of 
$\vphi$) discontinuously as $r$ decreases from large to small values. The demixing 
transition is then marked by a composition front whose thickness is much 
sharper than any length characterizing the field. At this state two phases coexist; 
both phases are nonuniform and the more polar one is located at the region of high 
electric field (small value of $r$).

\subsection{The pressure tensor and surface tension in vapor-liquid coexistence}

Electric field gradients modify the coexistence of pure fluids too. 
``Electro-prewetting'' occurs where an initially homogeneous vapor phase is brought under 
the influence of an electric field with gradients. For small gradients the molecules 
of the vapor are attracted to the region with large field and the density becomes 
larger there. At the critical voltage or field, the density becomes so large that liquid 
condenses with a well-defined liquid-vapor interface. This electroprewetting transition 
is different from sedimentation or condensation of particles in an external 
gravitational field or in centrifugation because, as mentioned above, in electric 
fields a small change in the composition or density affects the electric field even far 
away whereas the gravitational field is unaffected by the internal rearrangements of the 
liquid. 

Another simple system that allows analytical development, besides the wedge, is a 
charged wire. For an infinitely long wire the system is effectively two dimensional 
and depends on $r$ and on $\theta$. If azimuthal symmetry is preserved the system becomes 
effectively one-dimensional and the field oriented in the radial direction. If a wire 
of radius $R_1$ inside a container with density and pressure corresponding to the vapor 
phase is charged at small voltage, the fluid density increases at the vicinity of the 
wire. At large enough wire voltage a liquid phase will precipitate near the wire, 
coexisting with the vapor phase far from it. 

In Fig.~\ref{fig_pressure_profiles} we show the stress tensor as calculated for the wedge 
(a) and wire (b) cases. The two diagonal components $\overleftrightarrow{T}_{\rm rr}$ and 
$\overleftrightarrow{T}_{\rm 
\theta\theta}$ are shown vs scaled distance from the origin ($\tilde{r}=r/R_1$). In both 
panels dashed lines correspond to small potentials, solid lines are potentials sufficient 
for demixing, and dash-dot lines are even larger potentials. For the last two cases an 
interface is created by the field separating the liquid and vapor phases. After the phase 
transition occurs, the stress tensor is discontinuous in $r$ (the body force is 
continuous though).

An important difference between the wedge and the wire is that once the 
interface is created, in the wedge the electric field is tangential to the interface 
whereas in the wire the electric field is normal to the interface. This is evident in the 
plots -- in the wedge the $\theta\theta$ component of the pressure 
($-\overleftrightarrow{T}_{\theta\theta}$) is smaller than the $rr$ component while in the 
wire 
$p_{\theta\theta}$ is larger than $p_{rr}$. In both the pressure inside the liquid drop 
(small $r$'s) is {\it smaller} than outside, quite different from the ``regular'' 
pressure 
of droplets as given by Laplace's equation.

\subsection{Demixing dynamics in liquid mixtures}

The dynamics of field-induced phase separation can be described by a modified ``model H'' 
framework. In this model the total flux is the sum of an advection term and 
a thermodynamic flux which is a gradient of the mixture's chemical potential. This 
reaction-diffusion equation is accompanied by Navier-Stokes equation with a force 
depending on the field, and Laplace's equation for the field. In some cases the symmetry 
dictates that hydrodynamic flow vanish identically or ${\bf u}$ is small enough and can 
be neglected, then the governing equations reduce to
\begin{\ea}
\frac{\partial\phi}{\partial t}&=&L\nabla^2
\frac{\delta f}{\delta\phi}~,\nn\\
\nabla\cdot(\veps(\phi)\nabla\psi)&=&0~\label{modelB}~,
\end{\ea} 
where $L$ is a transport coefficient (assumed constant) and $\psi$ is the electrostatic 
potential (${\bf E}=-\nabla\psi$). 

Figure \ref{fig_stab_diagram} shows the stability diagram of a binary mixture confined in 
a cylindrical capacitor with inner and outer radii $R_1$ and $R_2$, respectively. The 
horizontal axis is the average mixture composition $\phi_0$. It is given for a specific 
value of the dimensionless electric field squared $M_c=\sigma^2v/(4k_BT_c\veps_0)$, where 
$\sigma$ is the surface charge density on the inner cylinder. Mixtures that are 
homogeneous in the absence of electric field were considered and hence only the area 
above 
the binodal curve $T_b(\phi_0)$ is relevant. The shaded region is unstable with respect 
to the field applied, namely it has a sharp composition gradient. If a point $(\phi_0,T)$ 
is outside of the shaded region the mixture still has composition variations but not a 
real interface. The unstable region is bounded by the binodal and by the lines marked 1 
and 3. Point 2 is a surface critical point and line 1 is obtained as the locus of points 
2 corresponding to electric field ($M_c$) increasing continuously from zero. Line 3 is 
the ``electrostatic binodal'', for every temperature below that of point 2 its 
composition is the smallest for which an interface appears. 
\begin{figure}[h!]
\includegraphics[scale=0.55,viewport=110 215 500 625,clip]{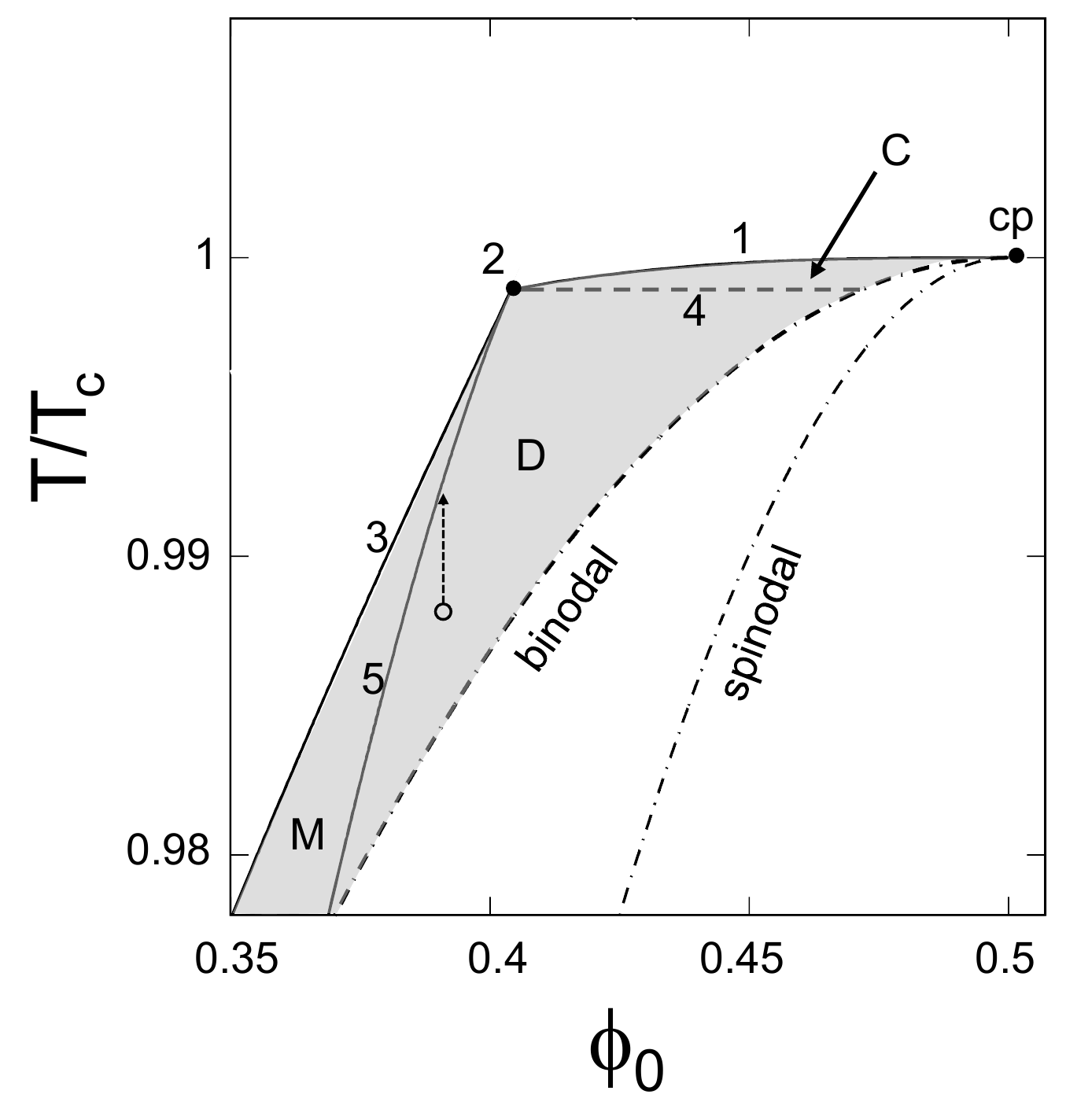}
\caption{\small Schematic stability diagram of nonpolar binary mixtures in a cylindrical 
capacitor. Above the binodal curve composition variations exists; only in the shaded 
region the electric field is strong enough to induce a real interface (defined as a jump 
in $\phi(r)$ in the limit of vanishingly small $(\nabla\phi)^2$ term in the free energy). 
In region C the interface appears continuously (Fig.~\ref{fig_phi_profiles}b) and in D 
it appears discontinuously (Fig.~\ref{fig_phi_profiles}a). M is the metastable region. 
Point 2 is a surface critical point and cp is the critical point of the mixture in the 
absence of field. Adapted from Ref. \cite{tsori_jcp_2014b}}
\label{fig_stab_diagram}
\end{figure}

The unstable region can be further divided to three parts: M, D, and C 
\cite{tsori_jcp_2014a}. At a fixed temperature above that of point 2, when the average 
composition $\phi_0$ increases and crosses line 1 into region C there is a continuous 
(second-order) transition where the equilibrium interface appears at a finite location 
$R$ 
intermediate between $R_1$ and $R_2$. The ``jump'' in $\phi$ across the interface, 
calculated in the sharp interface limit (i.e. without a $(\nabla\phi)^2$ term in the 
energy density \cite{safran_book}) vanishes at the transition line.
\begin{figure}[h!]
\includegraphics[scale=0.55,viewport=55 380 550 635,clip]{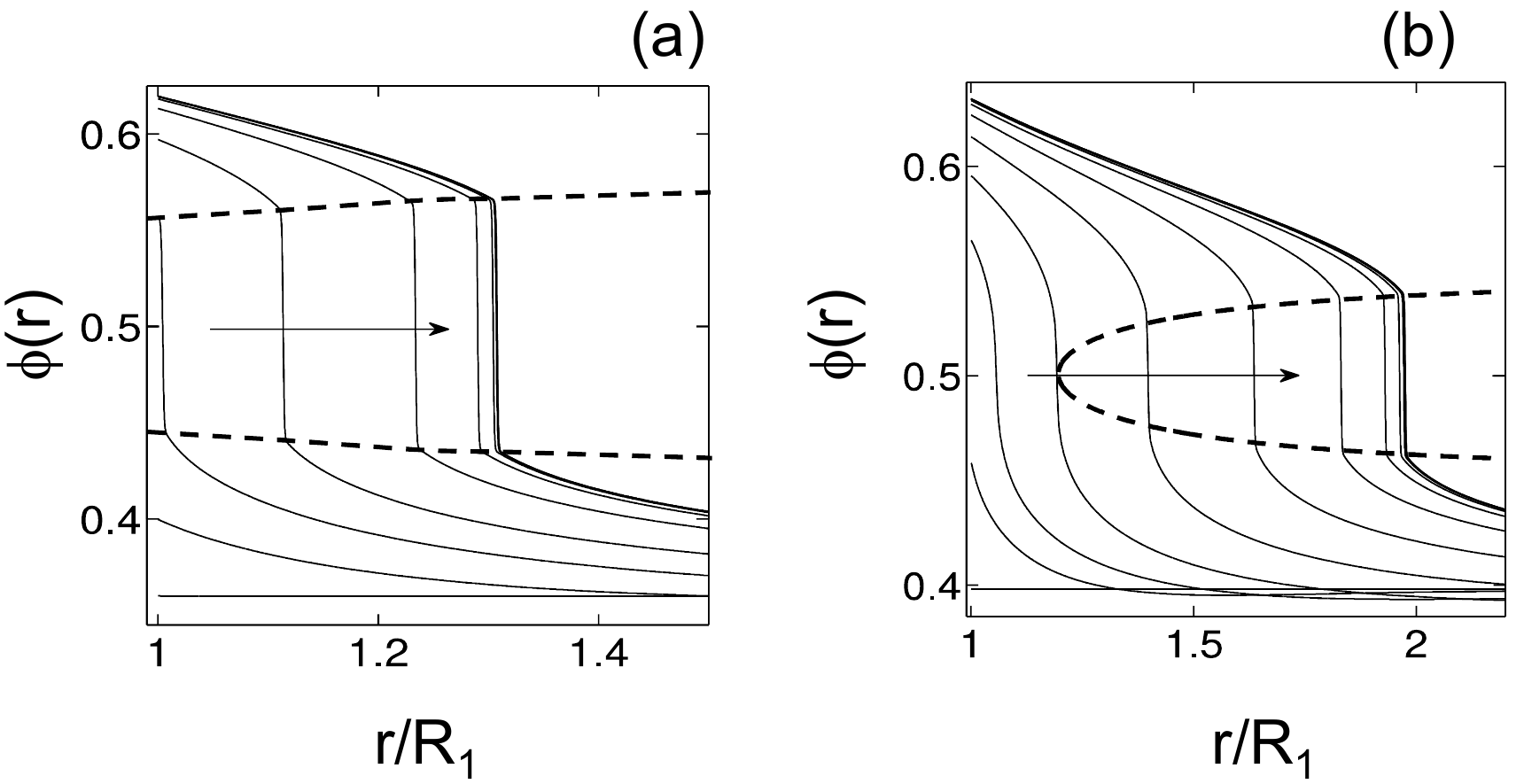}
\caption{\small Composition profiles $\phi(r)$ for two regions (a) discontinuous, region 
D 
($\phi_0=0.36$, $T/T_c=0.992$) and (b) continuous, region C ($\phi_0\simeq 
0.4$, $T/T_c=0.997$) in Fig.~\ref{fig_stab_diagram}. Each curve is a different 
snapshot in time taken at regular intervals on a logarithmic scale. Arrows indicate 
the direction of increasing time. Dashed lines indicate the composition where coexistence 
of two phases can exist based on a mixture free energy density $f_{\rm m}$ symmetric 
around 
$\phi=1/2$. In both parts $M_c=\sigma^2v/(4k_BT_c\veps_0)=0.069$ is the dimensionless 
field squared. Adapted from Ref. \cite{tsori_jcp_2014b}
}
\label{fig_phi_profiles}
\end{figure}

In region D, separated from region C by the line 4, the transition is first order, namely 
when $\phi_0$ increases at fixed $T$ and crosses line 3 the equilibrium interface appears 
at the minimal radius $R=R_1$ and the jump in $\phi(R)$ is finite. The difference in 
equilibrium behavior leads to a different dynamics. Fig.~\ref{fig_phi_profiles} shows the 
profiles $\phi(r)$ calculated from Eqs. (\ref{modelB}) at different times. The difference 
between the dynamics in region D (left) and C (right) is clearly seen when the interface 
appears at $R=R_1$ in D and at $R>R_1$ in C. 

Line 5 is the ``electrostatic spinodal'' $\phi_{\rm es}(T)$ differentiating between D and 
M and defined by $f''(\phi_0,T,r=R_1)=0$. In D $f''(\phi_0,r=R_1)$ is always positive 
whereas in M $f''$ can be negative. The 
convexity of $f$  has a kinetic meaning. In region D the interface appears after a 
finite lag time $t_L$. The closer $\phi_0$ is to line 5 the longer the lag time is. The 
lag time diverges as a power law \cite{tsori_jcp_2014b}
\begin{\ea}
t_L={\rm const.}\times(\Delta\phi)^\alpha
\end{\ea} 
where $\Delta\phi=(\phi_0-\phi_{\rm es})/\phi_{\rm es}$ is the scaled distance from the 
electrostatic spinodal and $\alpha\simeq -1.16$ is the exponent. This 
relation holds irrespective of the value of $M_c$ as long as $\Delta\phi$ is small. 
In region M the system is metastable and phase separates only with sufficiently strong
thermal noise or other nucleation event.

\begin{figure}[h!]
\includegraphics[scale=0.5,viewport=60 270 530 725,clip]{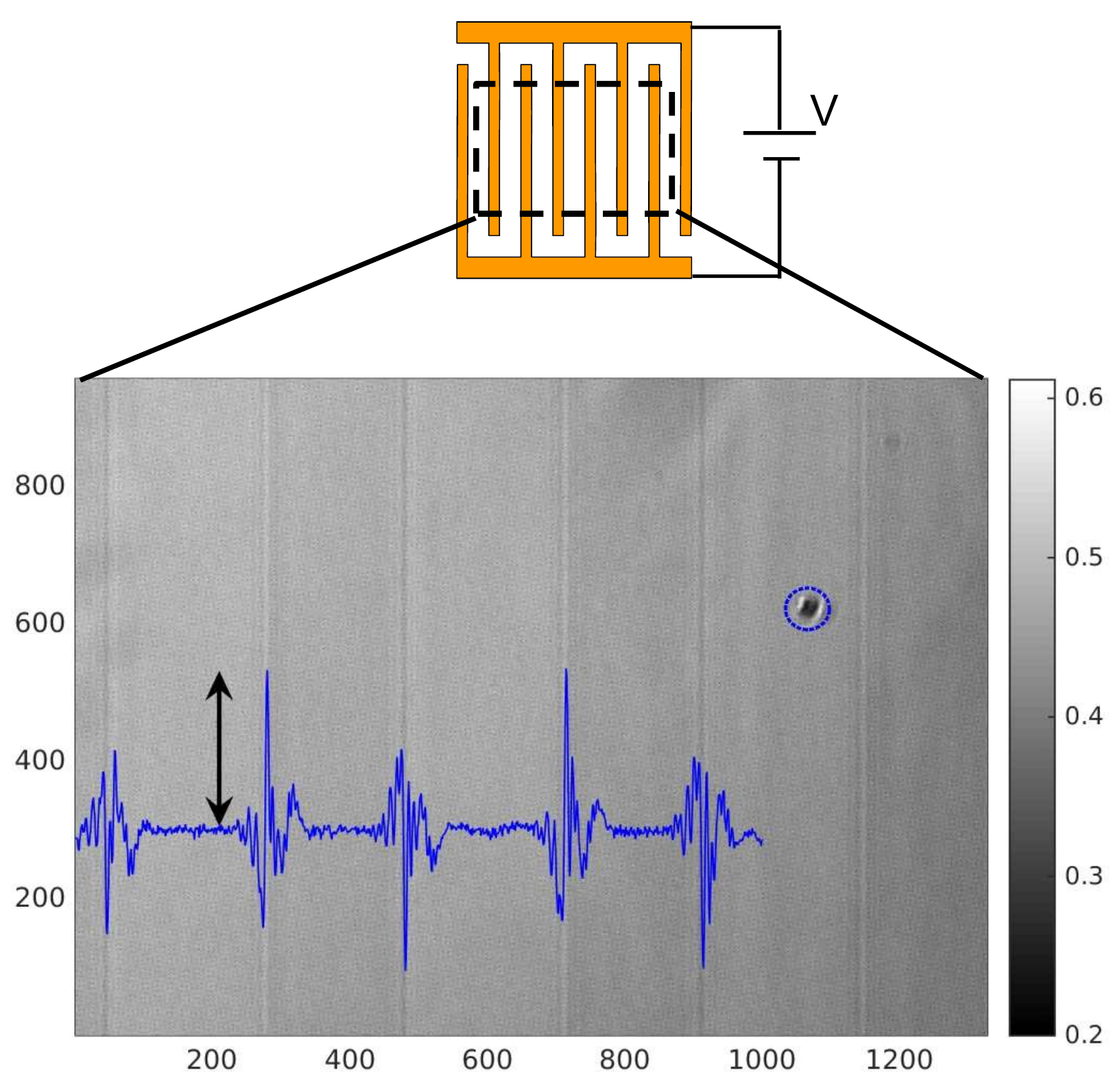}
\caption{\small Representative phase contrast microscopy image from a mixture of ethyl 
lactate and squalene (2:8 v:v ratio) near interdigitated ITO electrodes (shown 
schematically above). The flat electrodes (thickness $\simeq 25$nm) is seen as vertical 
dark stripes; the gap is brighter. The polarity of the stripes is alternating between 
positive and negative. The relative brightness indicates the mixture's composition.  The 
voltage between adjacent electrodes was $700$V DC and the temperature was regulated by a 
Linkam LTS120E temperature stage to be $0.2^\circ$K above the binodal temperature. 
Colorbar shows the gray levels. Overlaid curve (blue) was obtained by 
subtraction of the gray levels (black=0, white=1) of the image and the `control' (image 
without voltage), then averaging over the $y$-axis. Black arrow equals $0.3$ gray levels. 
Encircled dark spot is a dirty spot on the cover-slip glass and was not included in 
analysis. Images were cropped to size 1328x956.
}
\label{fig_interdig_with_profile}
\end{figure}

To achieve phase separation with field gradients experimentally, the electrode setup is 
straightforward since strong field gradients occur in electrodes whose 
typical size is on the micrometer scale with voltages on the order of $\sim 
100$V.
The general design of a liquid cell consisted of standard $2.5$cm$\times 7.5$cm 
glass slide as the substrate, covered by a $2$cm$\times 2$cm cover slip. The liquid 
mixture was put 
in the $\sim300\mu$m gap between the two layers. The edges of the cover slip were 
sealed by various glues or by a Teflon frame leading to actual cell size of 
$1.5$cm$\times 1.5$cm. Three types of electrodes were used. The first is wire 
electrodes, made from several metals with or without coating. Two wires were inserted 
into the cell prior to sealing and were connected to a DC or AC voltage supply. The 
wire's radius varied from $50\mu$m to $200\mu$m. The field gradients results from the 
cylindrical symmetry around the wires. The second design consisted of a thin metallic 
coating on the bottom slide (Indium Tin Oxide, platinum, gold, silver or other metals) 
fabricated so as to have two ``razor-blade'' parts with a gap (width $\sim 
50\mu$m) separating them. Here the field is mainly dictated by the thickness of the 
metallic layer ($\sim 25$nm) and not by the thickness of the gap. The field is largest 
close to the edges of the metal layer. In the third design the metallic layer had the 
shape of inter-digitated stripes in a comb-like manner. A voltage difference was 
imposed across any two adjacent stripes, as is schematically depicted in 
Fig.~\ref{fig_interdig_with_profile}. The whole cell was put in a carefully controlled 
temperature chamber and was observed with phase-contrast optical microscopy. 
The large number of stripes and large sample area allows to infer the influence of small 
local defects and on the same time to obtain better statistics.

Images of the sample prior and after application of voltage were recorded and analyzed. 
Due to local irregularities of the electrodes and to enhance the detection of 
the separation, analysis was done only after subtraction of the `control' image 
(corresponding to stable and thermally equilibrated sample without electric field). 
Fig.~\ref{fig_interdig_with_profile} is an example of a bare image. The blue  
superimposed curve is the gray level intensity averaged along the electrodes 
(y-direction) after the control image was subtracted.
\begin{figure}[tb]
\subfloat[\label{fig_interdig_dynam}]{\includegraphics[keepaspectratio=true,
width=0.5\textwidth]{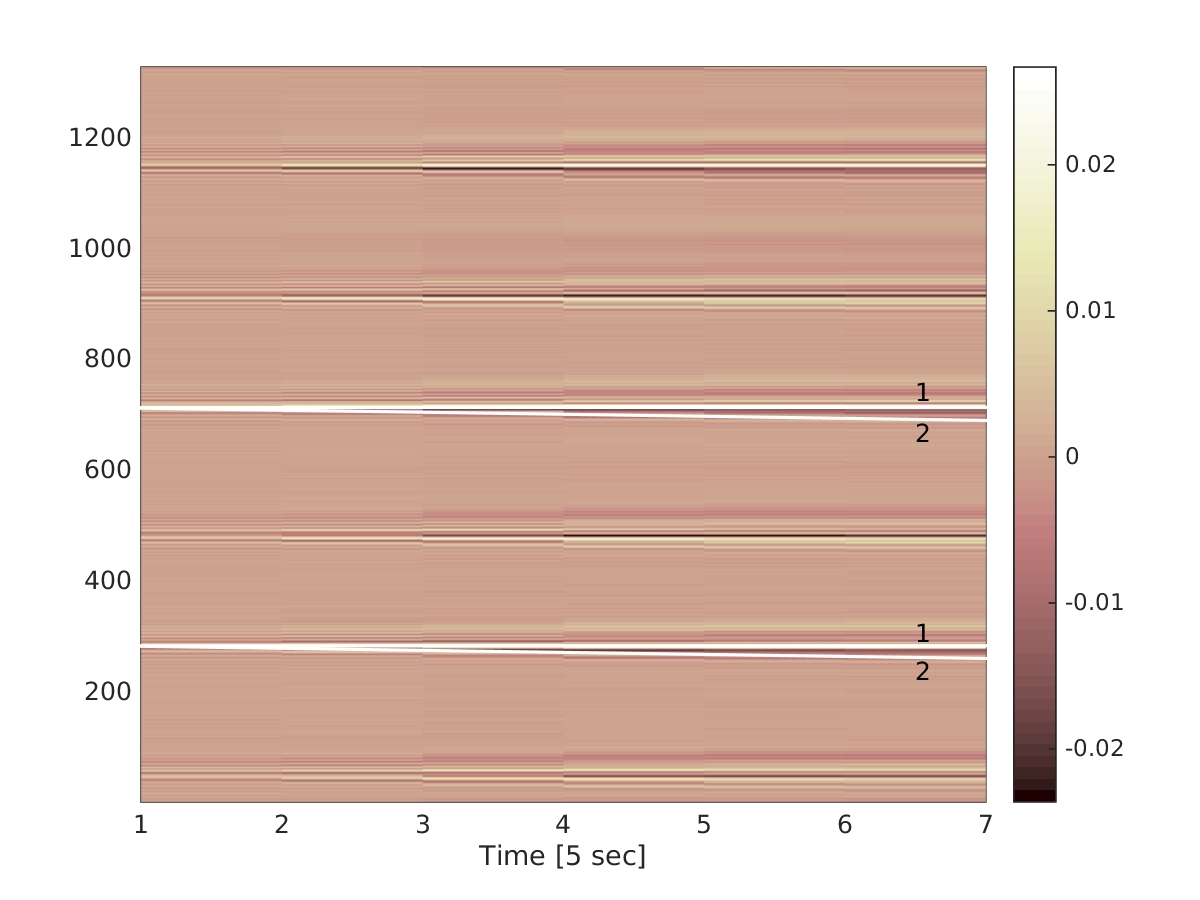}}
\subfloat[\label{fig_width_vs_voltage}]{\includegraphics[keepaspectratio=true,
width=0.4\textwidth,viewport=40 170 555 600,clip]{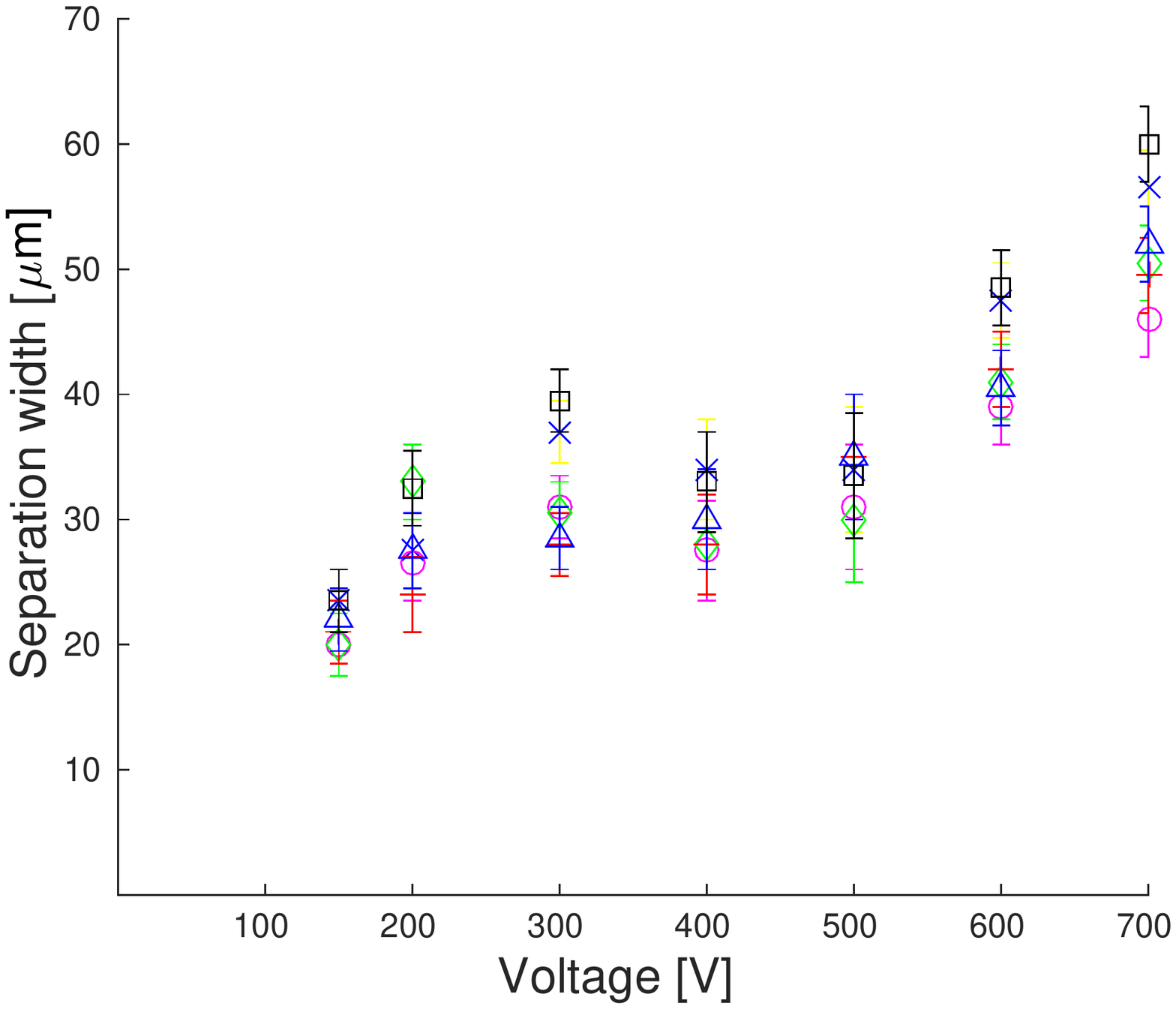}}
\caption{\small (a) Time progression of phase separation. Horizontal axis is time (in 
units of $5$sec) and vertical axis is the $x$-axis from 
Fig.~\ref{fig_interdig_with_profile}. Colors indicate the gray levels of profiles such as 
the one overlaid in Fig.~\ref{fig_interdig_with_profile}. White lines highlight the 
widening 
of the phase separation (only two pairs 1 and 2 are shown) -- line 1 is horizontal and 
static and line 2 deviates from it, approximately linearly at early times. The gap 
between them is the width of the phase-separated region as a function of time.
(b) The width of phase separated region at long times (15 mins) vs applied voltage. 
Measurements were taken from six different electrodes (different symbols). In (a) and (b) 
temperature, composition and other conditions are the same as in 
Fig.~\ref{fig_interdig_with_profile}.}
\end{figure}

The time progression of such intensity curves is shown in Fig.~\ref{fig_interdig_dynam}. 
The peaks and valleys of the curve in Fig.~\ref{fig_interdig_with_profile} correspond 
to red and blue tints, respectively. Out of the six edges shown in the images two were 
highlighted with white lines. The gap between them was defined as the thickness of the 
separated region. With increasing time the the phase separated regions near the edges of 
the electrodes widen.

The equilibrium thicknesses measured at long times are shown in 
Fig.~\ref{fig_width_vs_voltage} 
vs the applied voltage. For these 
liquids, mixture composition, and temperature, the thickness increases linearly with 
voltage until $V=300$V, it is approximately constant in the range $300$V-$500$V and 
increases again for $V>500$V. The plateau in intermediate voltages cannot be explained by 
the simple theory used so far. 

The simple theory based on model B dynamics predicts that if the point $(\phi_0,T)$ lies 
in region D of the stability diagram, Fig.~\ref{fig_stab_diagram}, then there is a lag 
time for the phase separation depending on the distance from the ``electrostatic 
spinodal'', line 5. Such lag time was observed experimentally as is shown in 
Fig.~\ref{fig_lagtime} for several temperatures above the field-free binodal curve $T_b$. 
For 
small values of $\Delta T=T-T_b$, the lag 
time is negligible compared to the error bars, recalling that the phase separation is 
quite faint and the onset of separation is not easy to detect. When $T$ increases closer 
to the ``electrostatic 
binodal'', $\Delta T\simeq 1^\circ$K, the lag time increases markedly. Despite the 
many obvious qualitative and quantitative differences between the experimental and 
theoretical phase diagrams and the shortcomings of the theory, it seems that the 
temperatures of Fig.~\ref{fig_lagtime} roughly correspond to the dashed vertical arrow in 
Fig.~\ref{fig_stab_diagram} and therefore they validate the theoretical lag time concept.

\begin{figure}[tb]
\includegraphics[scale=0.5,viewport=40 180 545 590,clip]{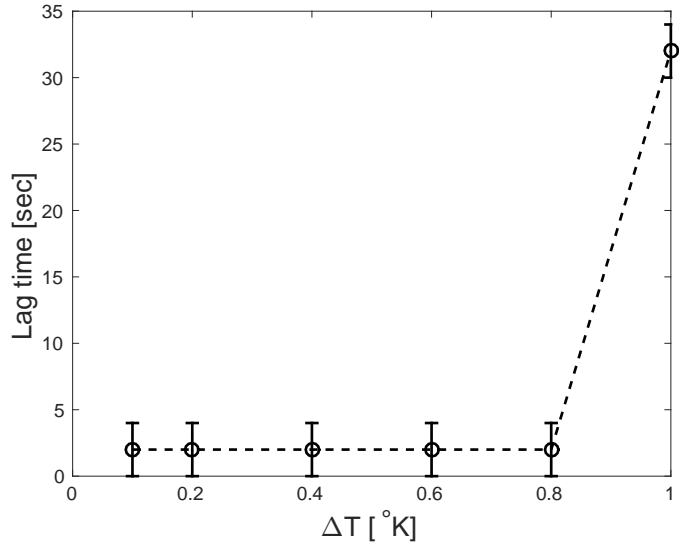}
\caption{\small Lag time for the appearance of phase separation vs temperature difference 
$\Delta T=T-T_b$. Increasing values of $\Delta T$ corresponds qualitatively to the 
upward pointing dashed arrow in Fig.~\ref{fig_stab_diagram}. The largest $\Delta T$ the 
closest $T$ is to the electrostatic binodal (line 5) and the larger the lag time is.
The mixture of ethyl lactate and squalene with compositions 2:8 v:v was subject to $400$V 
on interdigitated electrodes. 
}
\label{fig_lagtime}
\end{figure}
%

\section{Demixing in polar solutions}\label{chap5}

In Sec. \ref{chap2_leaky_dielectric} we discussed how, when two well defined phases are 
subject to electric fields, even a small amount of ions leads to marked differences as 
compared to pure dielectrics. The same is true when the initial state is mixed and there 
is no interface. In purely dielectric liquids the prewetting transition discussed in Sec. 
\ref{chap4}  is induced by a dielectrophoretic force and this force arise only in 
nontrivial electrode geometries (i.e. curved or finite size), and not in one dimensional 
systems. However, when ions exist field gradients due to screening occur irrespective of 
the geometry. The lengthscale associated with the field is no longer the typical 
electrode 
size, $~\sim 1\mu$m, but rather the Debye screening length $\lambda_D\sim 10$nm. Hence 
even small potentials of $\sim 0.1$V lead to large fields $E\sim V/\lambda_D=10^7$V/m.

But ions are not equally miscible in all solvents. Ionic specificity to liquids and 
surface has been known for a long time (See 
\cite{marcus_book,onuki_jcp_2004,onuki_review} and references therein). The ionic 
affinity 
to neutral or hydrophobic 
surfaces, the so-called Hofmeister series \cite{kunz_cocis_2004}, influences many 
physical properties, such as the water-air surface tension 
\cite{jungwirth_cr_2006,levin_prl_2009a,levin_prl_2009b}
\cite{andelman_epl_2014,andelman_jcp_2015}. It plays a central 
role in the solubility of proteins and underpins the ``salting out'' effect 
\cite{zhang_cocb_2006}, which is commonly used in protein separation techniques. 

\begin{figure}[h!]
\centering
\begin{minipage}{0.55\textwidth}
\caption{\small Schematic illustration of the solvation of Na$^+$ in (a) pure water 
and (b) water-nitrobenzene. Water molecules are preferentially attached to the 
hydrophilic 
ion. The ion's solvation energy is higher in (b) than in (a) and the difference exceeds 
the thermal energy. Adapted 
from Ref. \cite{onuki_cocis_2011}.
}
\label{fig_solvation}
\end{minipage}
\begin{minipage}{0.4\textwidth}
\includegraphics[width=\textwidth]{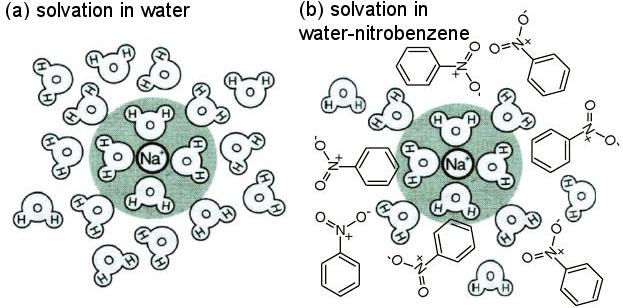}
\end{minipage}
\end{figure}
\begin{figure}[h!]
\centering
\begin{minipage}{0.58\textwidth}
\includegraphics[viewport=1 333 530 575,clip,width=\textwidth]{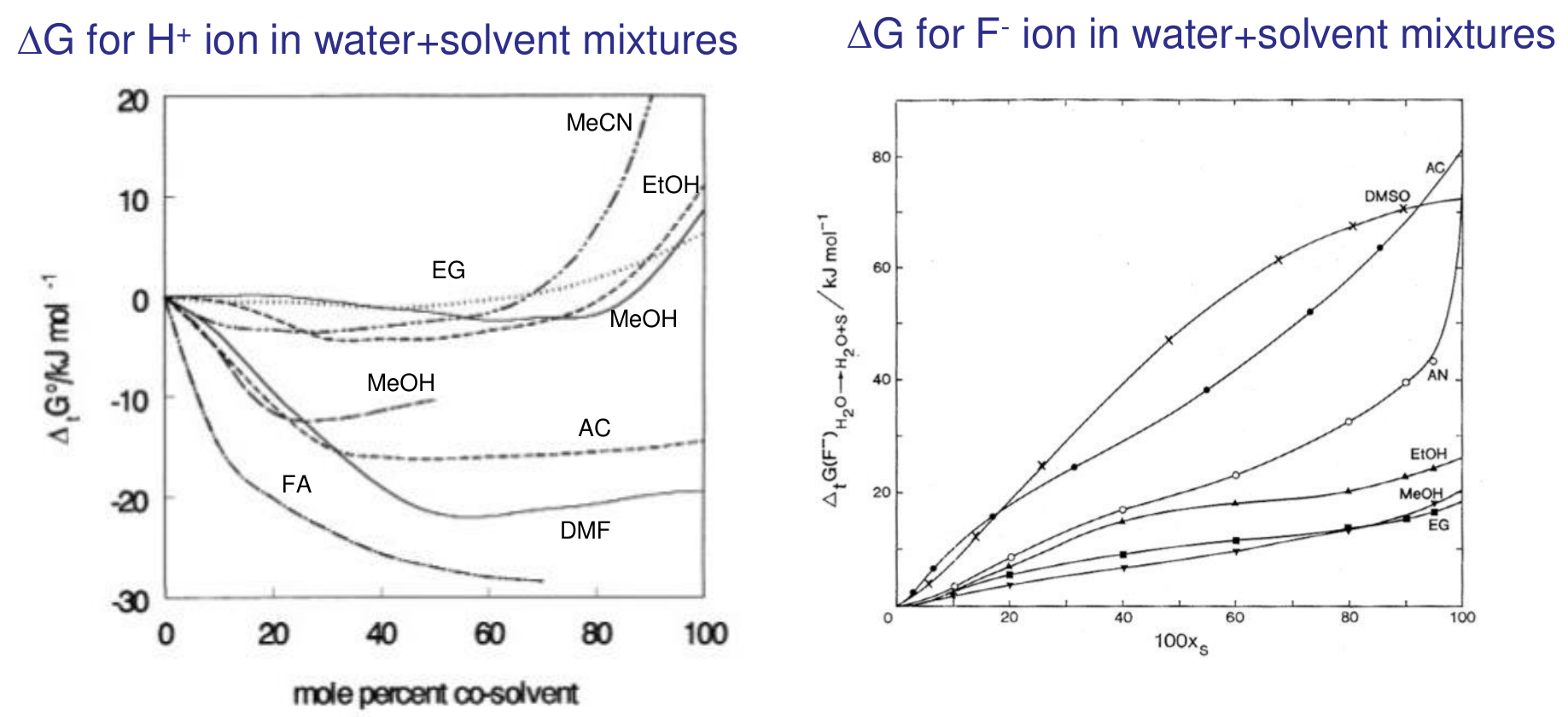}
\end{minipage}
\begin{minipage}{0.4\textwidth}
\caption{\small Gibbs transfer energy $\Delta G$ for moving an ion from pure water to 
mixtures with varying fractions of the co-solvent for H$^+$ (left) and F$^-$ (right) 
ions. Clearly the magnitude and trend (descending or ascending), or even the sign of the 
curves, is very ion-specific. Adapted from Ref. \cite{hefter_pac_2005}.
}
\label{fig_hefter}
\end{minipage}
\end{figure}
Ionic specificity is even more pronounced in liquid mixtures. For example, in mixture of 
water and a less polar co-solvent, water molecules selectively attach to hydrophilic 
ions [see illustration in Fig.~\ref{fig_solvation}]. The typical difference in the 
solvation energy of an ion between solvents is 
typically of the order of $5$--$10k_BT$ per ion and can even be much larger. 
This Gibbs transfer energy strongly depends on both the solvents and the chemical nature 
of the ion \cite{marcus_cation, marcus_anion,persson_jcsft_1990,osakai_jpcb_1998}. Hence, 
the magnitude and sometimes the 
sign of the Gibbs transfer energy for cations and anions may differ greatly between 
mixtures \cite{onuki_cocis_2011,marcus_book}, as is seen in the curves of 
Fig.~\ref{fig_hefter}.

In the next two section we describe shortly  two ``applications'' of liquid-liquid 
demixing in polar solutions.

\section{Colloidal stabilization by addition of salt}\label{chap6_colloids}

Steric stabilization of colloids 
against the attractive van der Waals forces can be achieved using surfactant or
polymer molecules that are physically or chemically attached to the colloid's surface.
Charged colloids can also be stabilized via the screened Coulomb repulsion, whose range
depends on the Debye length $\lambda_D$. In the celebrated Derjaguin, Landau,
Verwey, and Overbeek (DLVO) theory \cite{dlvo1,dlvo2}, addition of salt to the suspension 
decreases the 
Debye length and the electrostatic repulsion leading eventually to
coagulation and sedimentation of the colloids \cite{colloids_book}.

In immiscible solvents selective solvation leads to a 
large ion partitioning
between the liquid phases. This phenomenon underlies liquid-liquid extraction, a
widely used separation method in chemical laboratories and in industry
\cite{extraction_book}. But only few works investigate how selective solvation
affects the interaction between charged surfaces. Leunissen \textit{et al.}
\cite{leunissen_pnas_2007,leunissen_pccp_2007} and Zwanikken \textit{et al.} 
\cite{zwanikken_prl_2007} showed experimentally and theoretically that ion 
partitioning in an oil-water mixture can be used to tune the structure of colloidal 
suspensions and to produce
additive-free water-in-oil emulsions that can crystallize. 

Several studies investigated experimentally the
interaction between charged surfaces in
mixtures of partially miscible solvents
\cite{beysens_prl_1985,leunissen_pnas_2007,bechinger_nature_2008,nellen_softmatter_2011,
evans_jcp_2009}. In a binary mixture of water and 2,6-lutidine, a reversible flocculation 
of colloids occurs close to the demixing curve depending on the type and amount of salt 
\cite{beysens_prl_1985,beysens_jcp_1991,beysens_pre_1998,beysens_jsp_1999}. It was only 
recently understood that selective
solvation plays an important role in such experiments
\cite{andelman_cocis_2011,zwanikken_jpcm_2008,tsori_epl_2011,onuki_pre_2011}
\cite{tsori_jcp_2012,bier_epl_2011,sadakane_cpl_2006,sadakane_jpsj_2007}
\cite{sadakane_softmatter_2011,sadakane_prl_2009} due to the non trivial electrostatics. 

The bulk free energy density in polar mixtures is $f=f_{\rm m}+f_{\rm es}+f_{\rm ion}$, 
where 
the electrostatic $f_{\rm es}$ and ionic free energies are now 
\cite{tsori_pnas_2007,andelman_jpcm_2009}
\begin{\ea}\label{polar_mix_bulk_energy}
f_{\rm es}&=&-\frac12\veps(\phi)(\nabla\psi)^2+(n^+-n^-)e\psi~,\nn\\
f_{\rm ion}&=&k_BT[n^+\ln(v_0n^+)+n^-\ln(v_0n^-)]-(\Delta u^+n^++\Delta u^-n^-)\phi~.
\end{\ea}
Here $n^\pm$ are the ionic number densities and $\Delta u^\pm$ are the parameters 
proportional to the Gibbs transfer energies. This bilinear coupling of the ionic density 
to the mixtures composition is the lowest possible order for a generally complex 
interaction between the ions and mixture \cite{bier_jcp_2012}. For a mixture confined by 
hard walls the 
surface energy density $f_{\rm s}$ is 
\begin{\ea}\label{polar_mix_surf_energy}
f_s=\Delta\gamma\phi({\bf r}_s)+\sigma\psi({\bf r}_s),
\end{\ea}
where ${\bf r}_s$ is a vector on the surface. The first term 
models the short-range interaction between the fluid and the 
solid. The parameter $\Delta\gamma$ measures the difference between the 
solid-water and solid-cosolvent surface tensions. The second term 
is the electrostatic energy for a surface with charge density $\sigma$.

Due to preferential solvation, when ions move to the electrodes they also ``drag'' the 
liquid in which they are better solvated, leading to a force of electrophoretic origin. 
Both electrophoretic ($\propto\Delta u$) and dielectrophoretic ($\propto 
\veps'$) forces lead to a phase separation transitions in liquid mixtures near 
charged surfaces. The window of temperatures above the binodal $\Delta T$ in which the 
mixture is unstable near one chemically-neutral ($\Delta\gamma=0$) wall charged at 
potential $V$, analogous to the difference 
between curve 3 and the binodal in Fig.~\ref{fig_stab_diagram}, is given by 
\cite{tsori_pnas_2007}
\begin{\ea}
\frac{\Delta T}{T_c}\simeq \left(\frac{|\veps'|}{\veps_c}+\frac{\Delta 
u}{k_BT_c}\right)\frac{n_0v_0}{|\phi_0-\phi_c|}\exp\left(\frac{eV}{k_BT_c}\right)~.
\end{\ea}
Note the exponential dependence on $V$, rendering the demixing possible 
at virtually all temperatures above the binodal. The thickness of the demixing layer is a 
combination of the Debye length $\lambda_D$ and the correlation length of the mixture, 
and is of the order of $10$nm unless $T$ is very close to $T_c$.
\begin{figure}[th!]
\begin{center}
\includegraphics[clip,width=0.8\textwidth]{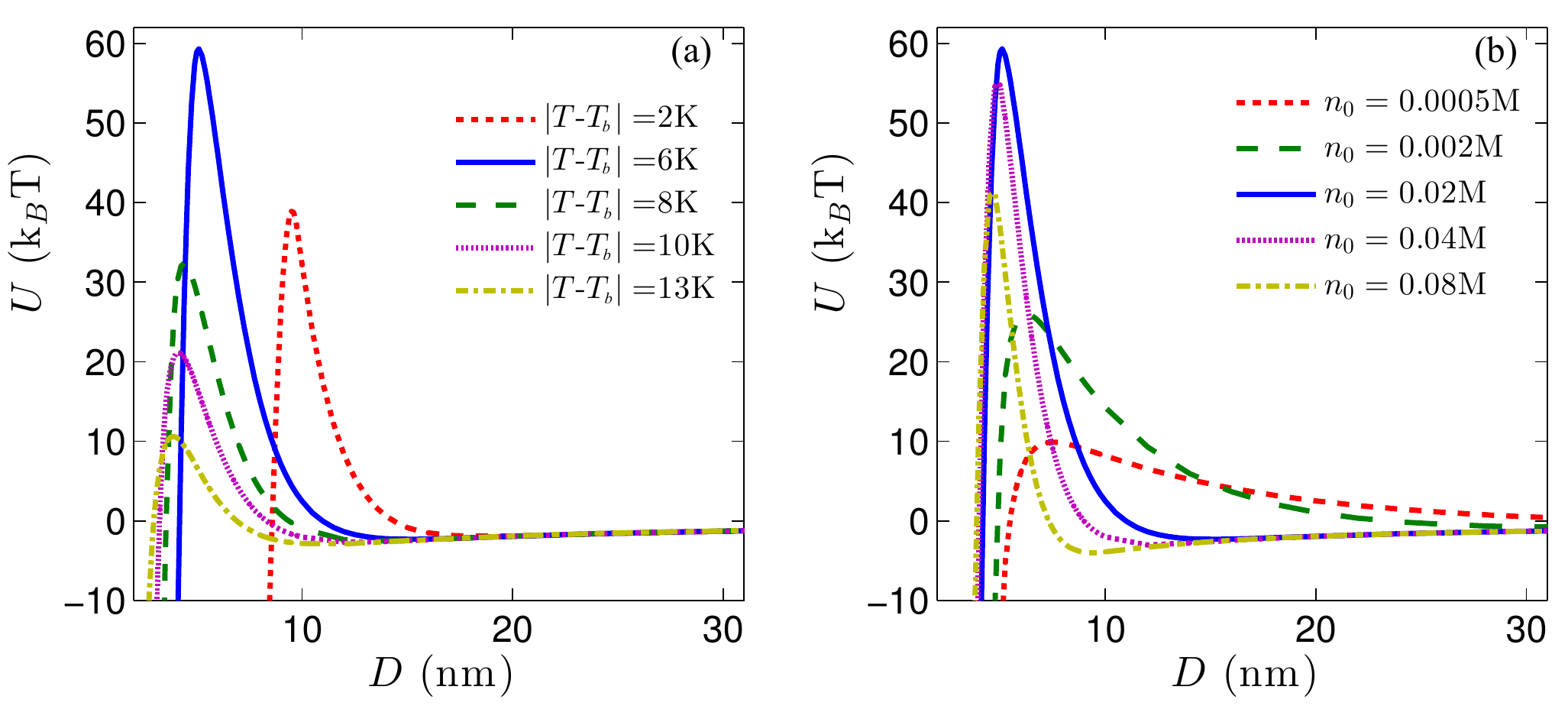}
\caption{\small Theoretical effective potential $U(D)$ between two colloids with
inter-surface separation $D$ immersed in a binary mixture with added antagonistic salt. 
The colloids repel at long distance and attract at short distance, with an energy barrier 
$U_{\rm max}$ at a distance $D_{\rm max}$. (a) Varying temperatures and fixed $n_0=20$mM 
(b) Varying salt amount and fixed $T=T_b+6$K. The location of the 
barrier at $D_{\rm max}$ decreases with increasing distance from the 
coexistence line $|T-T_b|$ or with increasing salt $n_0$ ($\lambda_D$ 
decreases). The surface of the colloids ($R=1\mu$m) was assumed hydrophobic and $\Delta 
u^+=-\Delta u^-=8k_BT$. Adapted from Ref. \cite{tsori_prapplied_2014}.
}
\label{fig_Ueff_profiles}
\end{center}
\end{figure}

Following the above insights it was suggested theoretically \cite{tsori_jcp_2013} and 
recently established experimentally \cite{tsori_prapplied_2014} that colloids that 
otherwise 
coagulate and sediment in mixtures can be stabilized by the {\it addition} of salt. The 
key idea is this: suppose a hydrophobic particle is immersed in a mixture of water and a 
co-solvent. The co-solvent will partially wet the particle while 
water will be depleted. Antagonistic ions are ion pairs where the cation is hydrophilic 
and the anion is hydrophobic (or vice versa) 
\cite{onuki_jcp_2004,pousaneh_softmatter_2014}. More generally, these are ions with a 
large 
difference in the Gibbs transfer energy, that is large $|\Delta u^+-\Delta u^-|$. 
If such ions are now added to the solution, 
the anions will be preferentially dissolved in the co-solvent (cation in the water) and 
hence the particles will be effectively charged and repelled from each other against van 
der Waals attraction. This repulsion is strong as long 
as the distance between them is not too small; at small separations the two co-solvent 
layers around the particle merge and capillary attraction together with van 
der Waals force become dominant leading to coagulation \cite{tsori_prapplied_2014}.

How to calculate the effective potential between two colloids with surface-to-surface 
separation $D$? One employs the variation principle for the total free energy with 
respect 
to the four fields: mixture composition $\delta f/\delta\phi=0$, electrostatic potential 
$\delta f/\delta\psi=0$ (yielding the Poisson equation), and the two ionic densities 
$\delta f/\delta n^\pm=0$ (yielding the Boltzmann's distribution for the ions). These 
equations are solved subject to the boundary conditions ${\bf 
n}\cdot\nabla\psi=\sigma/\veps(\phi)$ and ${\bf 
n}\cdot\nabla\phi=-\Delta\gamma/C$, where ${\bf n}$ is the unit vector normal to the 
colloid's surface, $\sigma$ is the surface charge density, $\Delta\gamma$ is the 
difference in the wettability of the two solvents at the colloid's surface, and $C$ is 
the prefactor of the $(1/2)(\nabla\phi)^2$ term in the free energy density. 
%
\begin{figure}[!th]
\subfloat[\label{fig_colloid_stab_contour_a}]{\includegraphics[keepaspectratio=true,
viewport=5 270 305 510,width=0.42\textwidth,clip]{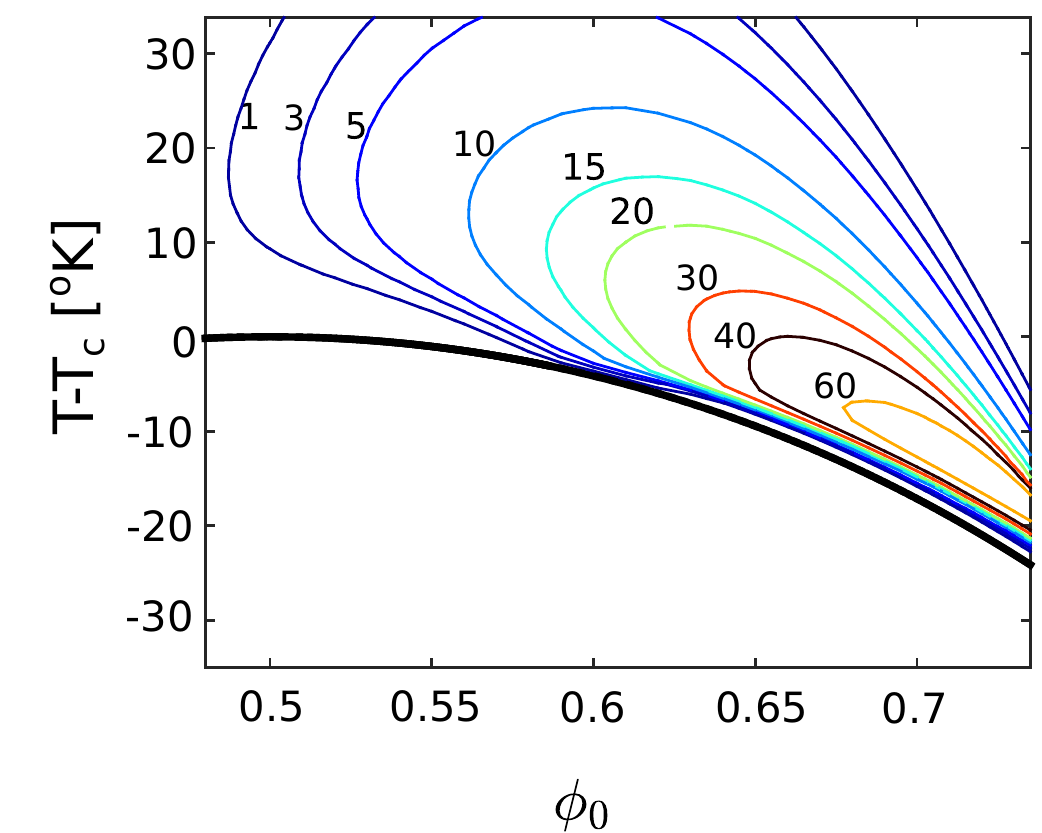}}
\subfloat[\label{fig_colloid_stab_contour_b}]{\includegraphics[keepaspectratio=true,
viewport=10 170 550 590,width=0.42\textwidth,clip]{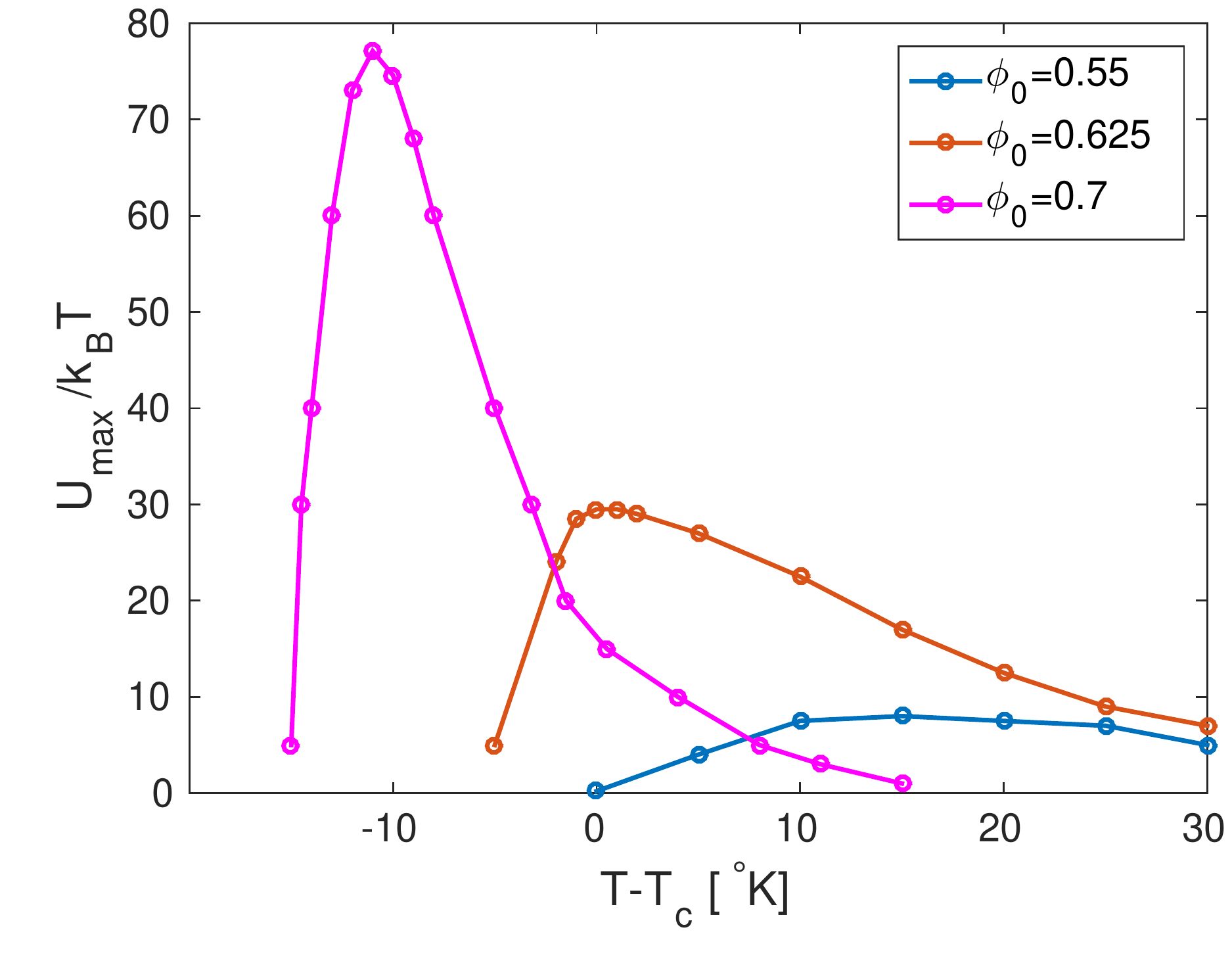}}
\caption{\small (a) A contour plot for the barrier height $U_{\rm max}$ in the 
$\phi_0$--$T$ plane in the Derjaguin's approximation including vdW attraction for a 
colloid radius $R=1\mu$m. (b) Vertical cross-sections of part (a) are shown as 
curves of $U_{\rm max}$ vs T for three different compositions . At $\phi_0=0.55$ the 
potential barrier is shallow and almost independent on $T$. At $\phi_0$ the barrier has a 
strong $T$ dependence only at $T$ close to the critical temperature $T_c$, while at 
$\phi_0$ the curve is steep at all temperatures. 
Adapted from Ref. \cite{tsori_jcp_2013}.}
\end{figure}

Once the equilibrium profiles are found all thermodynamic quantities are known. The 
pressure tensor is $p_{ij}=-\overleftrightarrow{T}_{ij}=\left(\phi \delta 
f/\delta\phi+n^+\delta f/\delta
n^++n^-\delta f/\delta n^--f\right)\delta_{ij}-\veps 
E_iE_j$\footnote{This expression for $p_{ij}$ is compatible with Eq. (\ref{stress_tensor}) 
: the terms in the brackets are $p_0+(1/2)\veps E^2$.}. 
The osmotic pressure is the 
difference between this expression and the bulk pressure, obtained when the composition 
and salt content equal their constant bulk values $\phi_0$ and $n_0$, respectively. 
If the distance between the colloids is smaller than their radius, $R\gg D$, the 
Derjaguin's approximation can be applied in this manner: the 
potential $\Omega(D)$ is defined as the integral of the osmotic pressure from $D$ 
to $\infty$. Then the effective inter-colloid potential, including the simplest form of 
van der Waals interaction with a Hamaker's constant $A$ is
\begin{\ea} \label{eq_U_of_D}
U(D)=\pi R\int_D^\infty\Omega(D')dD'-\frac{AR}{12D}
\end{\ea}

The resultant potential $U(D)$ is shown in Fig.~\ref{fig_Ueff_profiles} for various 
temperatures and salt concentrations. In all curves repulsion exists at large distances. 
The potential attains a maximum at a distance $D_{\rm max}\approx 5$nm, $U_{\rm 
max}=U(D_{\rm max})$. At short enough distances $U(D)$ becomes attractive and the 
colloids will stick to each other. The above calculation assumes reasonable numbers, such 
as colloid radius $R=1\mu$m, Hamaker constant $A\sim 10^{-21}$J and antagonistic ions 
with $|\Delta u^\pm|=8$ (other values in the caption). The qualitative behavior stays the 
same irrespective of the exact numerical values. As can be seen on the left panel, the 
dependence on $T$ is non-monotonic for fixed salt content. As $T$ approaches the 
binodal temperature $T_b$, $D_{\rm max}$ increases while the maximum 
$U_{\rm max}$ increases and then decreases. The behavior is also non-monotonic for fixed 
$T$ and increasing salt content -- in the right panel an increase in $n_0$ decreases 
$D_{\rm max}$ while $U_{\rm max}$ increases and then decreases. The location of the 
maximum at $D_{\rm max}$ is a nonlinear combination of the Debye length $\lambda_D$ and 
the correlation length of the mixture $\xi$. 

One can map the value of $D_{\rm max}$ onto the phase diagram in the $\phi_0$--$T$ plane. 
Fig.~\ref{fig_colloid_stab_contour_a} is such a contour plot  for a fixed amount of 
added salt. The highest inter-colloid barrier and the most effective colloidal 
stabilization is found asymmetrically for values of $\phi_0$ greater than $\phi_c$ 
($\phi_c=1/2$ in the model employed here) and for temperatures smaller than $T_c$. The 
experimental rule-of-thumb is that $U_{\rm max}$ needs to be larger than $\approx 3k_B$T 
in order to effectively prevent colloids from coagulation. The size of the area in the 
$\phi_0$--$T$ plane satisfying this requirement is generally large and independent of the 
exact numerical value of the parameters. 

This non-DLVO stabilization mechanism has been tested with two types of binary mixtures: 
water--2,6 lutidine (LCST) and water--acetonitrile (UCST). Using dynamic light 
scattering, visible-light transmission spectroscopy, and cryo-TEM it was verified 
that graphene flakes were indeed suspended in water--acetonitrile mixtures with 
antagonistic salt (NaBPh$_4$, Na$^+$ is hydrophilic, BPh$_4^-$ is hydrophobic) but not 
with ``regular'' salt (NaCl, both ions are hydrophilic). Cross-linked polystyrene 
colloidal spheres were successfully suspended in water--2,6 lutidine mixtures with 
antagonistic salt but not with NaCl \cite{tsori_prapplied_2014}.

An interesting opportunity arises for tuning the colloidal interactions with temperature. 
Fig.~\ref{fig_colloid_stab_contour_b} shows the temperature dependence of the 
potential barrier height for three different mixture compositions $\phi_0$. In the first 
curve, $\phi_0=0.55$, the system behavior is expected to be virtually independent of $T$ 
since $U_{\rm max}$ is shallow for all values of $T$. One could, however, work with 
a mixture with $\phi_0\approx 0.62$ (orange curve). In this case the system will be 
insensitive to temperature variations as long as $T\gtrsim T_c+10$K (small slope) but 
will be quite sensitive at lower temperatures. The third curve ($\phi_0=0.7$, magenta) is 
further off-critical, and here the curve is steep at all values of $T$, meaning that the 
system's stability is always sensitive to the temperature. The mixture composition is 
thus a valuable parameter for someone who wants to tune the stability of the dispersion 
against thermal variations between insensitive, partially sensitive, and sensitive. The 
additional boon of the method for certain applications is that the suspension is 
surfactant-free.

In the next section we investigate the role of electrostatics of liquid mixtures in pore 
filling transitions. 

\section{Pore filling transitions in membranes}\label{chap7_pore}

Membranes in liquids are ubiquitous in Nature and in technology. The function of 
the membrane's pore is to allow some species from one side to the other while blocking 
the rest 
\cite{katchalsky_bpa_1958,mackinnon_nature_2002,mackinnon_nature_2003,hoek_ees_2011}. 
It is advantageous to gate the pores to achieve better control of the dynamics 
and selectivity of the transport. Reversible pore gating has been achieved by 
temperature differences \cite{yang_jms_2003,azzaroni_small_2009}. An alternative 
strategy, providing good anti-fouling properties, has recently been 
demonstrated in liquid-gated pores controlled by pressure variations across the 
membrane \cite{aizenberg_nature_2015}.
Electric potentials hold great promise because they they can be implemented in 
most systems, they can easily be turned on or off, and the voltages required are 
quite low:  the field $E\sim V/\lambda_D$ is large even if $V$ small since $\lambda_D$ is 
on the nanometre scale. Indeed several works recently examined the effect of electric 
fields on the permeability of ions and solutes across the membranes and the influence on 
liquid-vapour coexistence near the pore 
\cite{hansen_jcp_2004,hansen_jcp_2005,siwy_nature_nanotech_2011,lavrik_acsnano_2011}. 
In these works the existence of a transmembrane potential means that the direction of the 
field is parallel to the pore's axis (and parallel to the pore walls).

A different direction for pore gating has been proposed recently, whereby the membrane is 
immersed in an aqueous mixture and the membrane is charged, leading to field in the 
direction perpendicular to the pore's axis. For hydrophobic membranes, the pore opens and 
fills with water when voltage is applied to the membrane. The pore closes by filling with 
the co-solvent which diffuses in after the voltage is removed \cite{tsori_cisc_2016}.
The filling transitions, expected to occur in both hydrophobic and hydrophilic membranes, 
can be triggered also by small changes in mixture composition or ambient 
temperature, and they can be continuous or abrupt. 
The role of preferential solvation of ions is crucial as it enables the 
transitions even for highly hydrophobic pores and at elevated temperatures above 
the critical temperature, unlike in regular capillary condensation \cite{onuki_pre_2010}. 
\begin{figure}
\centering
\begin{minipage}{12cm}
~~\includegraphics[viewport=5 185 435 
390,width=0.58\textwidth,clip]{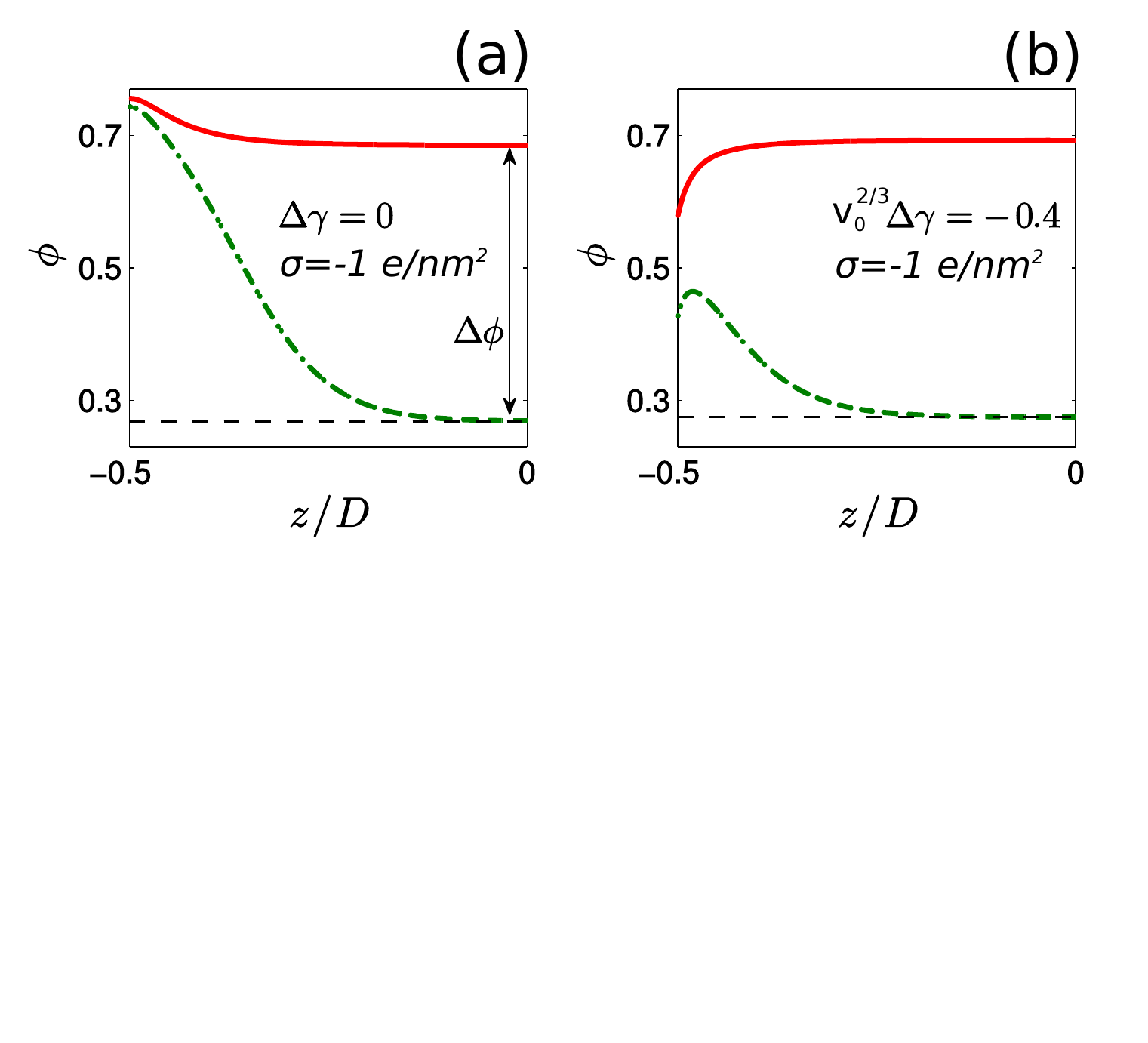}
\end{minipage}
\\ \vspace{0.3cm}
\begin{minipage}{12cm}
\includegraphics[viewport=0 0 420 
320,width=0.7\textwidth,clip]{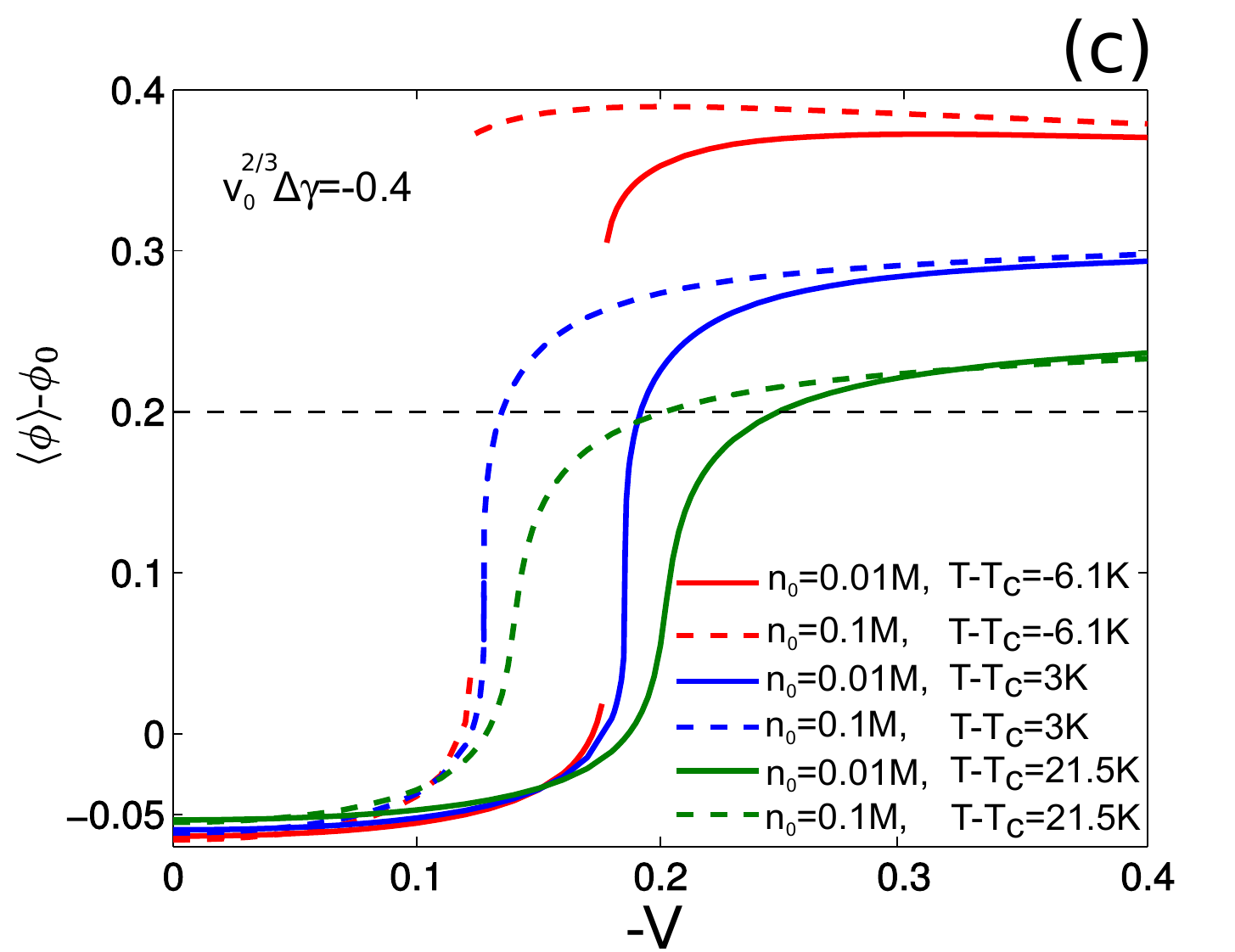}
\end{minipage}
\caption{\small (a) Composition profile for chemically neutral ($\Delta\gamma =0$) and 
electrically charged ($\sigma\neq 0$) pore of width $D$ as a function of the transverse 
coordinate $z$ ($-D/2\leq z\leq D/2$). Only negative values of $z$ are shown since the 
pore is symmetric ($\phi(z)=\phi(-z)$). The cation and anion are both hydrophilic and 
calculation were done within the Modified Poisson-Boltzmann framework 
\cite{borukhov_prl_1997}.
Dash-dot and solid curves are $\phi(z)$ just before and just after the filling 
composition. The horizontal dashed line is the bulk composition $\phi_0$. $\Delta \phi$, 
the difference between $\phi(z=0)$ and $\phi_0$, is vanishingly small before the filling 
transition and finite after it. (b) The same as in (a) but the pore is hydrophobic, 
$\Delta\gamma<0$. (c) Average water composition $\langle\phi\rangle$ in a hydrophobic 
pore vs negative pore potential $V$ (in Volts) for three different temperatures and two 
salt contents $n_0$. Adapted from Ref. \cite{tsori_cisc_2016} (values of parameters 
therein). 
}
\label{fig_pore_filling}
\end{figure}

The electrostatic potential, mixture composition, and ionic density profiles are 
calculated in the same manner as in Sec. \ref{chap6_colloids}, by employing the 
variational principle with respect to these fields, $\delta f/\delta\phi=0$, $\delta 
f/\delta\psi=0$, and $\delta f/\delta n^\pm=0$, where $f=f_{\rm m}+f_{\rm es}+f_{\rm 
ion}$, and 
$f_{\rm es}$ and $f_{\rm ion}$ are given in Eq. (\ref{polar_mix_bulk_energy}) and the 
mixture energy $f_{\rm m}$ includes a $(1/2)C(\nabla\phi)^2$ term. The equations are 
solved 
subject to the boundary conditions derivable from the surface energy in Eq. 
(\ref{polar_mix_surf_energy}): ${\bf n}\cdot\nabla\psi=\sigma/\veps(\phi)$ and ${\bf 
n}\cdot\nabla\phi=-\Delta\gamma/C$.

The short-range preference of the pore wall to water or to the co-solvent leads to 
gradients in $\phi(z)$; surface charge density leads to gradients too. How to define then 
whether the pore is filled or not? Denoting $\Delta \phi$ as the difference between the 
mid-pore composition $\phi(z=0)$ and the bulk value $\phi_0$, before the 
filling transition $\Delta\phi\approx 0$ whereas 
after it $\Delta\phi$ is finite. Fig.~\ref{fig_pore_filling} (a) shows $\phi(z)$ just 
before (dash-dot) and after (solid) the transition for a highly charged pore in a 
membrane that is neither hydrophilic nor hydrophobic. Both profiles are monotonously 
decreasing but the solid curve tends to $\approx 0.69>\phi_c$ in this case leading to a 
finite value of $\Delta\phi$. In part (b) of the Figure the corresponding 
profiles are shown with the change that now the pore is highly hydrophobic, as evidenced 
by the slope of both dash-dot and solid curves close to the left boundary ($z/D=-1/2$). 
The thickness of the adsorption layers in the curves corresponds to a modified Debye 
length $\lambda_D$ that depends on $n_0$, $\phi_0$, and $\Delta u^\pm$ 
\cite{tsori_cisc_2016}. 

For porous carbonaceous membranes and other types of membranes one can
control externally the surface potential within certain constraints. To quantify 
the filling transition as a function of external potential the average 
pore water composition is defined as 
$\langle\phi\rangle=(1/D)\int_{-D/2}^{D/2}\phi(z)dz$. The variation of 
$\langle\phi\rangle$ in a hydrophobic pore with increasing values of (negative) wall 
potential is shown in Fig.~\ref{fig_pore_filling} (b) for two salt contents and 
three temperatures. At zero potential, water is depleted from the pore, 
$\langle\phi\rangle<\phi_0$. The water content increases with increasing $|V|$.
Water is the majority in the pore once $\langle\phi\rangle-\phi_0$ exceeds $0.2$, 
since in this Figure $\phi_0=0.3$. The filling is continuous for the two high 
temperatures $T=T_c+21.5$K and $T=T_c+3$K and discontinuous at $T=T_c-6.1$K (the bulk 
mixture is homogeneous at this temperature). The voltage to (partially) fill the pore 
with water decreases with decreasing $T$ or with increasing $n_0$. 

Pore filling occurs also in regular capillary condensation phenomena with
short-range interactions. The selective solvation of the ions, being a volume 
contribution to the free energy, is dominant when either (i) the salt 
concentration is large, (ii) the membrane is highly charged, or (iii) the Gibbs transfer 
energy is large.

\section{Outlook}\label{chap8}

The dynamics and thermodynamics of liquids in electric fields is classical physics (pun 
intended). The elliptical nature of Laplace's equation means that boundaries dictate the 
electric forces that occur remotely from them. Despite our old and tested knowledge of 
the mathematical formulae for the energy densities and the stresses, in complex, 
multi-component liquids, the system's behavior is often difficult to predict or even 
describe. Enormous advances have been made in description of shape change, interfacial 
dynamics and related instabilities and the field is continuously evolving 
\cite{melcher_book,bazant_cocis_2010,chen_book_chapter, 
yariv_jfm_2015,zaltzman_prl_2015,cruz_pnas_2013,onuki_cocis_2016}. We have 
highlighted how the preferential solvation of ions leads to forces that change
dramatically the thermodynamics of liquid mixtures. This ingredient is expected to 
modify the classical models considerably. The addition of a new energy scale is especially 
intriguing in problems with time-varying fields where heating effects can be dominant. 
Researchers on these topics have plenty of work.

\section{Acknowledgments}

This work was supported by the European Research Council ``Starting Grant'' No. 259205, 
COST Action MP1106, and Israel Science Foundation Grant No. 56/14.

\bibliography{jpcm_refs}

\begin{thebibliography}{132}%
\makeatletter
\providecommand \@ifxundefined [1]{%
 \@ifx{#1\undefined}
}%
\providecommand \@ifnum [1]{%
 \ifnum #1\expandafter \@firstoftwo
 \else \expandafter \@secondoftwo
 \fi
}%
\providecommand \@ifx [1]{%
 \ifx #1\expandafter \@firstoftwo
 \else \expandafter \@secondoftwo
 \fi
}%
\providecommand \natexlab [1]{#1}%
\providecommand \enquote  [1]{``#1''}%
\providecommand \bibnamefont  [1]{#1}%
\providecommand \bibfnamefont [1]{#1}%
\providecommand \citenamefont [1]{#1}%
\providecommand \href@noop [0]{\@secondoftwo}%
\providecommand \href [0]{\begingroup \@sanitize@url \@href}%
\providecommand \@href[1]{\@@startlink{#1}\@@href}%
\providecommand \@@href[1]{\endgroup#1\@@endlink}%
\providecommand \@sanitize@url [0]{\catcode `\\12\catcode `\$12\catcode
  `\&12\catcode `\#12\catcode `\^12\catcode `\_12\catcode `\%12\relax}%
\providecommand \@@startlink[1]{}%
\providecommand \@@endlink[0]{}%
\providecommand \url  [0]{\begingroup\@sanitize@url \@url }%
\providecommand \@url [1]{\endgroup\@href {#1}{\urlprefix }}%
\providecommand \urlprefix  [0]{URL }%
\providecommand \Eprint [0]{\href }%
\providecommand \doibase [0]{http://dx.doi.org/}%
\providecommand \selectlanguage [0]{\@gobble}%
\providecommand \bibinfo  [0]{\@secondoftwo}%
\providecommand \bibfield  [0]{\@secondoftwo}%
\providecommand \translation [1]{[#1]}%
\providecommand \BibitemOpen [0]{}%
\providecommand \bibitemStop [0]{}%
\providecommand \bibitemNoStop [0]{.\EOS\space}%
\providecommand \EOS [0]{\spacefactor3000\relax}%
\providecommand \BibitemShut  [1]{\csname bibitem#1\endcsname}%
\let\auto@bib@innerbib\@empty
\bibitem [{\citenamefont {Landau}\ and\ \citenamefont
  {Lifshitz}(1957)}]{LL_book_elec}%
  \BibitemOpen
  \bibfield  {author} {\bibinfo {author} {\bibfnamefont {L.~D.}\ \bibnamefont
  {Landau}}\ and\ \bibinfo {author} {\bibfnamefont {E.~M.}\ \bibnamefont
  {Lifshitz}},\ }\href@noop {} {\emph {\bibinfo {title} {Elektrodinamika
  Sploshnykh Sred}}}\ (\bibinfo  {publisher} {Nauka, Moscow},\ \bibinfo {year}
  {1957})\ \bibinfo {note} {chap. II, Sec. 18, problem 1}\BibitemShut {NoStop}%
\bibitem [{\citenamefont {Melcher}(1963)}]{melcher_book}%
  \BibitemOpen
  \bibfield  {author} {\bibinfo {author} {\bibfnamefont {J.~R.}\ \bibnamefont
  {Melcher}},\ }\href@noop {} {\emph {\bibinfo {title} {Field-Coupled Surface
  Waves: A Comparative Study of Surface-Coupled Electrohydrodynamic and
  Magnetohydrodynamic Systems}}}\ (\bibinfo  {publisher} {M.I.T},\ \bibinfo
  {year} {1963})\BibitemShut {NoStop}%
\bibitem [{\citenamefont {Saville}(1997)}]{saville_arfm_1997}%
  \BibitemOpen
  \bibfield  {author} {\bibinfo {author} {\bibfnamefont {D.~A.}\ \bibnamefont
  {Saville}},\ }\href@noop {} {\bibfield  {journal} {\bibinfo  {journal} {Annu.
  Rev. Fluid Mech.}\ }\textbf {\bibinfo {volume} {29}},\ \bibinfo {pages} {27}
  (\bibinfo {year} {1997})}\BibitemShut {NoStop}%
\bibitem [{\citenamefont {K.~H.~Panofsky}\ and\ \citenamefont
  {Phillips}(2005)}]{panofsky_phillips_book}%
  \BibitemOpen
  \bibfield  {author} {\bibinfo {author} {\bibfnamefont {W.}~\bibnamefont
  {K.~H.~Panofsky}}\ and\ \bibinfo {author} {\bibfnamefont {M.}~\bibnamefont
  {Phillips}},\ }\href@noop {} {\emph {\bibinfo {title} {Classical Electricity
  and Magnetism}}},\ \bibinfo {edition} {2nd}\ ed.\ (\bibinfo  {publisher}
  {Dover Publications},\ \bibinfo {year} {2005})\BibitemShut {NoStop}%
\bibitem [{\citenamefont {Herminghaus}(1999)}]{hermin_prl_1999}%
  \BibitemOpen
  \bibfield  {author} {\bibinfo {author} {\bibfnamefont {S.}~\bibnamefont
  {Herminghaus}},\ }\href@noop {} {\bibfield  {journal} {\bibinfo  {journal}
  {Phys. Rev. Lett.}\ }\textbf {\bibinfo {volume} {83}},\ \bibinfo {pages}
  {2359} (\bibinfo {year} {1999})}\BibitemShut {NoStop}%
\bibitem [{\citenamefont {Pease}\ and\ \citenamefont
  {Russel}(2002)}]{russel_jnnfm_2002}%
  \BibitemOpen
  \bibfield  {author} {\bibinfo {author} {\bibfnamefont {L.~F.}\ \bibnamefont
  {Pease}}\ and\ \bibinfo {author} {\bibfnamefont {W.~B.}\ \bibnamefont
  {Russel}},\ }\href@noop {} {\bibfield  {journal} {\bibinfo  {journal} {J.
  Non-Newtonian Fluid Mech.}\ }\textbf {\bibinfo {volume} {102}},\ \bibinfo
  {pages} {233} (\bibinfo {year} {2002})}\BibitemShut {NoStop}%
\bibitem [{\citenamefont {Schaffer}\ \emph {et~al.}(2000)\citenamefont
  {Schaffer}, \citenamefont {Thurn-Albrecht}, \citenamefont {Russell},\ and\
  \citenamefont {Steiner}}]{russell_nature_2000}%
  \BibitemOpen
  \bibfield  {author} {\bibinfo {author} {\bibfnamefont {E.}~\bibnamefont
  {Schaffer}}, \bibinfo {author} {\bibfnamefont {T.}~\bibnamefont
  {Thurn-Albrecht}}, \bibinfo {author} {\bibfnamefont {T.}~\bibnamefont
  {Russell}}, \ and\ \bibinfo {author} {\bibfnamefont {U.}~\bibnamefont
  {Steiner}},\ }\href@noop {} {\bibfield  {journal} {\bibinfo  {journal}
  {Nature}\ }\textbf {\bibinfo {volume} {403}},\ \bibinfo {pages} {874}
  (\bibinfo {year} {2000})}\BibitemShut {NoStop}%
\bibitem [{\citenamefont {Lau}\ and\ \citenamefont
  {Russel}(2011)}]{russel_mm_2011}%
  \BibitemOpen
  \bibfield  {author} {\bibinfo {author} {\bibfnamefont {C.~Y.}\ \bibnamefont
  {Lau}}\ and\ \bibinfo {author} {\bibfnamefont {W.~B.}\ \bibnamefont
  {Russel}},\ }\href {\doibase 10.1021/ma200952u} {\bibfield  {journal}
  {\bibinfo  {journal} {Macromolecules}\ }\textbf {\bibinfo {volume} {44}},\
  \bibinfo {pages} {7746} (\bibinfo {year} {2011})}\BibitemShut {NoStop}%
\bibitem [{\citenamefont {Lin}\ \emph {et~al.}(2001)\citenamefont {Lin},
  \citenamefont {Kerle}, \citenamefont {Baker}, \citenamefont {Hoagland},
  \citenamefont {Schaffer}, \citenamefont {Steiner},\ and\ \citenamefont
  {Russell}}]{russell_jcp_2001}%
  \BibitemOpen
  \bibfield  {author} {\bibinfo {author} {\bibfnamefont {Z.}~\bibnamefont
  {Lin}}, \bibinfo {author} {\bibfnamefont {T.}~\bibnamefont {Kerle}}, \bibinfo
  {author} {\bibfnamefont {S.}~\bibnamefont {Baker}}, \bibinfo {author}
  {\bibfnamefont {D.}~\bibnamefont {Hoagland}}, \bibinfo {author}
  {\bibfnamefont {E.}~\bibnamefont {Schaffer}}, \bibinfo {author}
  {\bibfnamefont {U.}~\bibnamefont {Steiner}}, \ and\ \bibinfo {author}
  {\bibfnamefont {T.}~\bibnamefont {Russell}},\ }\href {\doibase
  {10.1063/1.1338125}} {\bibfield  {journal} {\bibinfo  {journal} {J. Chem.
  Phys.}\ }\textbf {\bibinfo {volume} {114}},\ \bibinfo {pages} {2377}
  (\bibinfo {year} {2001})}\BibitemShut {NoStop}%
\bibitem [{\citenamefont {Lin}\ \emph {et~al.}(2002)\citenamefont {Lin},
  \citenamefont {Kerle}, \citenamefont {Russell}, \citenamefont {Schaffer},\
  and\ \citenamefont {Steiner}}]{russell_mm_2002a}%
  \BibitemOpen
  \bibfield  {author} {\bibinfo {author} {\bibfnamefont {Z.}~\bibnamefont
  {Lin}}, \bibinfo {author} {\bibfnamefont {T.}~\bibnamefont {Kerle}}, \bibinfo
  {author} {\bibfnamefont {T.}~\bibnamefont {Russell}}, \bibinfo {author}
  {\bibfnamefont {E.}~\bibnamefont {Schaffer}}, \ and\ \bibinfo {author}
  {\bibfnamefont {U.}~\bibnamefont {Steiner}},\ }\href {\doibase
  {10.1021/ma0122425}} {\bibfield  {journal} {\bibinfo  {journal}
  {Macromolecules}\ }\textbf {\bibinfo {volume} {35}},\ \bibinfo {pages} {3971}
  (\bibinfo {year} {2002})}\BibitemShut {NoStop}%
\bibitem [{\citenamefont {Wu}\ \emph {et~al.}(2006)\citenamefont {Wu},
  \citenamefont {Pease},\ and\ \citenamefont {Russel}}]{russel_afm_2006}%
  \BibitemOpen
  \bibfield  {author} {\bibinfo {author} {\bibfnamefont {N.}~\bibnamefont
  {Wu}}, \bibinfo {author} {\bibfnamefont {L.~F.}\ \bibnamefont {Pease},
  \bibfnamefont {III}}, \ and\ \bibinfo {author} {\bibfnamefont {W.~B.}\
  \bibnamefont {Russel}},\ }\href {\doibase {10.1002/adfm.200600092}}
  {\bibfield  {journal} {\bibinfo  {journal} {Adv. Funct. Mater.}\ }\textbf
  {\bibinfo {volume} {16}},\ \bibinfo {pages} {1992} (\bibinfo {year}
  {2006})}\BibitemShut {NoStop}%
\bibitem [{\citenamefont {Goldberg-Oppenheimer}\ and\ \citenamefont
  {Steiner}(2010)}]{steiner_small_2010}%
  \BibitemOpen
  \bibfield  {author} {\bibinfo {author} {\bibfnamefont {P.}~\bibnamefont
  {Goldberg-Oppenheimer}}\ and\ \bibinfo {author} {\bibfnamefont
  {U.}~\bibnamefont {Steiner}},\ }\href {\doibase {10.1002/smll.201000060}}
  {\bibfield  {journal} {\bibinfo  {journal} {Small}\ }\textbf {\bibinfo
  {volume} {6}},\ \bibinfo {pages} {1248} (\bibinfo {year} {2010})}\BibitemShut
  {NoStop}%
\bibitem [{\citenamefont {O'Konski}\ and\ \citenamefont
  {Thacher}(1953)}]{okonski_jpc_1953}%
  \BibitemOpen
  \bibfield  {author} {\bibinfo {author} {\bibfnamefont {C.~T.}\ \bibnamefont
  {O'Konski}}\ and\ \bibinfo {author} {\bibfnamefont {H.~C.}\ \bibnamefont
  {Thacher}},\ }\href@noop {} {\bibfield  {journal} {\bibinfo  {journal} {J.
  Phys. Chem.}\ }\textbf {\bibinfo {volume} {57}},\ \bibinfo {pages} {955}
  (\bibinfo {year} {1953})}\BibitemShut {NoStop}%
\bibitem [{\citenamefont {Allan}\ and\ \citenamefont
  {Mason}(1962)}]{mason_prsl_1962}%
  \BibitemOpen
  \bibfield  {author} {\bibinfo {author} {\bibfnamefont {R.~S.}\ \bibnamefont
  {Allan}}\ and\ \bibinfo {author} {\bibfnamefont {S.~G.}\ \bibnamefont
  {Mason}},\ }\href@noop {} {\bibfield  {journal} {\bibinfo  {journal} {Proc.
  R. Soc. London, Ser. A}\ }\textbf {\bibinfo {volume} {267}},\ \bibinfo
  {pages} {45} (\bibinfo {year} {1962})}\BibitemShut {NoStop}%
\bibitem [{\citenamefont {Taylor}(1966)}]{taylor_prsl_1966}%
  \BibitemOpen
  \bibfield  {author} {\bibinfo {author} {\bibfnamefont {G.~I.}\ \bibnamefont
  {Taylor}},\ }\href@noop {} {\bibfield  {journal} {\bibinfo  {journal} {Proc.
  R. Soc. London, Ser. A}\ }\textbf {\bibinfo {volume} {291}},\ \bibinfo
  {pages} {159} (\bibinfo {year} {1966})}\BibitemShut {NoStop}%
\bibitem [{\citenamefont {Melcher}\ and\ \citenamefont
  {Taylor}(1969)}]{taylor_arfd_1969}%
  \BibitemOpen
  \bibfield  {author} {\bibinfo {author} {\bibfnamefont {J.~R.}\ \bibnamefont
  {Melcher}}\ and\ \bibinfo {author} {\bibfnamefont {G.~I.}\ \bibnamefont
  {Taylor}},\ }\href@noop {} {\bibfield  {journal} {\bibinfo  {journal} {Annual
  Reviews of Fluid Dynamics}\ }\textbf {\bibinfo {volume} {1}},\ \bibinfo
  {pages} {111} (\bibinfo {year} {1969})}\BibitemShut {NoStop}%
\bibitem [{\citenamefont {Schaffer}\ \emph {et~al.}(2001)\citenamefont
  {Schaffer}, \citenamefont {Thurn-Albrecht}, \citenamefont {Russell},\ and\
  \citenamefont {Steiner}}]{russell_epl_2001}%
  \BibitemOpen
  \bibfield  {author} {\bibinfo {author} {\bibfnamefont {E.}~\bibnamefont
  {Schaffer}}, \bibinfo {author} {\bibfnamefont {T.}~\bibnamefont
  {Thurn-Albrecht}}, \bibinfo {author} {\bibfnamefont {T.~P.}\ \bibnamefont
  {Russell}}, \ and\ \bibinfo {author} {\bibfnamefont {U.}~\bibnamefont
  {Steiner}},\ }\href@noop {} {\bibfield  {journal} {\bibinfo  {journal} {EPL}\
  }\textbf {\bibinfo {volume} {53}},\ \bibinfo {pages} {518} (\bibinfo {year}
  {2001})}\BibitemShut {NoStop}%
\bibitem [{\citenamefont {Morariu}\ \emph {et~al.}(2003)\citenamefont
  {Morariu}, \citenamefont {Voicu}, \citenamefont {Schaffer}, \citenamefont
  {Lin}, \citenamefont {Russell},\ and\ \citenamefont
  {Steiner}}]{russell_nature_mat_2003}%
  \BibitemOpen
  \bibfield  {author} {\bibinfo {author} {\bibfnamefont {M.}~\bibnamefont
  {Morariu}}, \bibinfo {author} {\bibfnamefont {N.}~\bibnamefont {Voicu}},
  \bibinfo {author} {\bibfnamefont {E.}~\bibnamefont {Schaffer}}, \bibinfo
  {author} {\bibfnamefont {Z.}~\bibnamefont {Lin}}, \bibinfo {author}
  {\bibfnamefont {T.}~\bibnamefont {Russell}}, \ and\ \bibinfo {author}
  {\bibfnamefont {U.}~\bibnamefont {Steiner}},\ }\href {\doibase
  {10.1038/nmat789}} {\bibfield  {journal} {\bibinfo  {journal} {Nature
  Materials}\ }\textbf {\bibinfo {volume} {2}},\ \bibinfo {pages} {48}
  (\bibinfo {year} {2003})}\BibitemShut {NoStop}%
\bibitem [{\citenamefont {Debye}\ and\ \citenamefont
  {Kleboth}(1965)}]{debye_jcp_1965}%
  \BibitemOpen
  \bibfield  {author} {\bibinfo {author} {\bibfnamefont {P.}~\bibnamefont
  {Debye}}\ and\ \bibinfo {author} {\bibfnamefont {K.}~\bibnamefont
  {Kleboth}},\ }\href@noop {} {\bibfield  {journal} {\bibinfo  {journal} {J.
  Chem. Phys.}\ }\textbf {\bibinfo {volume} {42}},\ \bibinfo {pages} {3155}
  (\bibinfo {year} {1965})}\BibitemShut {NoStop}%
\bibitem [{\citenamefont {Orzechowski}(1999)}]{orzech_chemphys_1999}%
  \BibitemOpen
  \bibfield  {author} {\bibinfo {author} {\bibfnamefont {K.}~\bibnamefont
  {Orzechowski}},\ }\href@noop {} {\bibfield  {journal} {\bibinfo  {journal}
  {Chem. Phys.}\ }\textbf {\bibinfo {volume} {240}},\ \bibinfo {pages} {275}
  (\bibinfo {year} {1999})}\BibitemShut {NoStop}%
\bibitem [{\citenamefont {Wirtz}\ and\ \citenamefont
  {Fuller}(1993)}]{wirtz_fuller_prl_1993}%
  \BibitemOpen
  \bibfield  {author} {\bibinfo {author} {\bibfnamefont {D.}~\bibnamefont
  {Wirtz}}\ and\ \bibinfo {author} {\bibfnamefont {G.~G.}\ \bibnamefont
  {Fuller}},\ }\href@noop {} {\bibfield  {journal} {\bibinfo  {journal} {Phys.
  Rev. Lett.}\ }\textbf {\bibinfo {volume} {71}},\ \bibinfo {pages} {2236}
  (\bibinfo {year} {1993})}\BibitemShut {NoStop}%
\bibitem [{\citenamefont {Beaglehole}(1981)}]{beaglehole_jcp_1981}%
  \BibitemOpen
  \bibfield  {author} {\bibinfo {author} {\bibfnamefont {D.}~\bibnamefont
  {Beaglehole}},\ }\href@noop {} {\bibfield  {journal} {\bibinfo  {journal} {J.
  Chem. Phys.}\ }\textbf {\bibinfo {volume} {74}},\ \bibinfo {pages} {5251}
  (\bibinfo {year} {1981})}\BibitemShut {NoStop}%
\bibitem [{\citenamefont {Early}(1992)}]{early_jcp_1992}%
  \BibitemOpen
  \bibfield  {author} {\bibinfo {author} {\bibfnamefont {M.~D.}\ \bibnamefont
  {Early}},\ }\href@noop {} {\bibfield  {journal} {\bibinfo  {journal} {J.
  Chem. Phys.}\ }\textbf {\bibinfo {volume} {96}},\ \bibinfo {pages} {641}
  (\bibinfo {year} {1992})}\BibitemShut {NoStop}%
\bibitem [{\citenamefont {Reich}\ and\ \citenamefont
  {Gordon}(1979)}]{reich_jpspp_1979}%
  \BibitemOpen
  \bibfield  {author} {\bibinfo {author} {\bibfnamefont {S.}~\bibnamefont
  {Reich}}\ and\ \bibinfo {author} {\bibfnamefont {J.~M.}\ \bibnamefont
  {Gordon}},\ }\href@noop {} {\bibfield  {journal} {\bibinfo  {journal} {J.
  Polym. Sci., Part B: Polym. Phys.}\ }\textbf {\bibinfo {volume} {17}},\
  \bibinfo {pages} {371} (\bibinfo {year} {1979})}\BibitemShut {NoStop}%
\bibitem [{\citenamefont {Kriisa}\ and\ \citenamefont
  {Roth}(2014)}]{roth_jcp_2014}%
  \BibitemOpen
  \bibfield  {author} {\bibinfo {author} {\bibfnamefont {A.}~\bibnamefont
  {Kriisa}}\ and\ \bibinfo {author} {\bibfnamefont {C.~B.}\ \bibnamefont
  {Roth}},\ }\href@noop {} {\bibfield  {journal} {\bibinfo  {journal} {J. Chem.
  Phys.}\ }\textbf {\bibinfo {volume} {141}},\ \bibinfo {pages} {134908}
  (\bibinfo {year} {2014})}\BibitemShut {NoStop}%
\bibitem [{\citenamefont {Amundson}\ \emph {et~al.}(1994)\citenamefont
  {Amundson}, \citenamefont {Helfand}, \citenamefont {Quan}, \citenamefont
  {Hudson},\ and\ \citenamefont {Smith}}]{ah_mm_1994}%
  \BibitemOpen
  \bibfield  {author} {\bibinfo {author} {\bibfnamefont {K.}~\bibnamefont
  {Amundson}}, \bibinfo {author} {\bibfnamefont {E.}~\bibnamefont {Helfand}},
  \bibinfo {author} {\bibfnamefont {X.}~\bibnamefont {Quan}}, \bibinfo {author}
  {\bibfnamefont {S.~D.}\ \bibnamefont {Hudson}}, \ and\ \bibinfo {author}
  {\bibfnamefont {S.~D.}\ \bibnamefont {Smith}},\ }\href@noop {} {\bibfield
  {journal} {\bibinfo  {journal} {Macromolecules}\ }\textbf {\bibinfo {volume}
  {27}},\ \bibinfo {pages} {6559} (\bibinfo {year} {1994})}\BibitemShut
  {NoStop}%
\bibitem [{\citenamefont {Tsori}(2009)}]{tsori_rmp_2009}%
  \BibitemOpen
  \bibfield  {author} {\bibinfo {author} {\bibfnamefont {Y.}~\bibnamefont
  {Tsori}},\ }\href {\doibase 10.1103/RevModPhys.81.1471} {\bibfield  {journal}
  {\bibinfo  {journal} {Rev. Mod. Phys.}\ }\textbf {\bibinfo {volume} {81}},\
  \bibinfo {pages} {1471} (\bibinfo {year} {2009})}\BibitemShut {NoStop}%
\bibitem [{\citenamefont {Onuki}\ and\ \citenamefont
  {Fukuda}(1995)}]{onuki_mm_1995}%
  \BibitemOpen
  \bibfield  {author} {\bibinfo {author} {\bibfnamefont {A.}~\bibnamefont
  {Onuki}}\ and\ \bibinfo {author} {\bibfnamefont {J.}~\bibnamefont {Fukuda}},\
  }\href {\doibase 10.1021/ma00130a011} {\bibfield  {journal} {\bibinfo
  {journal} {Macromolecules}\ }\textbf {\bibinfo {volume} {28}},\ \bibinfo
  {pages} {8788} (\bibinfo {year} {1995})}\BibitemShut {NoStop}%
\bibitem [{\citenamefont {Orzechowski}\ \emph {et~al.}(2014)\citenamefont
  {Orzechowski}, \citenamefont {Adamczyk}, \citenamefont {Wolny},\ and\
  \citenamefont {Tsori}}]{tsori_jpcb_2014}%
  \BibitemOpen
  \bibfield  {author} {\bibinfo {author} {\bibfnamefont {K.}~\bibnamefont
  {Orzechowski}}, \bibinfo {author} {\bibfnamefont {M.}~\bibnamefont
  {Adamczyk}}, \bibinfo {author} {\bibfnamefont {A.}~\bibnamefont {Wolny}}, \
  and\ \bibinfo {author} {\bibfnamefont {Y.}~\bibnamefont {Tsori}},\
  }\href@noop {} {\bibfield  {journal} {\bibinfo  {journal} {J. Phys. Chem. B}\
  }\textbf {\bibinfo {volume} {118}},\ \bibinfo {pages} {7187} (\bibinfo {year}
  {2014})}\BibitemShut {NoStop}%
\bibitem [{\citenamefont {Amundson}\ \emph {et~al.}(1991)\citenamefont
  {Amundson}, \citenamefont {Helfand}, \citenamefont {Davis}, \citenamefont
  {Quan}, \citenamefont {Patel},\ and\ \citenamefont {Smith}}]{ah_mm_1991}%
  \BibitemOpen
  \bibfield  {author} {\bibinfo {author} {\bibfnamefont {K.}~\bibnamefont
  {Amundson}}, \bibinfo {author} {\bibfnamefont {E.}~\bibnamefont {Helfand}},
  \bibinfo {author} {\bibfnamefont {D.~D.}\ \bibnamefont {Davis}}, \bibinfo
  {author} {\bibfnamefont {X.}~\bibnamefont {Quan}}, \bibinfo {author}
  {\bibfnamefont {S.~S.}\ \bibnamefont {Patel}}, \ and\ \bibinfo {author}
  {\bibfnamefont {S.~D.}\ \bibnamefont {Smith}},\ }\href@noop {} {\bibfield
  {journal} {\bibinfo  {journal} {Macromolecules}\ }\textbf {\bibinfo {volume}
  {24}},\ \bibinfo {pages} {6546} (\bibinfo {year} {1991})}\BibitemShut
  {NoStop}%
\bibitem [{\citenamefont {Amundson}\ \emph {et~al.}(1993)\citenamefont
  {Amundson}, \citenamefont {Helfand}, \citenamefont {Quan},\ and\
  \citenamefont {Smith}}]{ah_mm_1993}%
  \BibitemOpen
  \bibfield  {author} {\bibinfo {author} {\bibfnamefont {K.}~\bibnamefont
  {Amundson}}, \bibinfo {author} {\bibfnamefont {E.}~\bibnamefont {Helfand}},
  \bibinfo {author} {\bibfnamefont {X.}~\bibnamefont {Quan}}, \ and\ \bibinfo
  {author} {\bibfnamefont {S.~D.}\ \bibnamefont {Smith}},\ }\href@noop {}
  {\bibfield  {journal} {\bibinfo  {journal} {Macromolecules}\ }\textbf
  {\bibinfo {volume} {26}},\ \bibinfo {pages} {2698} (\bibinfo {year}
  {1993})}\BibitemShut {NoStop}%
\bibitem [{\citenamefont {Pereira}\ and\ \citenamefont
  {Williams}(1999)}]{per_williams_mm_1999}%
  \BibitemOpen
  \bibfield  {author} {\bibinfo {author} {\bibfnamefont {G.~G.}\ \bibnamefont
  {Pereira}}\ and\ \bibinfo {author} {\bibfnamefont {D.~R.~M.}\ \bibnamefont
  {Williams}},\ }\href@noop {} {\bibfield  {journal} {\bibinfo  {journal}
  {Macromolecules}\ }\textbf {\bibinfo {volume} {32}},\ \bibinfo {pages} {8115}
  (\bibinfo {year} {1999})}\BibitemShut {NoStop}%
\bibitem [{\citenamefont {Tsori}\ and\ \citenamefont
  {Andelman}(2002)}]{TA_mm_2002}%
  \BibitemOpen
  \bibfield  {author} {\bibinfo {author} {\bibfnamefont {Y.}~\bibnamefont
  {Tsori}}\ and\ \bibinfo {author} {\bibfnamefont {D.}~\bibnamefont
  {Andelman}},\ }\href@noop {} {\bibfield  {journal} {\bibinfo  {journal}
  {Macromolecules}\ }\textbf {\bibinfo {volume} {35}},\ \bibinfo {pages} {5161}
  (\bibinfo {year} {2002})}\BibitemShut {NoStop}%
\bibitem [{\citenamefont {Kyrylyuk}\ \emph {et~al.}(2002)\citenamefont
  {Kyrylyuk}, \citenamefont {Zvelindovsky}, \citenamefont {Sevink},\ and\
  \citenamefont {Fraaije}}]{fraaije_mm_2002}%
  \BibitemOpen
  \bibfield  {author} {\bibinfo {author} {\bibfnamefont {A.}~\bibnamefont
  {Kyrylyuk}}, \bibinfo {author} {\bibfnamefont {A.}~\bibnamefont
  {Zvelindovsky}}, \bibinfo {author} {\bibfnamefont {G.}~\bibnamefont
  {Sevink}}, \ and\ \bibinfo {author} {\bibfnamefont {J.}~\bibnamefont
  {Fraaije}},\ }\href {\doibase 10.1021/ma0110756} {\bibfield  {journal}
  {\bibinfo  {journal} {Macromolecules}\ }\textbf {\bibinfo {volume} {35}},\
  \bibinfo {pages} {1473} (\bibinfo {year} {2002})}\BibitemShut {NoStop}%
\bibitem [{\citenamefont {Tsori}\ \emph
  {et~al.}(2003{\natexlab{a}})\citenamefont {Tsori}, \citenamefont
  {Tournilhac},\ and\ \citenamefont {Leibler}}]{tsori_mm_2003}%
  \BibitemOpen
  \bibfield  {author} {\bibinfo {author} {\bibfnamefont {Y.}~\bibnamefont
  {Tsori}}, \bibinfo {author} {\bibfnamefont {F.}~\bibnamefont {Tournilhac}}, \
  and\ \bibinfo {author} {\bibfnamefont {L.}~\bibnamefont {Leibler}},\
  }\href@noop {} {\bibfield  {journal} {\bibinfo  {journal} {Macromolecules}\
  }\textbf {\bibinfo {volume} {36}},\ \bibinfo {pages} {5873} (\bibinfo {year}
  {2003}{\natexlab{a}})}\BibitemShut {NoStop}%
\bibitem [{\citenamefont {Tsori}\ \emph {et~al.}(2006)\citenamefont {Tsori},
  \citenamefont {Andelman}, \citenamefont {Lin},\ and\ \citenamefont
  {Schick}}]{tsori_mm_2006}%
  \BibitemOpen
  \bibfield  {author} {\bibinfo {author} {\bibfnamefont {Y.}~\bibnamefont
  {Tsori}}, \bibinfo {author} {\bibfnamefont {D.}~\bibnamefont {Andelman}},
  \bibinfo {author} {\bibfnamefont {C.-Y.}\ \bibnamefont {Lin}}, \ and\
  \bibinfo {author} {\bibfnamefont {M.}~\bibnamefont {Schick}},\ }\href@noop {}
  {\bibfield  {journal} {\bibinfo  {journal} {Macromolecules}\ }\textbf
  {\bibinfo {volume} {39}},\ \bibinfo {pages} {289} (\bibinfo {year}
  {2006})}\BibitemShut {NoStop}%
\bibitem [{\citenamefont {Morkved}\ \emph {et~al.}(1996)\citenamefont
  {Morkved}, \citenamefont {Lu}, \citenamefont {Urbas}, \citenamefont
  {Ehrichs}, \citenamefont {Jaeger}, \citenamefont {Mansky},\ and\
  \citenamefont {Russell}}]{russell_sci_1996}%
  \BibitemOpen
  \bibfield  {author} {\bibinfo {author} {\bibfnamefont {T.~L.}\ \bibnamefont
  {Morkved}}, \bibinfo {author} {\bibfnamefont {M.}~\bibnamefont {Lu}},
  \bibinfo {author} {\bibfnamefont {A.~M.}\ \bibnamefont {Urbas}}, \bibinfo
  {author} {\bibfnamefont {E.~E.}\ \bibnamefont {Ehrichs}}, \bibinfo {author}
  {\bibfnamefont {H.~M.}\ \bibnamefont {Jaeger}}, \bibinfo {author}
  {\bibfnamefont {P.}~\bibnamefont {Mansky}}, \ and\ \bibinfo {author}
  {\bibfnamefont {T.~P.}\ \bibnamefont {Russell}},\ }\href@noop {} {\bibfield
  {journal} {\bibinfo  {journal} {Science}\ }\textbf {\bibinfo {volume}
  {273}},\ \bibinfo {pages} {931} (\bibinfo {year} {1996})}\BibitemShut
  {NoStop}%
\bibitem [{\citenamefont {Thurn-Albrecht}\ \emph {et~al.}(2002)\citenamefont
  {Thurn-Albrecht}, \citenamefont {DeRouchey}, \citenamefont {Russell},\ and\
  \citenamefont {Kolb}}]{russell_mm_2002b}%
  \BibitemOpen
  \bibfield  {author} {\bibinfo {author} {\bibfnamefont {T.}~\bibnamefont
  {Thurn-Albrecht}}, \bibinfo {author} {\bibfnamefont {J.}~\bibnamefont
  {DeRouchey}}, \bibinfo {author} {\bibfnamefont {T.~P.}\ \bibnamefont
  {Russell}}, \ and\ \bibinfo {author} {\bibfnamefont {R.}~\bibnamefont
  {Kolb}},\ }\href {\doibase 10.1021/ma020567v} {\bibfield  {journal} {\bibinfo
   {journal} {Macromolecules}\ }\textbf {\bibinfo {volume} {35}},\ \bibinfo
  {pages} {8106} (\bibinfo {year} {2002})}\BibitemShut {NoStop}%
\bibitem [{\citenamefont {Xu}\ \emph {et~al.}(2003)\citenamefont {Xu},
  \citenamefont {Hawker},\ and\ \citenamefont {Russell}}]{russell_mm_2003}%
  \BibitemOpen
  \bibfield  {author} {\bibinfo {author} {\bibfnamefont {T.}~\bibnamefont
  {Xu}}, \bibinfo {author} {\bibfnamefont {C.~J.}\ \bibnamefont {Hawker}}, \
  and\ \bibinfo {author} {\bibfnamefont {T.~P.}\ \bibnamefont {Russell}},\
  }\href {\doibase 10.1021/ma034511s} {\bibfield  {journal} {\bibinfo
  {journal} {Macromolecules}\ }\textbf {\bibinfo {volume} {36}},\ \bibinfo
  {pages} {6178} (\bibinfo {year} {2003})}\BibitemShut {NoStop}%
\bibitem [{\citenamefont {Xu}\ \emph {et~al.}(2004)\citenamefont {Xu},
  \citenamefont {Zvelindovsky}, \citenamefont {Sevink}, \citenamefont {Gang},
  \citenamefont {Ocko}, \citenamefont {Zhu}, \citenamefont {Gido},\ and\
  \citenamefont {Russell}}]{russell_mm_2004a}%
  \BibitemOpen
  \bibfield  {author} {\bibinfo {author} {\bibfnamefont {T.}~\bibnamefont
  {Xu}}, \bibinfo {author} {\bibfnamefont {A.~V.}\ \bibnamefont
  {Zvelindovsky}}, \bibinfo {author} {\bibfnamefont {G.~J.~A.}\ \bibnamefont
  {Sevink}}, \bibinfo {author} {\bibfnamefont {O.}~\bibnamefont {Gang}},
  \bibinfo {author} {\bibfnamefont {B.}~\bibnamefont {Ocko}}, \bibinfo {author}
  {\bibfnamefont {Y.~Q.}\ \bibnamefont {Zhu}}, \bibinfo {author} {\bibfnamefont
  {S.~P.}\ \bibnamefont {Gido}}, \ and\ \bibinfo {author} {\bibfnamefont
  {T.~P.}\ \bibnamefont {Russell}},\ }\href {\doibase 10.1021/ma049235b}
  {\bibfield  {journal} {\bibinfo  {journal} {Macromolecules}\ }\textbf
  {\bibinfo {volume} {37}},\ \bibinfo {pages} {6980} (\bibinfo {year}
  {2004})}\BibitemShut {NoStop}%
\bibitem [{\citenamefont {Boker}\ \emph
  {et~al.}(2002{\natexlab{a}})\citenamefont {Boker}, \citenamefont {Elbs},
  \citenamefont {Hansel}, \citenamefont {Knoll}, \citenamefont {Ludwigs},
  \citenamefont {Zettl}, \citenamefont {Urban}, \citenamefont {Abetz},
  \citenamefont {Muller},\ and\ \citenamefont {Krausch}}]{boker_prl_2002}%
  \BibitemOpen
  \bibfield  {author} {\bibinfo {author} {\bibfnamefont {A.}~\bibnamefont
  {Boker}}, \bibinfo {author} {\bibfnamefont {H.}~\bibnamefont {Elbs}},
  \bibinfo {author} {\bibfnamefont {H.}~\bibnamefont {Hansel}}, \bibinfo
  {author} {\bibfnamefont {A.}~\bibnamefont {Knoll}}, \bibinfo {author}
  {\bibfnamefont {S.}~\bibnamefont {Ludwigs}}, \bibinfo {author} {\bibfnamefont
  {H.}~\bibnamefont {Zettl}}, \bibinfo {author} {\bibfnamefont
  {V.}~\bibnamefont {Urban}}, \bibinfo {author} {\bibfnamefont
  {V.}~\bibnamefont {Abetz}}, \bibinfo {author} {\bibfnamefont {A.~H.~E.}\
  \bibnamefont {Muller}}, \ and\ \bibinfo {author} {\bibfnamefont
  {G.}~\bibnamefont {Krausch}},\ }\href {\doibase
  10.1103/PhysRevLett.89.135502} {\bibfield  {journal} {\bibinfo  {journal}
  {Phys. Rev. Lett.}\ }\textbf {\bibinfo {volume} {89}},\ \bibinfo {pages}
  {135502} (\bibinfo {year} {2002}{\natexlab{a}})}\BibitemShut {NoStop}%
\bibitem [{\citenamefont {Boker}\ \emph {et~al.}(2003)\citenamefont {Boker},
  \citenamefont {Elbs}, \citenamefont {Hansel}, \citenamefont {Knoll},
  \citenamefont {Ludwigs}, \citenamefont {Zettl}, \citenamefont {Zvelindovsky},
  \citenamefont {Sevink}, \citenamefont {Urban}, \citenamefont {Abetz},
  \citenamefont {Muller},\ and\ \citenamefont {Krausch}}]{boker_mm_2003}%
  \BibitemOpen
  \bibfield  {author} {\bibinfo {author} {\bibfnamefont {A.}~\bibnamefont
  {Boker}}, \bibinfo {author} {\bibfnamefont {H.}~\bibnamefont {Elbs}},
  \bibinfo {author} {\bibfnamefont {H.}~\bibnamefont {Hansel}}, \bibinfo
  {author} {\bibfnamefont {A.}~\bibnamefont {Knoll}}, \bibinfo {author}
  {\bibfnamefont {S.}~\bibnamefont {Ludwigs}}, \bibinfo {author} {\bibfnamefont
  {H.}~\bibnamefont {Zettl}}, \bibinfo {author} {\bibfnamefont {A.~V.}\
  \bibnamefont {Zvelindovsky}}, \bibinfo {author} {\bibfnamefont {G.~J.~A.}\
  \bibnamefont {Sevink}}, \bibinfo {author} {\bibfnamefont {V.}~\bibnamefont
  {Urban}}, \bibinfo {author} {\bibfnamefont {V.}~\bibnamefont {Abetz}},
  \bibinfo {author} {\bibfnamefont {A.~H.~E.}\ \bibnamefont {Muller}}, \ and\
  \bibinfo {author} {\bibfnamefont {G.}~\bibnamefont {Krausch}},\ }\href
  {\doibase 10.1021/ma021347k} {\bibfield  {journal} {\bibinfo  {journal}
  {Macromolecules}\ }\textbf {\bibinfo {volume} {36}},\ \bibinfo {pages} {8078}
  (\bibinfo {year} {2003})}\BibitemShut {NoStop}%
\bibitem [{\citenamefont {Boker}\ \emph {et~al.}(2006)\citenamefont {Boker},
  \citenamefont {Schmidt}, \citenamefont {Knoll}, \citenamefont {Zettl},
  \citenamefont {Hansel}, \citenamefont {Urban}, \citenamefont {Abetz},\ and\
  \citenamefont {Krausch}}]{boker_polymer_2006}%
  \BibitemOpen
  \bibfield  {author} {\bibinfo {author} {\bibfnamefont {A.}~\bibnamefont
  {Boker}}, \bibinfo {author} {\bibfnamefont {K.}~\bibnamefont {Schmidt}},
  \bibinfo {author} {\bibfnamefont {A.}~\bibnamefont {Knoll}}, \bibinfo
  {author} {\bibfnamefont {H.}~\bibnamefont {Zettl}}, \bibinfo {author}
  {\bibfnamefont {H.}~\bibnamefont {Hansel}}, \bibinfo {author} {\bibfnamefont
  {V.}~\bibnamefont {Urban}}, \bibinfo {author} {\bibfnamefont
  {V.}~\bibnamefont {Abetz}}, \ and\ \bibinfo {author} {\bibfnamefont
  {G.}~\bibnamefont {Krausch}},\ }\href {\doibase
  10.1016/j.polymer.2005.11.069} {\bibfield  {journal} {\bibinfo  {journal}
  {Polymer}\ }\textbf {\bibinfo {volume} {47}},\ \bibinfo {pages} {849}
  (\bibinfo {year} {2006})}\BibitemShut {NoStop}%
\bibitem [{\citenamefont {Schmidt}\ \emph {et~al.}(2005)\citenamefont
  {Schmidt}, \citenamefont {Boker}, \citenamefont {Zettl}, \citenamefont
  {Schubert}, \citenamefont {Hansel}, \citenamefont {Fischer}, \citenamefont
  {Weiss}, \citenamefont {Abetz}, \citenamefont {Zvelindovsky}, \citenamefont
  {Sevink},\ and\ \citenamefont {Krausch}}]{boker_langmuir_2005}%
  \BibitemOpen
  \bibfield  {author} {\bibinfo {author} {\bibfnamefont {K.}~\bibnamefont
  {Schmidt}}, \bibinfo {author} {\bibfnamefont {A.}~\bibnamefont {Boker}},
  \bibinfo {author} {\bibfnamefont {H.}~\bibnamefont {Zettl}}, \bibinfo
  {author} {\bibfnamefont {F.}~\bibnamefont {Schubert}}, \bibinfo {author}
  {\bibfnamefont {H.}~\bibnamefont {Hansel}}, \bibinfo {author} {\bibfnamefont
  {F.}~\bibnamefont {Fischer}}, \bibinfo {author} {\bibfnamefont {T.~M.}\
  \bibnamefont {Weiss}}, \bibinfo {author} {\bibfnamefont {V.}~\bibnamefont
  {Abetz}}, \bibinfo {author} {\bibfnamefont {A.~V.}\ \bibnamefont
  {Zvelindovsky}}, \bibinfo {author} {\bibfnamefont {G.~J.~A.}\ \bibnamefont
  {Sevink}}, \ and\ \bibinfo {author} {\bibfnamefont {G.}~\bibnamefont
  {Krausch}},\ }\href {\doibase 10.1021/la051346w} {\bibfield  {journal}
  {\bibinfo  {journal} {Langmuir}\ }\textbf {\bibinfo {volume} {21}},\ \bibinfo
  {pages} {11974} (\bibinfo {year} {2005})}\BibitemShut {NoStop}%
\bibitem [{\citenamefont {Schmidt}\ \emph {et~al.}(2007)\citenamefont
  {Schmidt}, \citenamefont {Schoberth}, \citenamefont {Schubert}, \citenamefont
  {H{\"a}nsel}, \citenamefont {Fischer}, \citenamefont {Weiss}, \citenamefont
  {Sevink}, \citenamefont {Zvelindovsky}, \citenamefont {B{\"o}ker},\ and\
  \citenamefont {Krausch}}]{boker_softmatter_2007}%
  \BibitemOpen
  \bibfield  {author} {\bibinfo {author} {\bibfnamefont {K.}~\bibnamefont
  {Schmidt}}, \bibinfo {author} {\bibfnamefont {H.~G.}\ \bibnamefont
  {Schoberth}}, \bibinfo {author} {\bibfnamefont {F.}~\bibnamefont {Schubert}},
  \bibinfo {author} {\bibfnamefont {H.}~\bibnamefont {H{\"a}nsel}}, \bibinfo
  {author} {\bibfnamefont {F.}~\bibnamefont {Fischer}}, \bibinfo {author}
  {\bibfnamefont {T.~M.}\ \bibnamefont {Weiss}}, \bibinfo {author}
  {\bibfnamefont {G.~J.~A.}\ \bibnamefont {Sevink}}, \bibinfo {author}
  {\bibfnamefont {A.~V.}\ \bibnamefont {Zvelindovsky}}, \bibinfo {author}
  {\bibfnamefont {A.}~\bibnamefont {B{\"o}ker}}, \ and\ \bibinfo {author}
  {\bibfnamefont {G.}~\bibnamefont {Krausch}},\ }\href@noop {} {\bibfield
  {journal} {\bibinfo  {journal} {Soft Matter}\ }\textbf {\bibinfo {volume}
  {3}},\ \bibinfo {pages} {448} (\bibinfo {year} {2007})}\BibitemShut {NoStop}%
\bibitem [{\citenamefont {Boker}\ \emph
  {et~al.}(2002{\natexlab{b}})\citenamefont {Boker}, \citenamefont {Knoll},
  \citenamefont {Elbs}, \citenamefont {Abetz}, \citenamefont {Muller},\ and\
  \citenamefont {Krausch}}]{boker_mm_2002}%
  \BibitemOpen
  \bibfield  {author} {\bibinfo {author} {\bibfnamefont {A.}~\bibnamefont
  {Boker}}, \bibinfo {author} {\bibfnamefont {A.}~\bibnamefont {Knoll}},
  \bibinfo {author} {\bibfnamefont {H.}~\bibnamefont {Elbs}}, \bibinfo {author}
  {\bibfnamefont {V.}~\bibnamefont {Abetz}}, \bibinfo {author} {\bibfnamefont
  {A.}~\bibnamefont {Muller}}, \ and\ \bibinfo {author} {\bibfnamefont
  {G.}~\bibnamefont {Krausch}},\ }\href {\doibase {10.1021/ma0108113}}
  {\bibfield  {journal} {\bibinfo  {journal} {Macromolecules}\ }\textbf
  {\bibinfo {volume} {35}},\ \bibinfo {pages} {1319} (\bibinfo {year}
  {2002}{\natexlab{b}})}\BibitemShut {NoStop}%
\bibitem [{\citenamefont {Xu}\ \emph {et~al.}(2005)\citenamefont {Xu},
  \citenamefont {Zvelindovsky}, \citenamefont {Sevink}, \citenamefont
  {Lyakhova}, \citenamefont {Jinnai},\ and\ \citenamefont
  {Russell}}]{russell_mm_2005}%
  \BibitemOpen
  \bibfield  {author} {\bibinfo {author} {\bibfnamefont {T.}~\bibnamefont
  {Xu}}, \bibinfo {author} {\bibfnamefont {A.~V.}\ \bibnamefont
  {Zvelindovsky}}, \bibinfo {author} {\bibfnamefont {G.~J.~A.}\ \bibnamefont
  {Sevink}}, \bibinfo {author} {\bibfnamefont {K.~S.}\ \bibnamefont
  {Lyakhova}}, \bibinfo {author} {\bibfnamefont {H.}~\bibnamefont {Jinnai}}, \
  and\ \bibinfo {author} {\bibfnamefont {T.~P.}\ \bibnamefont {Russell}},\
  }\href {\doibase 10.1021/ma050221c} {\bibfield  {journal} {\bibinfo
  {journal} {Macromolecules}\ }\textbf {\bibinfo {volume} {38}},\ \bibinfo
  {pages} {10788} (\bibinfo {year} {2005})}\BibitemShut {NoStop}%
\bibitem [{\citenamefont {Wang}\ \emph {et~al.}(2006)\citenamefont {Wang},
  \citenamefont {Leiston-Belanger}, \citenamefont {Sievert},\ and\
  \citenamefont {Russell}}]{russell_mm_2006}%
  \BibitemOpen
  \bibfield  {author} {\bibinfo {author} {\bibfnamefont {J.-Y.}\ \bibnamefont
  {Wang}}, \bibinfo {author} {\bibfnamefont {J.~M.}\ \bibnamefont
  {Leiston-Belanger}}, \bibinfo {author} {\bibfnamefont {J.~D.}\ \bibnamefont
  {Sievert}}, \ and\ \bibinfo {author} {\bibfnamefont {T.~P.}\ \bibnamefont
  {Russell}},\ }\href {\doibase 10.1021/ma0614287} {\bibfield  {journal}
  {\bibinfo  {journal} {Macromolecules}\ }\textbf {\bibinfo {volume} {39}},\
  \bibinfo {pages} {8487} (\bibinfo {year} {2006})}\BibitemShut {NoStop}%
\bibitem [{\citenamefont {Liedel}\ \emph {et~al.}(2012)\citenamefont {Liedel},
  \citenamefont {Hund}, \citenamefont {Olszowka},\ and\ \citenamefont
  {Boeker}}]{boker_softmatter_2012}%
  \BibitemOpen
  \bibfield  {author} {\bibinfo {author} {\bibfnamefont {C.}~\bibnamefont
  {Liedel}}, \bibinfo {author} {\bibfnamefont {M.}~\bibnamefont {Hund}},
  \bibinfo {author} {\bibfnamefont {V.}~\bibnamefont {Olszowka}}, \ and\
  \bibinfo {author} {\bibfnamefont {A.}~\bibnamefont {Boeker}},\ }\href
  {\doibase 10.1039/c1sm06531a} {\bibfield  {journal} {\bibinfo  {journal}
  {Soft Matter}\ }\textbf {\bibinfo {volume} {8}},\ \bibinfo {pages} {995}
  (\bibinfo {year} {2012})}\BibitemShut {NoStop}%
\bibitem [{\citenamefont {Lyakhova}\ \emph {et~al.}(2006)\citenamefont
  {Lyakhova}, \citenamefont {Zvelindovsky},\ and\ \citenamefont
  {Sevink}}]{sevink_mm_2006}%
  \BibitemOpen
  \bibfield  {author} {\bibinfo {author} {\bibfnamefont {K.~S.}\ \bibnamefont
  {Lyakhova}}, \bibinfo {author} {\bibfnamefont {A.~V.}\ \bibnamefont
  {Zvelindovsky}}, \ and\ \bibinfo {author} {\bibfnamefont {G.~J.~A.}\
  \bibnamefont {Sevink}},\ }\href {\doibase 10.1021/ma060143r} {\bibfield
  {journal} {\bibinfo  {journal} {Macromolecules}\ }\textbf {\bibinfo {volume}
  {39}},\ \bibinfo {pages} {3024} (\bibinfo {year} {2006})}\BibitemShut
  {NoStop}%
\bibitem [{\citenamefont {Kyrylyuk}\ and\ \citenamefont
  {Fraaije}(2006)}]{fraaije_jcp_2006}%
  \BibitemOpen
  \bibfield  {author} {\bibinfo {author} {\bibfnamefont {A.~V.}\ \bibnamefont
  {Kyrylyuk}}\ and\ \bibinfo {author} {\bibfnamefont {J.~G. E.~M.}\
  \bibnamefont {Fraaije}},\ }\href {\doibase 10.1063/1.2360947} {\bibfield
  {journal} {\bibinfo  {journal} {J. Chem. Phys.}\ }\textbf {\bibinfo {volume}
  {125}},\ \bibinfo {pages} {164716} (\bibinfo {year} {2006})}\BibitemShut
  {NoStop}%
\bibitem [{\citenamefont {Tsori}\ \emph
  {et~al.}(2003{\natexlab{b}})\citenamefont {Tsori}, \citenamefont
  {Tournilhac}, \citenamefont {Andelman},\ and\ \citenamefont
  {Leibler}}]{tsori_prl_2003}%
  \BibitemOpen
  \bibfield  {author} {\bibinfo {author} {\bibfnamefont {Y.}~\bibnamefont
  {Tsori}}, \bibinfo {author} {\bibfnamefont {F.}~\bibnamefont {Tournilhac}},
  \bibinfo {author} {\bibfnamefont {D.}~\bibnamefont {Andelman}}, \ and\
  \bibinfo {author} {\bibfnamefont {L.}~\bibnamefont {Leibler}},\ }\href@noop
  {} {\bibfield  {journal} {\bibinfo  {journal} {Phys. Rev. Lett.}\ }\textbf
  {\bibinfo {volume} {90}},\ \bibinfo {pages} {145504} (\bibinfo {year}
  {2003}{\natexlab{b}})}\BibitemShut {NoStop}%
\bibitem [{\citenamefont {Giacomelli}\ \emph {et~al.}(2010)\citenamefont
  {Giacomelli}, \citenamefont {da~Silveira}, \citenamefont {Nallet},
  \citenamefont {Cernoch}, \citenamefont {Steinhart},\ and\ \citenamefont
  {Stepanek}}]{stepanek_mm_2010}%
  \BibitemOpen
  \bibfield  {author} {\bibinfo {author} {\bibfnamefont {F.~C.}\ \bibnamefont
  {Giacomelli}}, \bibinfo {author} {\bibfnamefont {N.~P.}\ \bibnamefont
  {da~Silveira}}, \bibinfo {author} {\bibfnamefont {F.}~\bibnamefont {Nallet}},
  \bibinfo {author} {\bibfnamefont {P.}~\bibnamefont {Cernoch}}, \bibinfo
  {author} {\bibfnamefont {M.}~\bibnamefont {Steinhart}}, \ and\ \bibinfo
  {author} {\bibfnamefont {P.}~\bibnamefont {Stepanek}},\ }\href {\doibase
  10.1021/ma1000817} {\bibfield  {journal} {\bibinfo  {journal}
  {Macromolecules}\ }\textbf {\bibinfo {volume} {43}},\ \bibinfo {pages} {4261}
  (\bibinfo {year} {2010})}\BibitemShut {NoStop}%
\bibitem [{\citenamefont {Matsen}(2006)}]{matsen_jcp_2006}%
  \BibitemOpen
  \bibfield  {author} {\bibinfo {author} {\bibfnamefont {M.~W.}\ \bibnamefont
  {Matsen}},\ }\href {\doibase 10.1063/1.2170082} {\bibfield  {journal}
  {\bibinfo  {journal} {J. Chem. Phys.}\ }\textbf {\bibinfo {volume} {124}},\
  \bibinfo {pages} {074906} (\bibinfo {year} {2006})}\BibitemShut {NoStop}%
\bibitem [{\citenamefont {Ly}\ \emph {et~al.}(2007)\citenamefont {Ly},
  \citenamefont {Honda}, \citenamefont {Kawakatsu},\ and\ \citenamefont
  {Zvelindovsky}}]{zvelin_mm_2007}%
  \BibitemOpen
  \bibfield  {author} {\bibinfo {author} {\bibfnamefont {D.~Q.}\ \bibnamefont
  {Ly}}, \bibinfo {author} {\bibfnamefont {T.}~\bibnamefont {Honda}}, \bibinfo
  {author} {\bibfnamefont {T.}~\bibnamefont {Kawakatsu}}, \ and\ \bibinfo
  {author} {\bibfnamefont {A.~V.}\ \bibnamefont {Zvelindovsky}},\ }\href
  {\doibase 10.1021/ma061875m} {\bibfield  {journal} {\bibinfo  {journal}
  {Macromolecules}\ }\textbf {\bibinfo {volume} {40}},\ \bibinfo {pages} {2928}
  (\bibinfo {year} {2007})}\BibitemShut {NoStop}%
\bibitem [{\citenamefont {Pinna}\ and\ \citenamefont
  {Zvelindovsky}(2008)}]{zvelin_softmatter_2008}%
  \BibitemOpen
  \bibfield  {author} {\bibinfo {author} {\bibfnamefont {M.}~\bibnamefont
  {Pinna}}\ and\ \bibinfo {author} {\bibfnamefont {A.~V.}\ \bibnamefont
  {Zvelindovsky}},\ }\href {\doibase 10.1039/b706815h} {\bibfield  {journal}
  {\bibinfo  {journal} {Soft Matter}\ }\textbf {\bibinfo {volume} {4}},\
  \bibinfo {pages} {316} (\bibinfo {year} {2008})}\BibitemShut {NoStop}%
\bibitem [{\citenamefont {Crossland}\ \emph {et~al.}(2010)\citenamefont
  {Crossland}, \citenamefont {Ludwigs}, \citenamefont {Hillmyer},\ and\
  \citenamefont {Steiner}}]{steiner_softmatter_2010}%
  \BibitemOpen
  \bibfield  {author} {\bibinfo {author} {\bibfnamefont {E.~J.~W.}\
  \bibnamefont {Crossland}}, \bibinfo {author} {\bibfnamefont {S.}~\bibnamefont
  {Ludwigs}}, \bibinfo {author} {\bibfnamefont {M.~A.}\ \bibnamefont
  {Hillmyer}}, \ and\ \bibinfo {author} {\bibfnamefont {U.}~\bibnamefont
  {Steiner}},\ }\href {\doibase 10.1039/b914421h} {\bibfield  {journal}
  {\bibinfo  {journal} {Soft Matter}\ }\textbf {\bibinfo {volume} {6}},\
  \bibinfo {pages} {670} (\bibinfo {year} {2010})}\BibitemShut {NoStop}%
\bibitem [{\citenamefont {Pohl}(1978)}]{pohl_book}%
  \BibitemOpen
  \bibfield  {author} {\bibinfo {author} {\bibfnamefont {H.~A.}\ \bibnamefont
  {Pohl}},\ }\href@noop {} {\emph {\bibinfo {title} {Dielectrophoresis: The
  Behavior of Neutral Matter in Nonuniform Electric Fields}}}\ (\bibinfo
  {publisher} {Cambridge University Press},\ \bibinfo {year}
  {1978})\BibitemShut {NoStop}%
\bibitem [{\citenamefont {Sullivan}\ \emph {et~al.}(2003)\citenamefont
  {Sullivan}, \citenamefont {Zhao}, \citenamefont {Harrison}, \citenamefont
  {Austin}, \citenamefont {Megens}, \citenamefont {Hollingsworth},
  \citenamefont {Russel}, \citenamefont {Cheng}, \citenamefont {Mason},\ and\
  \citenamefont {Chaikin}}]{chaikin_jpcm_2003}%
  \BibitemOpen
  \bibfield  {author} {\bibinfo {author} {\bibfnamefont {M.}~\bibnamefont
  {Sullivan}}, \bibinfo {author} {\bibfnamefont {K.}~\bibnamefont {Zhao}},
  \bibinfo {author} {\bibfnamefont {C.}~\bibnamefont {Harrison}}, \bibinfo
  {author} {\bibfnamefont {R.}~\bibnamefont {Austin}}, \bibinfo {author}
  {\bibfnamefont {M.}~\bibnamefont {Megens}}, \bibinfo {author} {\bibfnamefont
  {A.}~\bibnamefont {Hollingsworth}}, \bibinfo {author} {\bibfnamefont
  {W.}~\bibnamefont {Russel}}, \bibinfo {author} {\bibfnamefont
  {Z.}~\bibnamefont {Cheng}}, \bibinfo {author} {\bibfnamefont
  {T.}~\bibnamefont {Mason}}, \ and\ \bibinfo {author} {\bibfnamefont
  {P.}~\bibnamefont {Chaikin}},\ }\href {\doibase {10.1088/0953-8984/15/1/302}}
  {\bibfield  {journal} {\bibinfo  {journal} {J. Phys.: condens. Matter}\
  }\textbf {\bibinfo {volume} {15}},\ \bibinfo {pages} {S11} (\bibinfo {year}
  {2003})},\ \bibinfo {note} {5th Liquid Matter Conference, CONSTANCE, GERMANY,
  SEP 14-18, 2002}\BibitemShut {NoStop}%
\bibitem [{\citenamefont {Sullivan}\ \emph {et~al.}(2006)\citenamefont
  {Sullivan}, \citenamefont {Zhao}, \citenamefont {Hollingsworth},
  \citenamefont {Austin}, \citenamefont {Russel},\ and\ \citenamefont
  {Chaikin}}]{chaikin_prl_2006}%
  \BibitemOpen
  \bibfield  {author} {\bibinfo {author} {\bibfnamefont {M.}~\bibnamefont
  {Sullivan}}, \bibinfo {author} {\bibfnamefont {K.}~\bibnamefont {Zhao}},
  \bibinfo {author} {\bibfnamefont {A.}~\bibnamefont {Hollingsworth}}, \bibinfo
  {author} {\bibfnamefont {R.}~\bibnamefont {Austin}}, \bibinfo {author}
  {\bibfnamefont {W.}~\bibnamefont {Russel}}, \ and\ \bibinfo {author}
  {\bibfnamefont {P.}~\bibnamefont {Chaikin}},\ }\href {\doibase
  {10.1103/PhysRevLett.96.015703}} {\bibfield  {journal} {\bibinfo  {journal}
  {Phys. Rev. Lett.}\ }\textbf {\bibinfo {volume} {96}},\ \bibinfo {pages}
  {015703} (\bibinfo {year} {2006})}\BibitemShut {NoStop}%
\bibitem [{\citenamefont {Leunissen}\ \emph {et~al.}(2008)\citenamefont
  {Leunissen}, \citenamefont {Sullivan}, \citenamefont {Chaikin},\ and\
  \citenamefont {van Blaaderen}}]{chaikin_jcp_2008}%
  \BibitemOpen
  \bibfield  {author} {\bibinfo {author} {\bibfnamefont {M.~E.}\ \bibnamefont
  {Leunissen}}, \bibinfo {author} {\bibfnamefont {M.~T.}\ \bibnamefont
  {Sullivan}}, \bibinfo {author} {\bibfnamefont {P.~M.}\ \bibnamefont
  {Chaikin}}, \ and\ \bibinfo {author} {\bibfnamefont {A.}~\bibnamefont {van
  Blaaderen}},\ }\href {\doibase {10.1063/1.2909198}} {\bibfield  {journal}
  {\bibinfo  {journal} {J. Chem. Phys.}\ }\textbf {\bibinfo {volume} {128}},\
  \bibinfo {pages} {164508} (\bibinfo {year} {2008})}\BibitemShut {NoStop}%
\bibitem [{\citenamefont {Tsori}\ \emph {et~al.}(2004)\citenamefont {Tsori},
  \citenamefont {Tournilhac},\ and\ \citenamefont
  {Leibler}}]{tsori_nature_2004}%
  \BibitemOpen
  \bibfield  {author} {\bibinfo {author} {\bibfnamefont {Y.}~\bibnamefont
  {Tsori}}, \bibinfo {author} {\bibfnamefont {F.}~\bibnamefont {Tournilhac}}, \
  and\ \bibinfo {author} {\bibfnamefont {L.}~\bibnamefont {Leibler}},\ }\href
  {\doibase 10.1038/nature02758} {\bibfield  {journal} {\bibinfo  {journal}
  {Nature}\ }\textbf {\bibinfo {volume} {430}},\ \bibinfo {pages} {544}
  (\bibinfo {year} {2004})}\BibitemShut {NoStop}%
\bibitem [{\citenamefont {Samin}\ and\ \citenamefont
  {Tsori}(2011{\natexlab{a}})}]{tsori_jpcb_2011}%
  \BibitemOpen
  \bibfield  {author} {\bibinfo {author} {\bibfnamefont {S.}~\bibnamefont
  {Samin}}\ and\ \bibinfo {author} {\bibfnamefont {Y.}~\bibnamefont {Tsori}},\
  }\href {\doibase 10.1021/jp107529n} {\bibfield  {journal} {\bibinfo
  {journal} {J. Phys. Chem. B}\ }\textbf {\bibinfo {volume} {115}},\ \bibinfo
  {pages} {75} (\bibinfo {year} {2011}{\natexlab{a}})}\BibitemShut {NoStop}%
\bibitem [{\citenamefont {Galanis}\ and\ \citenamefont
  {Tsori}(2014{\natexlab{a}})}]{tsori_jcp_2014b}%
  \BibitemOpen
  \bibfield  {author} {\bibinfo {author} {\bibfnamefont {J.}~\bibnamefont
  {Galanis}}\ and\ \bibinfo {author} {\bibfnamefont {Y.}~\bibnamefont
  {Tsori}},\ }\href {\doibase 10.1063/1.4902406} {\bibfield  {journal}
  {\bibinfo  {journal} {J. Chem. Phys.}\ }\textbf {\bibinfo {volume} {141}},\
  \bibinfo {pages} {214506} (\bibinfo {year} {2014}{\natexlab{a}})}\BibitemShut
  {NoStop}%
\bibitem [{\citenamefont {Galanis}\ and\ \citenamefont
  {Tsori}(2014{\natexlab{b}})}]{tsori_jcp_2014a}%
  \BibitemOpen
  \bibfield  {author} {\bibinfo {author} {\bibfnamefont {J.}~\bibnamefont
  {Galanis}}\ and\ \bibinfo {author} {\bibfnamefont {Y.}~\bibnamefont
  {Tsori}},\ }\href {\doibase 10.1063/1.4869113} {\bibfield  {journal}
  {\bibinfo  {journal} {J. Chem. Phys.}\ }\textbf {\bibinfo {volume} {140}},\
  \bibinfo {pages} {124505} (\bibinfo {year} {2014}{\natexlab{b}})}\BibitemShut
  {NoStop}%
\bibitem [{\citenamefont {Safran}(1994)}]{safran_book}%
  \BibitemOpen
  \bibfield  {author} {\bibinfo {author} {\bibfnamefont {S.}~\bibnamefont
  {Safran}},\ }\href@noop {} {\emph {\bibinfo {title} {Statistical
  Thermodynamics of Surfaces, Interfaces, and Membranes}}}\ (\bibinfo
  {publisher} {Westview Press},\ \bibinfo {address} {New York},\ \bibinfo
  {year} {1994})\BibitemShut {NoStop}%
\bibitem [{\citenamefont {Marcus}(2002)}]{marcus_book}%
  \BibitemOpen
  \bibfield  {author} {\bibinfo {author} {\bibfnamefont {Y.}~\bibnamefont
  {Marcus}},\ }\href@noop {} {\emph {\bibinfo {title} {{Solvent Mixtures:
  Properties and Selective Solvation}}}}\ (\bibinfo  {publisher} {CRC Press},\
  \bibinfo {year} {2002})\BibitemShut {NoStop}%
\bibitem [{\citenamefont {Onuki}\ and\ \citenamefont
  {Kitamura}(2004)}]{onuki_jcp_2004}%
  \BibitemOpen
  \bibfield  {author} {\bibinfo {author} {\bibfnamefont {A.}~\bibnamefont
  {Onuki}}\ and\ \bibinfo {author} {\bibfnamefont {H.}~\bibnamefont
  {Kitamura}},\ }\href {\doibase 10.1063/1.1769357} {\bibfield  {journal}
  {\bibinfo  {journal} {J. Chem. Phys.}\ }\textbf {\bibinfo {volume} {121}},\
  \bibinfo {pages} {3143} (\bibinfo {year} {2004})}\BibitemShut {NoStop}%
\bibitem [{\citenamefont {Onuki}\ \emph {et~al.}(2011)\citenamefont {Onuki},
  \citenamefont {Okamoto},\ and\ \citenamefont {Araki}}]{onuki_review}%
  \BibitemOpen
  \bibfield  {author} {\bibinfo {author} {\bibfnamefont {A.}~\bibnamefont
  {Onuki}}, \bibinfo {author} {\bibfnamefont {R.}~\bibnamefont {Okamoto}}, \
  and\ \bibinfo {author} {\bibfnamefont {T.}~\bibnamefont {Araki}},\
  }\href@noop {} {\bibfield  {journal} {\bibinfo  {journal} {Bull. Chem. Soc.
  Jpn.}\ }\textbf {\bibinfo {volume} {84}},\ \bibinfo {pages} {569} (\bibinfo
  {year} {2011})}\BibitemShut {NoStop}%
\bibitem [{\citenamefont {Kunz}\ \emph {et~al.}(2004)\citenamefont {Kunz},
  \citenamefont {Henle},\ and\ \citenamefont {Ninham}}]{kunz_cocis_2004}%
  \BibitemOpen
  \bibfield  {author} {\bibinfo {author} {\bibfnamefont {W.}~\bibnamefont
  {Kunz}}, \bibinfo {author} {\bibfnamefont {J.}~\bibnamefont {Henle}}, \ and\
  \bibinfo {author} {\bibfnamefont {B.}~\bibnamefont {Ninham}},\ }\href
  {\doibase 10.1016/j.cocis.2004.05.005} {\bibfield  {journal} {\bibinfo
  {journal} {Curr. Opin. Colloid Interface Sci.}\ }\textbf {\bibinfo {volume}
  {9}},\ \bibinfo {pages} {19} (\bibinfo {year} {2004})}\BibitemShut {NoStop}%
\bibitem [{\citenamefont {Jungwirth}\ and\ \citenamefont
  {Tobias}(2006)}]{jungwirth_cr_2006}%
  \BibitemOpen
  \bibfield  {author} {\bibinfo {author} {\bibfnamefont {P.}~\bibnamefont
  {Jungwirth}}\ and\ \bibinfo {author} {\bibfnamefont {D.~J.}\ \bibnamefont
  {Tobias}},\ }\href {\doibase 10.1021/cr0403741} {\bibfield  {journal}
  {\bibinfo  {journal} {Chem. Rev.}\ }\textbf {\bibinfo {volume} {106}},\
  \bibinfo {pages} {1259} (\bibinfo {year} {2006})},\ \Eprint
  {http://arxiv.org/abs/http://pubs.acs.org/doi/pdf/10.1021/cr0403741}
  {http://pubs.acs.org/doi/pdf/10.1021/cr0403741} \BibitemShut {NoStop}%
\bibitem [{\citenamefont {Levin}(2009)}]{levin_prl_2009a}%
  \BibitemOpen
  \bibfield  {author} {\bibinfo {author} {\bibfnamefont {Y.}~\bibnamefont
  {Levin}},\ }\href {\doibase 10.1103/PhysRevLett.102.147803} {\bibfield
  {journal} {\bibinfo  {journal} {Phys. Rev. Lett.}\ }\textbf {\bibinfo
  {volume} {102}},\ \bibinfo {pages} {147803} (\bibinfo {year}
  {2009})}\BibitemShut {NoStop}%
\bibitem [{\citenamefont {Levin}\ \emph {et~al.}(2009)\citenamefont {Levin},
  \citenamefont {dos Santos},\ and\ \citenamefont {Diehl}}]{levin_prl_2009b}%
  \BibitemOpen
  \bibfield  {author} {\bibinfo {author} {\bibfnamefont {Y.}~\bibnamefont
  {Levin}}, \bibinfo {author} {\bibfnamefont {A.~P.}\ \bibnamefont {dos
  Santos}}, \ and\ \bibinfo {author} {\bibfnamefont {A.}~\bibnamefont
  {Diehl}},\ }\href {\doibase 10.1103/PhysRevLett.103.257802} {\bibfield
  {journal} {\bibinfo  {journal} {Phys. Rev. Lett.}\ }\textbf {\bibinfo
  {volume} {103}},\ \bibinfo {pages} {257802} (\bibinfo {year}
  {2009})}\BibitemShut {NoStop}%
\bibitem [{\citenamefont {Markovich}\ \emph {et~al.}(2014)\citenamefont
  {Markovich}, \citenamefont {Andelman},\ and\ \citenamefont
  {Podgornik}}]{andelman_epl_2014}%
  \BibitemOpen
  \bibfield  {author} {\bibinfo {author} {\bibfnamefont {T.}~\bibnamefont
  {Markovich}}, \bibinfo {author} {\bibfnamefont {D.}~\bibnamefont {Andelman}},
  \ and\ \bibinfo {author} {\bibfnamefont {R.}~\bibnamefont {Podgornik}},\
  }\href@noop {} {\bibfield  {journal} {\bibinfo  {journal} {Europhys. Lett.}\
  }\textbf {\bibinfo {volume} {106}},\ \bibinfo {pages} {16002} (\bibinfo
  {year} {2014})}\BibitemShut {NoStop}%
\bibitem [{\citenamefont {Markovich}\ \emph {et~al.}(2015)\citenamefont
  {Markovich}, \citenamefont {Andelman},\ and\ \citenamefont
  {Podgornik}}]{andelman_jcp_2015}%
  \BibitemOpen
  \bibfield  {author} {\bibinfo {author} {\bibfnamefont {T.}~\bibnamefont
  {Markovich}}, \bibinfo {author} {\bibfnamefont {D.}~\bibnamefont {Andelman}},
  \ and\ \bibinfo {author} {\bibfnamefont {R.}~\bibnamefont {Podgornik}},\
  }\href@noop {} {\bibfield  {journal} {\bibinfo  {journal} {J. Chem. Phys.}\
  }\textbf {\bibinfo {volume} {142}},\ \bibinfo {pages} {044702.1} (\bibinfo
  {year} {2015})}\BibitemShut {NoStop}%
\bibitem [{\citenamefont {Zhang}\ and\ \citenamefont
  {Cremer}(2006)}]{zhang_cocb_2006}%
  \BibitemOpen
  \bibfield  {author} {\bibinfo {author} {\bibfnamefont {Y.}~\bibnamefont
  {Zhang}}\ and\ \bibinfo {author} {\bibfnamefont {P.~S.}\ \bibnamefont
  {Cremer}},\ }\href {\doibase 10.1016/j.cbpa.2006.09.020} {\bibfield
  {journal} {\bibinfo  {journal} {Curr. Opin. Chem. Biol.}\ }\textbf {\bibinfo
  {volume} {10}},\ \bibinfo {pages} {658} (\bibinfo {year} {2006})}\BibitemShut
  {NoStop}%
\bibitem [{\citenamefont {Onuki}\ and\ \citenamefont
  {Okamoto}(2011)}]{onuki_cocis_2011}%
  \BibitemOpen
  \bibfield  {author} {\bibinfo {author} {\bibfnamefont {A.}~\bibnamefont
  {Onuki}}\ and\ \bibinfo {author} {\bibfnamefont {R.}~\bibnamefont
  {Okamoto}},\ }\href {\doibase 10.1016/j.cocis.2011.04.002} {\bibfield
  {journal} {\bibinfo  {journal} {Curr. Opin. Colloid Interface Sci.}\ }\textbf
  {\bibinfo {volume} {16}},\ \bibinfo {pages} {525} (\bibinfo {year}
  {2011})}\BibitemShut {NoStop}%
\bibitem [{\citenamefont {Hefter}(2005)}]{hefter_pac_2005}%
  \BibitemOpen
  \bibfield  {author} {\bibinfo {author} {\bibfnamefont {G.}~\bibnamefont
  {Hefter}},\ }\href@noop {} {\bibfield  {journal} {\bibinfo  {journal} {Pure
  Appl. Chem.}\ }\textbf {\bibinfo {volume} {77}},\ \bibinfo {pages} {605}
  (\bibinfo {year} {2005})}\BibitemShut {NoStop}%
\bibitem [{\citenamefont {Kalidas}\ \emph {et~al.}(2000)\citenamefont
  {Kalidas}, \citenamefont {Hefter},\ and\ \citenamefont
  {Marcus}}]{marcus_cation}%
  \BibitemOpen
  \bibfield  {author} {\bibinfo {author} {\bibfnamefont {C.}~\bibnamefont
  {Kalidas}}, \bibinfo {author} {\bibfnamefont {G.}~\bibnamefont {Hefter}}, \
  and\ \bibinfo {author} {\bibfnamefont {Y.}~\bibnamefont {Marcus}},\ }\href
  {\doibase 10.1021/cr980144k} {\bibfield  {journal} {\bibinfo  {journal}
  {Chem. Rev.}\ }\textbf {\bibinfo {volume} {100}},\ \bibinfo {pages} {819}
  (\bibinfo {year} {2000})}\BibitemShut {NoStop}%
\bibitem [{\citenamefont {Marcus}(2007)}]{marcus_anion}%
  \BibitemOpen
  \bibfield  {author} {\bibinfo {author} {\bibfnamefont {Y.}~\bibnamefont
  {Marcus}},\ }\href {\doibase 10.1021/cr068045r} {\bibfield  {journal}
  {\bibinfo  {journal} {Chem. Rev.}\ }\textbf {\bibinfo {volume} {107}},\
  \bibinfo {pages} {3880} (\bibinfo {year} {2007})}\BibitemShut {NoStop}%
\bibitem [{\citenamefont {Inerowicz}\ \emph {et~al.}(1994)\citenamefont
  {Inerowicz}, \citenamefont {Li},\ and\ \citenamefont
  {Persson}}]{persson_jcsft_1990}%
  \BibitemOpen
  \bibfield  {author} {\bibinfo {author} {\bibfnamefont {H.~D.}\ \bibnamefont
  {Inerowicz}}, \bibinfo {author} {\bibfnamefont {W.}~\bibnamefont {Li}}, \
  and\ \bibinfo {author} {\bibfnamefont {I.}~\bibnamefont {Persson}},\ }\href
  {\doibase 10.1039/FT9949002223} {\bibfield  {journal} {\bibinfo  {journal}
  {J. Chem. Soc.{,} Faraday Trans.}\ }\textbf {\bibinfo {volume} {90}},\
  \bibinfo {pages} {2223} (\bibinfo {year} {1994})}\BibitemShut {NoStop}%
\bibitem [{\citenamefont {Osakai}\ and\ \citenamefont
  {Ebina}(1998)}]{osakai_jpcb_1998}%
  \BibitemOpen
  \bibfield  {author} {\bibinfo {author} {\bibfnamefont {T.}~\bibnamefont
  {Osakai}}\ and\ \bibinfo {author} {\bibfnamefont {K.}~\bibnamefont {Ebina}},\
  }\href@noop {} {\bibfield  {journal} {\bibinfo  {journal} {J. Phys. Chem. B}\
  }\textbf {\bibinfo {volume} {102}},\ \bibinfo {pages} {5691} (\bibinfo {year}
  {1998})}\BibitemShut {NoStop}%
\bibitem [{\citenamefont {Derjaguin}\ and\ \citenamefont
  {Landau}(1941)}]{dlvo1}%
  \BibitemOpen
  \bibfield  {author} {\bibinfo {author} {\bibfnamefont {B.~V.}\ \bibnamefont
  {Derjaguin}}\ and\ \bibinfo {author} {\bibfnamefont {L.~D.}\ \bibnamefont
  {Landau}},\ }\href@noop {} {\bibfield  {journal} {\bibinfo  {journal} {Acta
  Physicochim (USSR)}\ }\textbf {\bibinfo {volume} {14}},\ \bibinfo {pages}
  {633} (\bibinfo {year} {1941})}\BibitemShut {NoStop}%
\bibitem [{\citenamefont {Verwey}\ and\ \citenamefont
  {Overbeek}(1948)}]{dlvo2}%
  \BibitemOpen
  \bibfield  {author} {\bibinfo {author} {\bibfnamefont {E.~J.~W.}\
  \bibnamefont {Verwey}}\ and\ \bibinfo {author} {\bibfnamefont {J.~T.~G.}\
  \bibnamefont {Overbeek}},\ }\href@noop {} {\emph {\bibinfo {title} {{Theory
  of the Stability of Lyophobic Colloids}}}}\ (\bibinfo  {publisher}
  {Elsevier},\ \bibinfo {address} {Amsterdam},\ \bibinfo {year}
  {1948})\BibitemShut {NoStop}%
\bibitem [{\citenamefont {Russel}\ \emph {et~al.}(1992)\citenamefont {Russel},
  \citenamefont {Saville},\ and\ \citenamefont {Schowalter}}]{colloids_book}%
  \BibitemOpen
  \bibfield  {author} {\bibinfo {author} {\bibfnamefont {W.~B.}\ \bibnamefont
  {Russel}}, \bibinfo {author} {\bibfnamefont {D.~A.}\ \bibnamefont {Saville}},
  \ and\ \bibinfo {author} {\bibfnamefont {W.~R.}\ \bibnamefont {Schowalter}},\
  }\href@noop {} {\emph {\bibinfo {title} {Colloidal Dispersions}}}\ (\bibinfo
  {publisher} {Cambridge University Press},\ \bibinfo {year}
  {1992})\BibitemShut {NoStop}%
\bibitem [{\citenamefont {Hatti-Kaul}(2000)}]{extraction_book}%
  \BibitemOpen
  \bibfield  {author} {\bibinfo {author} {\bibfnamefont {R.}~\bibnamefont
  {Hatti-Kaul}},\ }\href@noop {} {\emph {\bibinfo {title} {{Aqueous Two-Phase
  Systems}}}}\ (\bibinfo  {publisher} {Humana Press},\ \bibinfo {year}
  {2000})\BibitemShut {NoStop}%
\bibitem [{\citenamefont {Leunissen}\ \emph
  {et~al.}(2007{\natexlab{a}})\citenamefont {Leunissen}, \citenamefont {van
  Blaaderen}, \citenamefont {Hollingsworth}, \citenamefont {Sullivan},\ and\
  \citenamefont {Chaikin}}]{leunissen_pnas_2007}%
  \BibitemOpen
  \bibfield  {author} {\bibinfo {author} {\bibfnamefont {M.~E.}\ \bibnamefont
  {Leunissen}}, \bibinfo {author} {\bibfnamefont {A.}~\bibnamefont {van
  Blaaderen}}, \bibinfo {author} {\bibfnamefont {A.~D.}\ \bibnamefont
  {Hollingsworth}}, \bibinfo {author} {\bibfnamefont {M.~T.}\ \bibnamefont
  {Sullivan}}, \ and\ \bibinfo {author} {\bibfnamefont {P.~M.}\ \bibnamefont
  {Chaikin}},\ }\href {\doibase 10.1073/pnas.0610589104} {\bibfield  {journal}
  {\bibinfo  {journal} {Proc. Natl. Acad. Sci. U.S.A.}\ }\textbf {\bibinfo
  {volume} {104}},\ \bibinfo {pages} {2585} (\bibinfo {year}
  {2007}{\natexlab{a}})}\BibitemShut {NoStop}%
\bibitem [{\citenamefont {Leunissen}\ \emph
  {et~al.}(2007{\natexlab{b}})\citenamefont {Leunissen}, \citenamefont
  {Zwanikken}, \citenamefont {van Roij}, \citenamefont {Chaikin},\ and\
  \citenamefont {van Blaaderen}}]{leunissen_pccp_2007}%
  \BibitemOpen
  \bibfield  {author} {\bibinfo {author} {\bibfnamefont {M.~E.}\ \bibnamefont
  {Leunissen}}, \bibinfo {author} {\bibfnamefont {J.}~\bibnamefont
  {Zwanikken}}, \bibinfo {author} {\bibfnamefont {R.}~\bibnamefont {van Roij}},
  \bibinfo {author} {\bibfnamefont {P.~M.}\ \bibnamefont {Chaikin}}, \ and\
  \bibinfo {author} {\bibfnamefont {A.}~\bibnamefont {van Blaaderen}},\ }\href
  {\doibase 10.1039/B711300E} {\bibfield  {journal} {\bibinfo  {journal} {Phys.
  Chem. Chem. Phys.}\ }\textbf {\bibinfo {volume} {9}},\ \bibinfo {pages}
  {6405} (\bibinfo {year} {2007}{\natexlab{b}})}\BibitemShut {NoStop}%
\bibitem [{\citenamefont {Zwanikken}\ and\ \citenamefont {van
  Roij}(2007)}]{zwanikken_prl_2007}%
  \BibitemOpen
  \bibfield  {author} {\bibinfo {author} {\bibfnamefont {J.}~\bibnamefont
  {Zwanikken}}\ and\ \bibinfo {author} {\bibfnamefont {R.}~\bibnamefont {van
  Roij}},\ }\href {\doibase 10.1103/PhysRevLett.99.178301} {\bibfield
  {journal} {\bibinfo  {journal} {Phys. Rev. Lett.}\ }\textbf {\bibinfo
  {volume} {99}},\ \bibinfo {pages} {178301} (\bibinfo {year}
  {2007})}\BibitemShut {NoStop}%
\bibitem [{\citenamefont {Beysens}\ and\ \citenamefont
  {Est{\`e}ve}(1985)}]{beysens_prl_1985}%
  \BibitemOpen
  \bibfield  {author} {\bibinfo {author} {\bibfnamefont {D.}~\bibnamefont
  {Beysens}}\ and\ \bibinfo {author} {\bibfnamefont {D.}~\bibnamefont
  {Est{\`e}ve}},\ }\href {\doibase 10.1103/PhysRevLett.54.2123} {\bibfield
  {journal} {\bibinfo  {journal} {Phys. Rev. Lett.}\ }\textbf {\bibinfo
  {volume} {54}},\ \bibinfo {pages} {2123} (\bibinfo {year}
  {1985})}\BibitemShut {NoStop}%
\bibitem [{\citenamefont {Hertlein}\ \emph {et~al.}(2008)\citenamefont
  {Hertlein}, \citenamefont {Helden}, \citenamefont {Gambassi}, \citenamefont
  {Dietrich},\ and\ \citenamefont {Bechinger}}]{bechinger_nature_2008}%
  \BibitemOpen
  \bibfield  {author} {\bibinfo {author} {\bibfnamefont {C.}~\bibnamefont
  {Hertlein}}, \bibinfo {author} {\bibfnamefont {L.}~\bibnamefont {Helden}},
  \bibinfo {author} {\bibfnamefont {A.}~\bibnamefont {Gambassi}}, \bibinfo
  {author} {\bibfnamefont {S.}~\bibnamefont {Dietrich}}, \ and\ \bibinfo
  {author} {\bibfnamefont {C.}~\bibnamefont {Bechinger}},\ }\href@noop {}
  {\bibfield  {journal} {\bibinfo  {journal} {Nature (London)}\ }\textbf
  {\bibinfo {volume} {451}},\ \bibinfo {pages} {172} (\bibinfo {year}
  {2008})}\BibitemShut {NoStop}%
\bibitem [{\citenamefont {Nellen}\ \emph {et~al.}(2011)\citenamefont {Nellen},
  \citenamefont {Dietrich}, \citenamefont {Helden}, \citenamefont {Chodankar},
  \citenamefont {Nyg{\aa}rd}, \citenamefont {van~der Veen},\ and\ \citenamefont
  {Bechinger}}]{nellen_softmatter_2011}%
  \BibitemOpen
  \bibfield  {author} {\bibinfo {author} {\bibfnamefont {U.}~\bibnamefont
  {Nellen}}, \bibinfo {author} {\bibfnamefont {J.}~\bibnamefont {Dietrich}},
  \bibinfo {author} {\bibfnamefont {L.}~\bibnamefont {Helden}}, \bibinfo
  {author} {\bibfnamefont {S.}~\bibnamefont {Chodankar}}, \bibinfo {author}
  {\bibfnamefont {K.}~\bibnamefont {Nyg{\aa}rd}}, \bibinfo {author}
  {\bibfnamefont {J.~F.}\ \bibnamefont {van~der Veen}}, \ and\ \bibinfo
  {author} {\bibfnamefont {C.}~\bibnamefont {Bechinger}},\ }\href@noop {}
  {\bibfield  {journal} {\bibinfo  {journal} {Soft Matter}\ }\textbf {\bibinfo
  {volume} {7}},\ \bibinfo {pages} {5360} (\bibinfo {year} {2011})}\BibitemShut
  {NoStop}%
\bibitem [{\citenamefont {Hopkins}\ \emph {et~al.}(2009)\citenamefont
  {Hopkins}, \citenamefont {Archer},\ and\ \citenamefont
  {Evans}}]{evans_jcp_2009}%
  \BibitemOpen
  \bibfield  {author} {\bibinfo {author} {\bibfnamefont {P.}~\bibnamefont
  {Hopkins}}, \bibinfo {author} {\bibfnamefont {A.~J.}\ \bibnamefont {Archer}},
  \ and\ \bibinfo {author} {\bibfnamefont {R.}~\bibnamefont {Evans}},\ }\href
  {\doibase 10.1063/1.3212888} {\bibfield  {journal} {\bibinfo  {journal} {J.
  Chem. Phys.}\ }\textbf {\bibinfo {volume} {131}},\ \bibinfo {eid} {124704}
  (\bibinfo {year} {2009})}\BibitemShut {NoStop}%
\bibitem [{\citenamefont {van Duijneveldt}\ and\ \citenamefont
  {Beysens}(1991)}]{beysens_jcp_1991}%
  \BibitemOpen
  \bibfield  {author} {\bibinfo {author} {\bibfnamefont {J.~S.}\ \bibnamefont
  {van Duijneveldt}}\ and\ \bibinfo {author} {\bibfnamefont {D.}~\bibnamefont
  {Beysens}},\ }\href {\doibase 10.1063/1.460526} {\bibfield  {journal}
  {\bibinfo  {journal} {J. Chem. Phys.}\ }\textbf {\bibinfo {volume} {94}},\
  \bibinfo {pages} {5222} (\bibinfo {year} {1991})}\BibitemShut {NoStop}%
\bibitem [{\citenamefont {Law}\ \emph {et~al.}(1998)\citenamefont {Law},
  \citenamefont {Petit},\ and\ \citenamefont {Beysens}}]{beysens_pre_1998}%
  \BibitemOpen
  \bibfield  {author} {\bibinfo {author} {\bibfnamefont {B.~M.}\ \bibnamefont
  {Law}}, \bibinfo {author} {\bibfnamefont {J.-M.}\ \bibnamefont {Petit}}, \
  and\ \bibinfo {author} {\bibfnamefont {D.}~\bibnamefont {Beysens}},\
  }\href@noop {} {\bibfield  {journal} {\bibinfo  {journal} {Phys. Rev. E}\
  }\textbf {\bibinfo {volume} {57}},\ \bibinfo {pages} {5782} (\bibinfo {year}
  {1998})}\BibitemShut {NoStop}%
\bibitem [{\citenamefont {Beysens}\ and\ \citenamefont
  {Narayanan}(1999)}]{beysens_jsp_1999}%
  \BibitemOpen
  \bibfield  {author} {\bibinfo {author} {\bibfnamefont {D.}~\bibnamefont
  {Beysens}}\ and\ \bibinfo {author} {\bibfnamefont {T.}~\bibnamefont
  {Narayanan}},\ }\href@noop {} {\bibfield  {journal} {\bibinfo  {journal} {J.
  Stat. Phys.}\ }\textbf {\bibinfo {volume} {95}},\ \bibinfo {pages} {997}
  (\bibinfo {year} {1999})}\BibitemShut {NoStop}%
\bibitem [{\citenamefont {Ben-Yaakov}\ \emph {et~al.}(2011)\citenamefont
  {Ben-Yaakov}, \citenamefont {Andelman}, \citenamefont {Podgornik},\ and\
  \citenamefont {Harries}}]{andelman_cocis_2011}%
  \BibitemOpen
  \bibfield  {author} {\bibinfo {author} {\bibfnamefont {D.}~\bibnamefont
  {Ben-Yaakov}}, \bibinfo {author} {\bibfnamefont {D.}~\bibnamefont
  {Andelman}}, \bibinfo {author} {\bibfnamefont {R.}~\bibnamefont {Podgornik}},
  \ and\ \bibinfo {author} {\bibfnamefont {D.}~\bibnamefont {Harries}},\
  }\href@noop {} {\bibfield  {journal} {\bibinfo  {journal} {Curr. Opin. Coll.
  \& Interface Sci.}\ }\textbf {\bibinfo {volume} {16}},\ \bibinfo {pages}
  {542} (\bibinfo {year} {2011})}\BibitemShut {NoStop}%
\bibitem [{\citenamefont {Zwanikken}\ \emph {et~al.}(2008)\citenamefont
  {Zwanikken}, \citenamefont {de~Graaf}, \citenamefont {Bier},\ and\
  \citenamefont {van Roij}}]{zwanikken_jpcm_2008}%
  \BibitemOpen
  \bibfield  {author} {\bibinfo {author} {\bibfnamefont {J.}~\bibnamefont
  {Zwanikken}}, \bibinfo {author} {\bibfnamefont {J.}~\bibnamefont {de~Graaf}},
  \bibinfo {author} {\bibfnamefont {M.}~\bibnamefont {Bier}}, \ and\ \bibinfo
  {author} {\bibfnamefont {R.}~\bibnamefont {van Roij}},\ }\href
  {http://stacks.iop.org/0953-8984/20/i=49/a=494238} {\bibfield  {journal}
  {\bibinfo  {journal} {J. Phys.: Condens. Matter}\ }\textbf {\bibinfo {volume}
  {20}},\ \bibinfo {pages} {494238} (\bibinfo {year} {2008})}\BibitemShut
  {NoStop}%
\bibitem [{\citenamefont {Samin}\ and\ \citenamefont
  {Tsori}(2011{\natexlab{b}})}]{tsori_epl_2011}%
  \BibitemOpen
  \bibfield  {author} {\bibinfo {author} {\bibfnamefont {S.}~\bibnamefont
  {Samin}}\ and\ \bibinfo {author} {\bibfnamefont {Y.}~\bibnamefont {Tsori}},\
  }\href@noop {} {\bibfield  {journal} {\bibinfo  {journal} {EPL}\ }\textbf
  {\bibinfo {volume} {95}},\ \bibinfo {pages} {36002} (\bibinfo {year}
  {2011}{\natexlab{b}})}\BibitemShut {NoStop}%
\bibitem [{\citenamefont {Okamoto}\ and\ \citenamefont
  {Onuki}(2011)}]{onuki_pre_2011}%
  \BibitemOpen
  \bibfield  {author} {\bibinfo {author} {\bibfnamefont {R.}~\bibnamefont
  {Okamoto}}\ and\ \bibinfo {author} {\bibfnamefont {A.}~\bibnamefont
  {Onuki}},\ }\href@noop {} {\bibfield  {journal} {\bibinfo  {journal} {Phys.
  Rev. E}\ }\textbf {\bibinfo {volume} {84}},\ \bibinfo {pages} {051401}
  (\bibinfo {year} {2011})}\BibitemShut {NoStop}%
\bibitem [{\citenamefont {Samin}\ and\ \citenamefont
  {Tsori}(2012)}]{tsori_jcp_2012}%
  \BibitemOpen
  \bibfield  {author} {\bibinfo {author} {\bibfnamefont {S.}~\bibnamefont
  {Samin}}\ and\ \bibinfo {author} {\bibfnamefont {Y.}~\bibnamefont {Tsori}},\
  }\href@noop {} {\bibfield  {journal} {\bibinfo  {journal} {J. Chem. Phys.}\
  }\textbf {\bibinfo {volume} {136}},\ \bibinfo {eid} {154908} (\bibinfo {year}
  {2012})}\BibitemShut {NoStop}%
\bibitem [{\citenamefont {Bier}\ \emph {et~al.}(2011)\citenamefont {Bier},
  \citenamefont {Gambassi}, \citenamefont {Oettel},\ and\ \citenamefont
  {Dietrich}}]{bier_epl_2011}%
  \BibitemOpen
  \bibfield  {author} {\bibinfo {author} {\bibfnamefont {M.}~\bibnamefont
  {Bier}}, \bibinfo {author} {\bibfnamefont {A.}~\bibnamefont {Gambassi}},
  \bibinfo {author} {\bibfnamefont {M.}~\bibnamefont {Oettel}}, \ and\ \bibinfo
  {author} {\bibfnamefont {S.}~\bibnamefont {Dietrich}},\ }\href@noop {}
  {\bibfield  {journal} {\bibinfo  {journal} {EPL}\ }\textbf {\bibinfo {volume}
  {95}},\ \bibinfo {pages} {60001} (\bibinfo {year} {2011})}\BibitemShut
  {NoStop}%
\bibitem [{\citenamefont {Sadakane}\ \emph {et~al.}(2006)\citenamefont
  {Sadakane}, \citenamefont {Seto},\ and\ \citenamefont
  {Nagao}}]{sadakane_cpl_2006}%
  \BibitemOpen
  \bibfield  {author} {\bibinfo {author} {\bibfnamefont {K.}~\bibnamefont
  {Sadakane}}, \bibinfo {author} {\bibfnamefont {H.}~\bibnamefont {Seto}}, \
  and\ \bibinfo {author} {\bibfnamefont {M.}~\bibnamefont {Nagao}},\ }\href
  {\doibase 10.1016/j.cplett.2006.05.028} {\bibfield  {journal} {\bibinfo
  {journal} {Chem. Phys. Lett.}\ }\textbf {\bibinfo {volume} {426}},\ \bibinfo
  {pages} {61} (\bibinfo {year} {2006})}\BibitemShut {NoStop}%
\bibitem [{\citenamefont {Sadakane}\ \emph {et~al.}(2007)\citenamefont
  {Sadakane}, \citenamefont {Seto}, \citenamefont {Endo},\ and\ \citenamefont
  {Shibayama}}]{sadakane_jpsj_2007}%
  \BibitemOpen
  \bibfield  {author} {\bibinfo {author} {\bibfnamefont {K.}~\bibnamefont
  {Sadakane}}, \bibinfo {author} {\bibfnamefont {H.}~\bibnamefont {Seto}},
  \bibinfo {author} {\bibfnamefont {H.}~\bibnamefont {Endo}}, \ and\ \bibinfo
  {author} {\bibfnamefont {M.}~\bibnamefont {Shibayama}},\ }\href@noop {}
  {\bibfield  {journal} {\bibinfo  {journal} {J. Phys. Soc. Jpn.}\ }\textbf
  {\bibinfo {volume} {76}},\ \bibinfo {pages} {113602} (\bibinfo {year}
  {2007})}\BibitemShut {NoStop}%
\bibitem [{\citenamefont {Sadakane}\ \emph {et~al.}(2011)\citenamefont
  {Sadakane}, \citenamefont {Iguchi}, \citenamefont {Nagao}, \citenamefont
  {Endo}, \citenamefont {Melnichenko},\ and\ \citenamefont
  {Seto}}]{sadakane_softmatter_2011}%
  \BibitemOpen
  \bibfield  {author} {\bibinfo {author} {\bibfnamefont {K.}~\bibnamefont
  {Sadakane}}, \bibinfo {author} {\bibfnamefont {N.}~\bibnamefont {Iguchi}},
  \bibinfo {author} {\bibfnamefont {M.}~\bibnamefont {Nagao}}, \bibinfo
  {author} {\bibfnamefont {H.}~\bibnamefont {Endo}}, \bibinfo {author}
  {\bibfnamefont {Y.~B.}\ \bibnamefont {Melnichenko}}, \ and\ \bibinfo {author}
  {\bibfnamefont {H.}~\bibnamefont {Seto}},\ }\href@noop {} {\bibfield
  {journal} {\bibinfo  {journal} {Soft Matter}\ }\textbf {\bibinfo {volume}
  {7}},\ \bibinfo {pages} {1334} (\bibinfo {year} {2011})}\BibitemShut
  {NoStop}%
\bibitem [{\citenamefont {Sadakane}\ \emph {et~al.}(2009)\citenamefont
  {Sadakane}, \citenamefont {Onuki}, \citenamefont {Nishida}, \citenamefont
  {Koizumi},\ and\ \citenamefont {Seto}}]{sadakane_prl_2009}%
  \BibitemOpen
  \bibfield  {author} {\bibinfo {author} {\bibfnamefont {K.}~\bibnamefont
  {Sadakane}}, \bibinfo {author} {\bibfnamefont {A.}~\bibnamefont {Onuki}},
  \bibinfo {author} {\bibfnamefont {K.}~\bibnamefont {Nishida}}, \bibinfo
  {author} {\bibfnamefont {S.}~\bibnamefont {Koizumi}}, \ and\ \bibinfo
  {author} {\bibfnamefont {H.}~\bibnamefont {Seto}},\ }\href {\doibase
  10.1103/PhysRevLett.103.167803} {\bibfield  {journal} {\bibinfo  {journal}
  {Phys. Rev. Lett.}\ }\textbf {\bibinfo {volume} {103}},\ \bibinfo {pages}
  {167803} (\bibinfo {year} {2009})}\BibitemShut {NoStop}%
\bibitem [{\citenamefont {Tsori}\ and\ \citenamefont
  {Leibler}(2007)}]{tsori_pnas_2007}%
  \BibitemOpen
  \bibfield  {author} {\bibinfo {author} {\bibfnamefont {Y.}~\bibnamefont
  {Tsori}}\ and\ \bibinfo {author} {\bibfnamefont {L.}~\bibnamefont
  {Leibler}},\ }\href {\doibase 10.1073/pnas.0607746104} {\bibfield  {journal}
  {\bibinfo  {journal} {Proc. Nat. Acad. Sci.}\ }\textbf {\bibinfo {volume}
  {104}},\ \bibinfo {pages} {7348} (\bibinfo {year} {2007})}\BibitemShut
  {NoStop}%
\bibitem [{\citenamefont {Ben-Yaakov}\ \emph {et~al.}(2009)\citenamefont
  {Ben-Yaakov}, \citenamefont {Andelman}, \citenamefont {Harries},\ and\
  \citenamefont {Podgornik}}]{andelman_jpcm_2009}%
  \BibitemOpen
  \bibfield  {author} {\bibinfo {author} {\bibfnamefont {D.}~\bibnamefont
  {Ben-Yaakov}}, \bibinfo {author} {\bibfnamefont {D.}~\bibnamefont
  {Andelman}}, \bibinfo {author} {\bibfnamefont {D.}~\bibnamefont {Harries}}, \
  and\ \bibinfo {author} {\bibfnamefont {R.}~\bibnamefont {Podgornik}},\
  }\href@noop {} {\bibfield  {journal} {\bibinfo  {journal} {J. Phys.: Condens.
  Matter}\ }\textbf {\bibinfo {volume} {21}},\ \bibinfo {pages} {424106.1}
  (\bibinfo {year} {2009})}\BibitemShut {NoStop}%
\bibitem [{\citenamefont {Bier}\ \emph {et~al.}(2012)\citenamefont {Bier},
  \citenamefont {Gambassi},\ and\ \citenamefont {Dietrich}}]{bier_jcp_2012}%
  \BibitemOpen
  \bibfield  {author} {\bibinfo {author} {\bibfnamefont {M.}~\bibnamefont
  {Bier}}, \bibinfo {author} {\bibfnamefont {A.}~\bibnamefont {Gambassi}}, \
  and\ \bibinfo {author} {\bibfnamefont {S.}~\bibnamefont {Dietrich}},\ }\href
  {\doibase 10.1063/1.4733973} {\bibfield  {journal} {\bibinfo  {journal} {J.
  Chem. Phys.}\ }\textbf {\bibinfo {volume} {137}},\ \bibinfo {eid} {034504}
  (\bibinfo {year} {2012})}\BibitemShut {NoStop}%
\bibitem [{\citenamefont {Samin}\ \emph {et~al.}(2014)\citenamefont {Samin},
  \citenamefont {Hod}, \citenamefont {Melamed}, \citenamefont {Gottlieb},\ and\
  \citenamefont {Tsori}}]{tsori_prapplied_2014}%
  \BibitemOpen
  \bibfield  {author} {\bibinfo {author} {\bibfnamefont {S.}~\bibnamefont
  {Samin}}, \bibinfo {author} {\bibfnamefont {M.}~\bibnamefont {Hod}}, \bibinfo
  {author} {\bibfnamefont {E.}~\bibnamefont {Melamed}}, \bibinfo {author}
  {\bibfnamefont {M.}~\bibnamefont {Gottlieb}}, \ and\ \bibinfo {author}
  {\bibfnamefont {Y.}~\bibnamefont {Tsori}},\ }\href {\doibase
  10.1103/physrevapplied.2.024008} {\bibfield  {journal} {\bibinfo  {journal}
  {Phys. Rev. Appl}\ }\textbf {\bibinfo {volume} {2}},\ \bibinfo {pages}
  {024008} (\bibinfo {year} {2014})}\BibitemShut {NoStop}%
\bibitem [{\citenamefont {Samin}\ and\ \citenamefont
  {Tsori}(2013)}]{tsori_jcp_2013}%
  \BibitemOpen
  \bibfield  {author} {\bibinfo {author} {\bibfnamefont {S.}~\bibnamefont
  {Samin}}\ and\ \bibinfo {author} {\bibfnamefont {Y.}~\bibnamefont {Tsori}},\
  }\href {\doibase 10.1063/1.4851477} {\bibfield  {journal} {\bibinfo
  {journal} {J. Chem. Phys.}\ }\textbf {\bibinfo {volume} {139}},\ \bibinfo
  {pages} {244905} (\bibinfo {year} {2013})}\BibitemShut {NoStop}%
\bibitem [{\citenamefont {Pousaneh}\ and\ \citenamefont
  {Ciach}(2014)}]{pousaneh_softmatter_2014}%
  \BibitemOpen
  \bibfield  {author} {\bibinfo {author} {\bibfnamefont {F.}~\bibnamefont
  {Pousaneh}}\ and\ \bibinfo {author} {\bibfnamefont {A.}~\bibnamefont
  {Ciach}},\ }\href {\doibase 10.1039/C4SM01264J} {\bibfield  {journal}
  {\bibinfo  {journal} {Soft Matter}\ }\textbf {\bibinfo {volume} {10}},\
  \bibinfo {pages} {8188} (\bibinfo {year} {2014})}\BibitemShut {NoStop}%
\bibitem [{\citenamefont {Kedem}\ and\ \citenamefont
  {Katchalsky}(1958)}]{katchalsky_bpa_1958}%
  \BibitemOpen
  \bibfield  {author} {\bibinfo {author} {\bibfnamefont {O.}~\bibnamefont
  {Kedem}}\ and\ \bibinfo {author} {\bibfnamefont {A.}~\bibnamefont
  {Katchalsky}},\ }\href {\doibase 10.1016/0006-3002(58)90330-5} {\bibfield
  {journal} {\bibinfo  {journal} {Biochim. Biophys. Acta}\ }\textbf {\bibinfo
  {volume} {27}},\ \bibinfo {pages} {229} (\bibinfo {year} {1958})}\BibitemShut
  {NoStop}%
\bibitem [{\citenamefont {Jiang}\ \emph {et~al.}(2002)\citenamefont {Jiang},
  \citenamefont {Lee}, \citenamefont {Chen}, \citenamefont {Cadene},
  \citenamefont {Chait},\ and\ \citenamefont
  {MacKinnon}}]{mackinnon_nature_2002}%
  \BibitemOpen
  \bibfield  {author} {\bibinfo {author} {\bibfnamefont {Y.}~\bibnamefont
  {Jiang}}, \bibinfo {author} {\bibfnamefont {A.}~\bibnamefont {Lee}}, \bibinfo
  {author} {\bibfnamefont {J.}~\bibnamefont {Chen}}, \bibinfo {author}
  {\bibfnamefont {M.}~\bibnamefont {Cadene}}, \bibinfo {author} {\bibfnamefont
  {B.}~\bibnamefont {Chait}}, \ and\ \bibinfo {author} {\bibfnamefont
  {R.}~\bibnamefont {MacKinnon}},\ }\href {\doibase {10.1038/417523a}}
  {\bibfield  {journal} {\bibinfo  {journal} {Nature}\ }\textbf {\bibinfo
  {volume} {417}},\ \bibinfo {pages} {523} (\bibinfo {year}
  {2002})}\BibitemShut {NoStop}%
\bibitem [{\citenamefont {Jiang}\ \emph {et~al.}(2003)\citenamefont {Jiang},
  \citenamefont {Lee}, \citenamefont {Chen}, \citenamefont {Ruta},
  \citenamefont {Cadene}, \citenamefont {Chait},\ and\ \citenamefont
  {MacKinnon}}]{mackinnon_nature_2003}%
  \BibitemOpen
  \bibfield  {author} {\bibinfo {author} {\bibfnamefont {Y.}~\bibnamefont
  {Jiang}}, \bibinfo {author} {\bibfnamefont {A.}~\bibnamefont {Lee}}, \bibinfo
  {author} {\bibfnamefont {J.}~\bibnamefont {Chen}}, \bibinfo {author}
  {\bibfnamefont {V.}~\bibnamefont {Ruta}}, \bibinfo {author} {\bibfnamefont
  {M.}~\bibnamefont {Cadene}}, \bibinfo {author} {\bibfnamefont
  {B.}~\bibnamefont {Chait}}, \ and\ \bibinfo {author} {\bibfnamefont
  {R.}~\bibnamefont {MacKinnon}},\ }\href {\doibase {10.1038/nature01580}}
  {\bibfield  {journal} {\bibinfo  {journal} {Nature}\ }\textbf {\bibinfo
  {volume} {423}},\ \bibinfo {pages} {33} (\bibinfo {year} {2003})}\BibitemShut
  {NoStop}%
\bibitem [{\citenamefont {Pendergast}\ and\ \citenamefont
  {Hoek}(2011)}]{hoek_ees_2011}%
  \BibitemOpen
  \bibfield  {author} {\bibinfo {author} {\bibfnamefont {M.~M.}\ \bibnamefont
  {Pendergast}}\ and\ \bibinfo {author} {\bibfnamefont {E.~M.~V.}\ \bibnamefont
  {Hoek}},\ }\href {\doibase 10.1039/c0ee00541j} {\bibfield  {journal}
  {\bibinfo  {journal} {Energy \& Environmental Science}\ }\textbf {\bibinfo
  {volume} {4}},\ \bibinfo {pages} {1946} (\bibinfo {year} {2011})}\BibitemShut
  {NoStop}%
\bibitem [{\citenamefont {Yang}\ and\ \citenamefont
  {Yang}(2003)}]{yang_jms_2003}%
  \BibitemOpen
  \bibfield  {author} {\bibinfo {author} {\bibfnamefont {B.}~\bibnamefont
  {Yang}}\ and\ \bibinfo {author} {\bibfnamefont {W.~T.}\ \bibnamefont
  {Yang}},\ }\href {\doibase 10.1016/s0376-7388(03)00182-0} {\bibfield
  {journal} {\bibinfo  {journal} {J. Membr. Sci.}\ }\textbf {\bibinfo {volume}
  {218}},\ \bibinfo {pages} {247} (\bibinfo {year} {2003})},\ \bibinfo {note}
  {86}\BibitemShut {NoStop}%
\bibitem [{\citenamefont {Yameen}\ \emph {et~al.}(2009)\citenamefont {Yameen},
  \citenamefont {Ali}, \citenamefont {Neumann}, \citenamefont {Ensinger},
  \citenamefont {Knoll},\ and\ \citenamefont {Azzaroni}}]{azzaroni_small_2009}%
  \BibitemOpen
  \bibfield  {author} {\bibinfo {author} {\bibfnamefont {B.}~\bibnamefont
  {Yameen}}, \bibinfo {author} {\bibfnamefont {M.}~\bibnamefont {Ali}},
  \bibinfo {author} {\bibfnamefont {R.}~\bibnamefont {Neumann}}, \bibinfo
  {author} {\bibfnamefont {W.}~\bibnamefont {Ensinger}}, \bibinfo {author}
  {\bibfnamefont {W.}~\bibnamefont {Knoll}}, \ and\ \bibinfo {author}
  {\bibfnamefont {O.}~\bibnamefont {Azzaroni}},\ }\href {\doibase
  10.1002/smll.200801318} {\bibfield  {journal} {\bibinfo  {journal} {Small}\
  }\textbf {\bibinfo {volume} {5}},\ \bibinfo {pages} {1287} (\bibinfo {year}
  {2009})},\ \bibinfo {note} {87}\BibitemShut {NoStop}%
\bibitem [{\citenamefont {Hou}\ \emph {et~al.}(2015)\citenamefont {Hou},
  \citenamefont {Hu}, \citenamefont {Grinthal}, \citenamefont {Khan},\ and\
  \citenamefont {Aizenberg}}]{aizenberg_nature_2015}%
  \BibitemOpen
  \bibfield  {author} {\bibinfo {author} {\bibfnamefont {X.}~\bibnamefont
  {Hou}}, \bibinfo {author} {\bibfnamefont {Y.}~\bibnamefont {Hu}}, \bibinfo
  {author} {\bibfnamefont {A.}~\bibnamefont {Grinthal}}, \bibinfo {author}
  {\bibfnamefont {M.}~\bibnamefont {Khan}}, \ and\ \bibinfo {author}
  {\bibfnamefont {J.}~\bibnamefont {Aizenberg}},\ }\href {\doibase
  10.1038/nature14253} {\bibfield  {journal} {\bibinfo  {journal} {Nature}\
  }\textbf {\bibinfo {volume} {519}},\ \bibinfo {pages} {70} (\bibinfo {year}
  {2015})},\ \bibinfo {note} {10}\BibitemShut {NoStop}%
\bibitem [{\citenamefont {Dzubiella}\ \emph {et~al.}(2004)\citenamefont
  {Dzubiella}, \citenamefont {Allen},\ and\ \citenamefont
  {Hansen}}]{hansen_jcp_2004}%
  \BibitemOpen
  \bibfield  {author} {\bibinfo {author} {\bibfnamefont {J.}~\bibnamefont
  {Dzubiella}}, \bibinfo {author} {\bibfnamefont {R.~J.}\ \bibnamefont
  {Allen}}, \ and\ \bibinfo {author} {\bibfnamefont {J.~P.}\ \bibnamefont
  {Hansen}},\ }\href {\doibase 10.1063/1.1665656} {\bibfield  {journal}
  {\bibinfo  {journal} {J. Chem. Phys.}\ }\textbf {\bibinfo {volume} {120}},\
  \bibinfo {pages} {5001} (\bibinfo {year} {2004})}\BibitemShut {NoStop}%
\bibitem [{\citenamefont {Dzubiella}\ and\ \citenamefont
  {Hansen}(2005)}]{hansen_jcp_2005}%
  \BibitemOpen
  \bibfield  {author} {\bibinfo {author} {\bibfnamefont {J.}~\bibnamefont
  {Dzubiella}}\ and\ \bibinfo {author} {\bibfnamefont {J.~P.}\ \bibnamefont
  {Hansen}},\ }\href {\doibase 10.1063/1.1927514} {\bibfield  {journal}
  {\bibinfo  {journal} {J. Chem. Phys.}\ }\textbf {\bibinfo {volume} {122}}
  (\bibinfo {year} {2005}),\ 10.1063/1.1927514}\BibitemShut {NoStop}%
\bibitem [{\citenamefont {Powell}\ \emph {et~al.}(2011)\citenamefont {Powell},
  \citenamefont {Cleary}, \citenamefont {Davenport}, \citenamefont {Shea},\
  and\ \citenamefont {Siwy}}]{siwy_nature_nanotech_2011}%
  \BibitemOpen
  \bibfield  {author} {\bibinfo {author} {\bibfnamefont {M.~R.}\ \bibnamefont
  {Powell}}, \bibinfo {author} {\bibfnamefont {L.}~\bibnamefont {Cleary}},
  \bibinfo {author} {\bibfnamefont {M.}~\bibnamefont {Davenport}}, \bibinfo
  {author} {\bibfnamefont {K.~J.}\ \bibnamefont {Shea}}, \ and\ \bibinfo
  {author} {\bibfnamefont {Z.~S.}\ \bibnamefont {Siwy}},\ }\href {\doibase
  10.1038/nnano.2011.189} {\bibfield  {journal} {\bibinfo  {journal} {Nature
  Nanotechnology}\ }\textbf {\bibinfo {volume} {6}},\ \bibinfo {pages} {798}
  (\bibinfo {year} {2011})},\ \bibinfo {note} {81}\BibitemShut {NoStop}%
\bibitem [{\citenamefont {Smirnov}\ \emph {et~al.}(2011)\citenamefont
  {Smirnov}, \citenamefont {Vlassiouk},\ and\ \citenamefont
  {Lavrik}}]{lavrik_acsnano_2011}%
  \BibitemOpen
  \bibfield  {author} {\bibinfo {author} {\bibfnamefont {S.~N.}\ \bibnamefont
  {Smirnov}}, \bibinfo {author} {\bibfnamefont {I.~V.}\ \bibnamefont
  {Vlassiouk}}, \ and\ \bibinfo {author} {\bibfnamefont {N.~V.}\ \bibnamefont
  {Lavrik}},\ }\href {\doibase 10.1021/nn202392d} {\bibfield  {journal}
  {\bibinfo  {journal} {Acs Nano}\ }\textbf {\bibinfo {volume} {5}},\ \bibinfo
  {pages} {7453} (\bibinfo {year} {2011})}\BibitemShut {NoStop}%
\bibitem [{\citenamefont {Samin}\ and\ \citenamefont
  {Tsori}(2016)}]{tsori_cisc_2016}%
  \BibitemOpen
  \bibfield  {author} {\bibinfo {author} {\bibfnamefont {S.}~\bibnamefont
  {Samin}}\ and\ \bibinfo {author} {\bibfnamefont {Y.}~\bibnamefont {Tsori}},\
  }\href@noop {} {\bibfield  {journal} {\bibinfo  {journal} {Colloid Interface
  Sci. Commun.}\ }\textbf {\bibinfo {volume} {12}},\ \bibinfo {pages} {9}
  (\bibinfo {year} {2016})}\BibitemShut {NoStop}%
\bibitem [{\citenamefont {Okamoto}\ and\ \citenamefont
  {Onuki}(2010)}]{onuki_pre_2010}%
  \BibitemOpen
  \bibfield  {author} {\bibinfo {author} {\bibfnamefont {R.}~\bibnamefont
  {Okamoto}}\ and\ \bibinfo {author} {\bibfnamefont {A.}~\bibnamefont
  {Onuki}},\ }\href {\doibase 10.1103/PhysRevE.82.051501} {\bibfield  {journal}
  {\bibinfo  {journal} {Phys. Rev. E}\ }\textbf {\bibinfo {volume} {82}},\
  \bibinfo {pages} {051501} (\bibinfo {year} {2010})}\BibitemShut {NoStop}%
\bibitem [{\citenamefont {Borukhov}\ \emph {et~al.}(1997)\citenamefont
  {Borukhov}, \citenamefont {Andelman},\ and\ \citenamefont
  {Orland}}]{borukhov_prl_1997}%
  \BibitemOpen
  \bibfield  {author} {\bibinfo {author} {\bibfnamefont {I.}~\bibnamefont
  {Borukhov}}, \bibinfo {author} {\bibfnamefont {D.}~\bibnamefont {Andelman}},
  \ and\ \bibinfo {author} {\bibfnamefont {H.}~\bibnamefont {Orland}},\
  }\href@noop {} {\bibfield  {journal} {\bibinfo  {journal} {Phys. Rev. Lett.}\
  }\textbf {\bibinfo {volume} {79}},\ \bibinfo {pages} {435} (\bibinfo {year}
  {1997})}\BibitemShut {NoStop}%
\bibitem [{\citenamefont {Bazant}\ and\ \citenamefont
  {Squires}(2010)}]{bazant_cocis_2010}%
  \BibitemOpen
  \bibfield  {author} {\bibinfo {author} {\bibfnamefont {M.~Z.}\ \bibnamefont
  {Bazant}}\ and\ \bibinfo {author} {\bibfnamefont {T.~M.}\ \bibnamefont
  {Squires}},\ }\href {\doibase {10.1016/j.cocis.2010.01.003}} {\bibfield
  {journal} {\bibinfo  {journal} {Current Opinion in Colloid \& Interface
  Science}\ }\textbf {\bibinfo {volume} {15}},\ \bibinfo {pages} {203}
  (\bibinfo {year} {2010})}\BibitemShut {NoStop}%
\bibitem [{\citenamefont {Chen}(2011)}]{chen_book_chapter}%
  \BibitemOpen
  \bibfield  {author} {\bibinfo {author} {\bibfnamefont {C.-H.}\ \bibnamefont
  {Chen}},\ }in\ \href@noop {} {\emph {\bibinfo {booktitle} {Electrokinetics
  And Electrohydrodynamics In Microsystems}}},\ \bibinfo {series} {CISM Courses
  and Lectures}, Vol.\ \bibinfo {volume} {530},\ \bibinfo {editor} {edited by\
  \bibinfo {editor} {\bibfnamefont {A.}~\bibnamefont {Ramos}}}\ (\bibinfo
  {publisher} {Springer-Verlag, Wien},\ \bibinfo {year} {2011})\ pp.\ \bibinfo
  {pages} {177--220},\ \bibinfo {note} {conference on Electrokinetics and
  Electrohydrodynamics in Microsystems, Udine, ITALY, JUN 22-26,
  2009}\BibitemShut {NoStop}%
\bibitem [{\citenamefont {Schnitzer}\ and\ \citenamefont
  {Yariv}(2015)}]{yariv_jfm_2015}%
  \BibitemOpen
  \bibfield  {author} {\bibinfo {author} {\bibfnamefont {O.}~\bibnamefont
  {Schnitzer}}\ and\ \bibinfo {author} {\bibfnamefont {E.}~\bibnamefont
  {Yariv}},\ }\href@noop {} {\bibfield  {journal} {\bibinfo  {journal} {J.
  Fluid Mech.}\ }\textbf {\bibinfo {volume} {773}},\ \bibinfo {pages} {1}
  (\bibinfo {year} {2015})}\BibitemShut {NoStop}%
\bibitem [{\citenamefont {Rubinstein}\ and\ \citenamefont
  {Zaltzman}(2015)}]{zaltzman_prl_2015}%
  \BibitemOpen
  \bibfield  {author} {\bibinfo {author} {\bibfnamefont {I.}~\bibnamefont
  {Rubinstein}}\ and\ \bibinfo {author} {\bibfnamefont {B.}~\bibnamefont
  {Zaltzman}},\ }\href {\doibase 10.1103/PhysRevLett.114.114502} {\bibfield
  {journal} {\bibinfo  {journal} {Phys. Rev. Lett.}\ }\textbf {\bibinfo
  {volume} {114}},\ \bibinfo {pages} {114502} (\bibinfo {year}
  {2015})}\BibitemShut {NoStop}%
\bibitem [{\citenamefont {Zwanikken}\ and\ \citenamefont {{Olvera de la
  Cruz}}(2013)}]{cruz_pnas_2013}%
  \BibitemOpen
  \bibfield  {author} {\bibinfo {author} {\bibfnamefont {J.~W.}\ \bibnamefont
  {Zwanikken}}\ and\ \bibinfo {author} {\bibfnamefont {M.}~\bibnamefont
  {{Olvera de la Cruz}}},\ }\href {\doibase 10.1073/pnas.1302406110} {\bibfield
   {journal} {\bibinfo  {journal} {Proc. Nat. Acad. Sci.}\ }\textbf {\bibinfo
  {volume} {110}},\ \bibinfo {pages} {5301} (\bibinfo {year}
  {2013})}\BibitemShut {NoStop}%
\bibitem [{\citenamefont {Onuki}\ \emph {et~al.}(2016)\citenamefont {Onuki},
  \citenamefont {Yabunaka}, \citenamefont {Araki},\ and\ \citenamefont
  {Okamoto}}]{onuki_cocis_2016}%
  \BibitemOpen
  \bibfield  {author} {\bibinfo {author} {\bibfnamefont {A.}~\bibnamefont
  {Onuki}}, \bibinfo {author} {\bibfnamefont {S.}~\bibnamefont {Yabunaka}},
  \bibinfo {author} {\bibfnamefont {T.}~\bibnamefont {Araki}}, \ and\ \bibinfo
  {author} {\bibfnamefont {R.}~\bibnamefont {Okamoto}},\ }\href {\doibase
  {10.1016/j.cocis.2016.02.007}} {\bibfield  {journal} {\bibinfo  {journal}
  {Current Opinion In Colloid \& Interface Science}\ }\textbf {\bibinfo
  {volume} {22}},\ \bibinfo {pages} {59} (\bibinfo {year} {2016})}\BibitemShut
  {NoStop}%
\end{thebibliography}%

\end{document}